\documentclass[11pt]{article}

\usepackage{amscd}
\usepackage{latexsym}
\usepackage{amsthm}
\usepackage{verbatim}
\usepackage[dutch,english]{babel}
\usepackage{amsfonts}
\usepackage{a4wide}
\usepackage{dsfont}
\usepackage{mathtools}
\usepackage{bbm}
\usepackage{lscape}
\usepackage{amsmath,amsfonts,amssymb}
\usepackage{amsmath,amsbsy,amssymb,latexsym, graphics, graphicx, tabularx}
\usepackage{multirow,multicol,subfigure,booktabs}
\usepackage{algorithm}
\usepackage{adjustbox}
\usepackage{lscape}
\usepackage[noend]{algpseudocode}
\usepackage{color}
\usepackage{url}
\usepackage{rotating}
\usepackage{yfonts}
\usepackage{mathrsfs}
\usepackage{enumitem}
\usepackage{booktabs}
\usepackage{caption}
\usepackage{graphicx}
\usepackage{subfig}
\usepackage[abbrev,backrefs]{amsrefs}
\usepackage{cite}
\usepackage{longtable}

\usepackage{fancyhdr}
\DeclareMathOperator*{\argmax}{arg\,max}
\DeclareMathOperator*{\e}{\text{e}}
\DeclareMathOperator*{\I}{\text{I}}
\DeclareMathOperator*{\Proba}{\text{P}}

\newcommand{\C}{{\cal C}}

\newtheorem{Theorem}{Theorem}[section] 

\makeatletter

\@addtoreset{equation}{section}
\makeatother

\begin{document}

\title{\bf Copula based dependent censoring in cure models}

\bigskip
\author{Morine Delhelle$\,^a$
    and 
    Ingrid Van Keilegom$\,^{a,b}$ \footnote{E-mail addresses: morine.delhelle@uclouvain.be, ingrid.vankeilegom@kuleuven.be.} \\
\vspace{0.3cm}\\
    $^a$ Institute of Statistics, Biostatistics and Actuarial Sciences (ISBA),\\
UCLouvain, Voie du Roman Pays 20, bte L1.04.01, Louvain-la-Neuve, 1348, Belgium\\
	$^b$ Research Centre for Operations Research and Statistics (ORSTAT),\\
KU Leuven, Naamsestraat 69 - bus 3555, Leuven, 3000, Belgium}
\date{}
   
\maketitle

\begin{abstract}
In this paper we consider a time-to-event variable $T$ that is subject to random right censoring, and we assume that the censoring time $C$ is stochastically dependent on $T$ and that there is a positive probability of not observing the event. There are various situations in practice where this happens, and appropriate models and methods need to be considered to avoid biased estimators of the survival function or incorrect conclusions in clinical trials. We consider a fully parametric model for the bivariate distribution of $(T,C)$, that takes these features into account. The model depends on a parametric copula (with unknown association parameter) and on parametric marginal distributions for $T$ and $C$. Sufficient conditions are developed under which the model is identified, and an estimation procedure is proposed. In particular, our model allows to identify and estimate the association between $T$ and $C$, even though only the smallest of these variables is observable. The finite sample performance of the estimated parameters is illustrated by means of a thorough simulation study and the analysis of breast cancer data.

\end{abstract}

\bigskip
\noindent%

\def\baselinestretch{1.2}
\newpage
\normalsize
\setcounter{footnote}{0}
\setcounter{equation}{0}

\section{Introduction}\label{sec1}

Survival analysis examines and models the time it takes for events to occur. In the medical area and in clinical trials, the event of interest is often the death of a patient, from which the name \textit{survival analysis} is derived. As data can only be collected over a finite period of time, the \textit{time-to-event} may not be observed for all individuals. This is the case for example if a patient leaves a clinical trial prematurely or if they are still alive at the end of the study. In such a case, the time of death (time to event) for that individual is unknown. This phenomenon, called \textit{censoring}, creates some unusual difficulties in the analysis of survival data that cannot be properly handled by standard statistical methods. Most of the time, for the sake of simplicity, \textit{independent censoring} is assumed, but this can lead to bias when there is actually \textit{dependent censoring}, that is, when the survival time and the censoring time are stochastically dependent on each other. This happens, for example, when the event of interest is the time to death from a particular disease, and censoring occurs when a patient leaves the study due to deteriorating health.
In this situation, the survival and censoring times are likely to be positively dependent. Conversely, if a patient leaves the clinical trial because their health has improved significantly, we can expect that the survival and censoring times are negatively dependent. Several authors have dealt with dependent censoring using different approaches. Some have used bivariate distributions, such as Emoto and Matthews \cite{EmotoMatthews} with their bivariate Weibull model. Others have used copulas to model the dependence, but with the drawback of assuming that the copula is fully known (including the dependence parameter), see for example Rivest and Wells \cite{RivestWells}, who did a sensitivity analysis. Recently, a number of authors showed that in some cases the dependence parameter can be identified. We cite, among others, the papers of Deresa and Van Keilegom \cite{DeresaVK1}, in which the model is based on parametric transformations of the survival and censoring times as well as on a bivariate normal distribution for the errors, Czado and Van Keilegom \cite{CzadoVanKeilegom}, who use a parametric copula with unknown dependence parameter, or Deresa and Van Keilegom \cite{DeresaVK_2023}, in which the authors combine a parametric copula and the Cox proportional hazards model in order to obtain a semiparametric model.

Traditional survival analysis assumes that all individuals in the population are susceptible to the event of interest, i.e. each individual has either already experienced the event or will experience it in the future. However, in many situations it may happen that a fraction of the individuals (long-term survivors) will never experience the event and are considered to be event free. For example, if we observe patients with breast cancer, some of them will fortunately never die of their cancer and can thus be considered as \textit{cured individuals}. Or, if we assign patients to a treatment to assess its effect on disease recurrence, some individuals will never experience a recurrence and can also be considered cured or immune. Models that deal with such data are called \textit{cure models}. In the literature on cure regression models a popular one is the \textit{mixture cure model}, which models the survival function by assuming that the underlying population is a mixture of two subpopulations: the subpopulation of \textit{susceptibles} (i.e. those who will experience the event and have finite survival time) and the subpopulation of \textit{non-susceptibles} (i.e. those who are event free and have an infinite survival time). There is an abundance of literature on cure models. Among the many existing references, we refer to the books by Maller and Zhou \cite{MallerZhou_1996} and Peng and Yu \cite{PengYu_2022} as well as the review papers by Amico and Van Keilegom \cite{AmicoVK} and Peng and Taylor \cite{PengTaylor_2014}, and Chapter 4 of the book by Legrand \cite{Legrand2021} for an overview on the topic of cure models in survival analysis.

Several authors have already demonstrated the importance of taking into account the presence of a cure fraction and dependent censoring to avoid biased results (some of the references that mention this are Li et al. \cite{LiAl}, Rondeau et al. \cite{RondeauAl2011} and Huang and Wolfe \cite{HuangWolfe}). Although these two features have been studied separately by many authors, the situation where both characteristics are present has not received a lot of attention in the literature so far. However, it is clear that there are many situations where we have both dependent censoring and a cure fraction. For example, in certain cancer trials some patients will fortunately never die of their cancer and/or may leave the study for health-related reasons. Or, in a study on unemployment, some individuals may never find a job and/or may leave the study because of reasons related to their job search (e.g.\ they might go back studying or move to a region where chances of finding a job are better, in which case they are censored). To the best of our knowledge, there are only a few articles that deal both with dependent censoring and the presence of a cure fraction, see Li et al. \cite{LiAl}, Othus et al. \cite{OthusAl}, Bernhardt \cite{Bernhardt} and Liu et al. \cite{LiuAl}. Some of them make prior assumptions on the dependence structure (Li et al. \cite{LiAl}) or on the cure threshold that is supposed to be known (Bernhardt \cite{Bernhardt}), where the dependence between the survival and the censoring time is not general but restricted to a dependence on the cure threshold). Moreover, for these two as well as for Liu et al. \cite{LiuAl}, there is no proof of the identifiability of the model, which is an important characteristic of any statistical model. The paper by Othus et al. \cite{OthusAl} does establish the identifiability of the model, but it requires the presence of covariates in the two parts of their mixture cure model. Also note that Liu et al. \cite{LiuAl} studies interval censoring instead of right censoring, which we are considering.

All this motivates us to study the conditions under which a model that allows for a cure fraction and dependent censoring can be identified without strong prior assumptions and without requiring covariates. To do so, we will propose a fully parametric mixture cure model that allows for possible dependent censoring. The model builds further on the model of Czado and Van Keilegom \cite{CzadoVanKeilegom} by allowing for a cure fraction. This extension to cure models, presented in Section \ref{sect2}, is however far from trivial, as it leads to an additional parameter in the model (namely the cure fraction), which turns out to be a disturbing element in the model when showing identifiability, as will be explained in Section \ref{sect3}. A major advantage of our model is that the copula that models the dependence between $T$ and $C$ is not assumed to be known. We will show that the model is identified under rather general conditions, and, in Section \ref{sect4}, that it provides asymptotically normal estimates of (among others) the strength of the dependence and the cure fraction. Then we will present some simulation results in Section \ref{sect5} and test our model on a real dataset in Section \ref{sect6}. Finally, Section \ref{sect7} is about some discussions, a self-criticism of the model and ideas for future work.

\section{The model}  \label{sect2}

Let $T$ be the survival time, and $C$ be the censoring time. Both variables are supposed to be non-negative and continuous. Due to random right censoring, we observe $Y=\min(T, C)$ and $\Delta=\I(T\leq C)$. We also assume that there is a positive probability $1-p$ that the event of interest does not occur, i.e.\ $1-p=\Proba(T = \infty)>0$. Note that we can write $T=UB + \infty (1-B)$, where $U$ is the event time for the susceptible (uncured) individuals, and $B=\I(T<\infty)$.

With these notations we can write the (improper) survival function of $T$ as follows: 
\begin{equation}
S_T(t)=\Proba(T>t)=1-p+pS_U(t), \label{eq:model}
\end{equation}
where $S_U(t)=\Proba(T>t|T<\infty)$ is the proper survival function of the susceptibles, called the latency, and $p=\Proba(T<\infty)$ is called the incidence. We can equivalently write our model as $F_T(t)=pF_U(t)$, 
with $F_T= 1-S_T$ and $F_U=1-S_U$. Let $F_C(t) = \Proba(C \le t)$ be the (proper) distribution function of $C$, which we assume to have support $[0,\infty)$. The support of $U$, on the other hand, is bounded; details of this are given in Section \ref{sect3}. We assume that $F_U$ and $F_C$ belong to parametric families of distributions:
$$ F_U \in \{F_{U,\theta_U} : \theta_U\in\Theta_U\} \quad \text{and} \quad F_C\in\{F_{C,\theta_C} : \theta_C\in\Theta_C\} $$
for certain finite dimensional parameter spaces $\Theta_U$ and $\Theta_C$. 

To model the possible stochastic dependence between $T$ and $C$, we will use copulas which are a very useful mathematical tool for modelling the dependence between random variables. They allow to write the bivariate distribution $F_{T,C}(t,c)=\Proba(T \le t,C \le c)$ as a function of the marginal distributions:
\begin{equation} \label{copul}
F_{T,C}(t,c)=\C\left\{F_T(t),F_C(c)\right\},
\end{equation}
where the copula $\C$ is a bivariate distribution defined on the unit square with uniform marginals. In addition to the parametric models for $F_T$ and $F_C$, we also assume that the copula belongs to a parametric family:
$$ \C\in\{\C_{\theta} : \theta\in\Theta\}, $$ 
for a finite dimensional parameter space $\Theta$. So, the parameters in our model are $(\theta,\theta_U,\theta_C,p)^T \in A = \Theta \times \Theta_U \times \Theta_C \times (0,1)$.

Common copula families are the Archimedean copulas and the elliptical copulas, of which the Gaussian copula is a special case. Archimedean copulas can be written as
$$\C(u, v)=\varphi^{[-1]}\left\{\varphi(u)+\varphi(v)\right\}, \quad u, v \in [0, 1], $$
where $\varphi$ is called the \textit{generator} of the copula and is a continuous, convex and strictly decreasing function from $[0, 1]$ to $[0, \infty]$ such that $\varphi(1)=0$, and $\varphi^{[-1]}$ is the pseudo-inverse of $\varphi$. On the other hand, the Gaussian copula is defined as
$$\C(u, v)=\Phi_{\theta}\left\{\Phi^{-1}(u), \Phi^{-1}(v)\right\}, $$
where $\Phi$ is the distribution function of a standard normal random variable and $\Phi_{\theta}$ is the distribution of a bivariate standard normal random vector with correlation $\theta$. For more background on copulas, we refer to the monograph by Nelsen \cite{Nelsen}.

Next, we introduce some notation that will be used to develop the likelihood of our model. We use lower case letters to denote densities, so $f_U\in\{f_{U,\theta_U} : \theta_U\in\Theta_U\}$ and $f_C\in\{f_{C,\theta_C} : \theta_C\in\Theta_C\}$. We also need the following partial derivatives of the copula, commonly called the $h$-functions:
$$ h_{T|C}(u|v)=\frac{\partial}{\partial v}\mathcal{C}(u,v) \quad \text{and} \quad 
h_{C|T}(v|u)=\frac{\partial}{\partial u}\mathcal{C}(u,v).$$
This allows to write the conditional distributions as 
$$ F_{T|C}(t|c)=\Proba(T \le t|C=c)=h_{T|C}\{F_T(t)|F_C(c)\}, $$
and
$$F_{C|T}(c|t)=\Proba(C \le c|T=t)=h_{C|T}\{F_C(c)|F_T(t)\}.$$
It follows that
$$ f_{Y,\Delta}(y,1) = \frac{\text{d}}{\text{dy}} \Proba(Y \le y, \Delta=1) = pf_U(y) [1-h_{C|T}\{F_C(y)|pF_U(y)\}], $$
and 
$$ f_{Y,\Delta}(y,0) = \frac{\text{d}}{\text{dy}} \Proba(Y \le y, \Delta=0) = f_C(y) [1-h_{T|C}\{pF_U(y)|F_C(y)\}]. $$
Finally, in the above formulas parameters can be added whenever confusion is possible, like e.g.\ $f_{Y,\Delta,\alpha}(y,1) = pf_{U,\theta_U}(y) [1-h_{C|T,\theta}\{F_{C,\theta_C}(y)|pF_{U,\theta_U}(y)\}]$, where $\alpha=(\theta,\theta_U,\theta_C,p)^T$.

\section{Identifiability} \label{sect3}

The proof of the identifiability of the model is presented in this section. We will first develop high-level conditions under which the model is identifiable (Theorem \ref{IdentMod}), and then show when these high-level conditions are satisfied (Theorem \ref{IdentCop}). The proofs are partially given in the Appendix, and partially in the Supplementary Material.

We assume throughout the paper that the support of $U$ is bounded. This is a classical assumption in the literature on cure models. The right endpoint of the support of $U$ is often called the cure threshold, which is nothing more than the time after which individuals who are still alive are known to be cured. We denote it by $\tau=\inf\{y : F_{U, \theta_U}(y)=1\}$.

Recall that the identification based on the observable variables is a fundamental step that consists in showing that different values of the parameters give different distributions. In other words, having identical densities for the observed variables must imply that the parameters are also identical.
In our case, identifying our model is equivalent to showing that if
\begin{align}
    & \left[pf_{U, \theta_U}(y)\left\{1-h_{C|T, \alpha}(y)\right\}\right]^{\delta} \left[f_{C, \theta_C}(y) \left\{1-h_{T|C, \alpha}(y)\right\}\right]^{1-\delta} \nonumber \\
    & \quad\quad = \left[\tilde pf_{U, \tilde\theta_U}(y)\left\{1-h_{C|T, \tilde\alpha}(y)\right\}\right]^{\delta} \left[f_{C, \tilde\theta_C}(y)\left\{1-h_{T|C, \tilde\alpha}(y)\right\}\right]^{1-\delta},\label{eq:identification}
\end{align}
for all $(y, \delta)$, then $p=\tilde p, \theta=\tilde\theta, \theta_U=\tilde\theta_U$ and $\theta_C=\tilde\theta_C$. Here, we have used the abbreviated notation $h_{C|T,\alpha}(y)=h_{C|T, \theta}\{F_{C, \theta_C}(y)|pF_{U, \theta_U}(y)\}$, similarly for $h_{T|C,\alpha}(y)$, and $\alpha=(\theta,\theta_U,\theta_C,p)^T$.

The following theorem gives us sufficient conditions on the marginal densities and the copula under which our model is identified. Contrary to the case where there is either only dependent censoring or only a cure fraction, proving the identifiability of a model in which both features are present, is not an easy task. This is because one feature is masking or disturbing the appearance of the other, which makes it difficult to identify both the cure fraction $1-p$ and the copula parameter $\theta$. 

\medskip

\begin{Theorem}[Identification conditions]
\label{IdentMod}
    Assume that for $\theta_{U}, \tilde\theta_{U} \in \Theta_U$ and $\theta_{C}, \tilde\theta_{C} \in \Theta_C$,
	\begin{equation}
		\lim_{y \rightarrow 0}\frac{f_{U, \theta_{U}}(y)}{f_{U, \tilde\theta_{U}}(y)}=1 \Longleftrightarrow \theta_{U}=\tilde\theta_{U}, \label{eq:C11}
	\end{equation}
	\begin{equation}
		\lim_{y \rightarrow 0}\frac{f_{C, \theta_{C}}(y)}{f_{C, \tilde\theta_{C}}(y)}=1 \Longleftrightarrow \theta_{C}=\tilde\theta_{C} \quad \mbox{and} \quad \lim_{y \rightarrow \infty}\frac{f_{C, \theta_{C}}(y)}{f_{C, \tilde\theta_{C}}(y)}=1 \Longleftrightarrow \theta_{C}=\tilde\theta_{C}. \label{eq:C12}
	\end{equation}
	Assume also that either
	\begin{equation}
		\lim_{y\rightarrow 0} h_{T|C, \alpha}(y) = 0 \quad \forall \alpha \in A \quad \mbox{or} \quad	\lim_{y \rightarrow \infty} h_{T|C, \alpha}(y) = 0 \quad \forall \alpha \in A, \label{eq:star1}
	\end{equation}
	and that
	\begin{equation}
		\lim_{y\rightarrow 0} h_{C|T, \alpha}(y) = 0 \quad \forall \alpha \in A. \label{eq:star2}
	\end{equation}
	Finally assume that
	\begin{equation}
	h_{T|C, \theta}\{p|F_{C, \theta_C}(y)\}=h_{T|C, \tilde\theta}\{\tilde p|F_{C, \theta_C}(y)\} \hspace{0.2cm} \forall y > \tau \text{ implies } p=\tilde p \text{ and } \theta=\tilde\theta. \label{eq:star3}
	\end{equation}
	Then model (\ref{eq:model}) is identified.
\end{Theorem}

\medskip

Note that condition \eqref{eq:C11} is an adaptation of Theorem 1 in Czado and Van Keilegom \cite{CzadoVanKeilegom}. In particular, since a right truncated distribution is used for $U$, only the limit when $y$ tends to zero is considered in condition \eqref{eq:C11}. This complicates the proof, since we have to get around the fact that the limit when $y$ tends to infinity cannot be used. Another complication is the fact that contrary to Theorem 1 in Czado and Van Keilegom \cite{CzadoVanKeilegom}, our condition \eqref{eq:C11} deals with the variable $U$ instead of $T$. This is because for $T$ the property is not valid for a large class of models. In the case of the exponential density e.g., $\lim_{y \rightarrow 0} f_{T,\theta_U,p}(y) / f_{T, \tilde\theta_U,\tilde p}(y) = \lim_{y \rightarrow 0} \{p f_{U,\theta_U}(y)\} / \{\tilde p f_{U, \tilde\theta_U}(y)\} = (p \theta_U)/(\tilde p \tilde \theta_U)$, and if this limit equals 1 it does not follow that $p=\tilde p$ and $\theta_U=\tilde\theta_U$. Therefore we impose condition \eqref{eq:C11} on the density of $U$ instead of $T$, and we identify $p$ thanks to condition \eqref{eq:star3}.

It is also worth noting that in condition \eqref{eq:star1} only one of the two limits needs to be zero. In fact, as can be seen from the proof, the identifiability of the model is shown by taking either the limit when $y$ tends to 0 or $\infty$, and the choice between these two limits is made according to which limit in \eqref{eq:star1} equals zero (since for some copulas, only one of the two limits is zero). Also note that in \eqref{eq:C12} we require both properties to hold. However, this could be slightly weakened, as it is enough to impose \eqref{eq:C12} for the limit for which \eqref{eq:star1} holds.

Let Clayton($\alpha$) be the Clayton copula rotated by $\alpha$ degrees and restricted to the case where the copula is strict, so $\theta>0$. Note that Clayton(90) and Clayton(270) can be used to model negative associations. More information on rotated copulas can be found in the dedicated section in the Supplementary Material. We now have the following result.

\begin{Theorem}[Identifiability] 
\label{IdentCop}
The following holds:
\begin{itemize}
    \item[(a)] Condition \eqref{eq:C11} is satisfied for the families of truncated log-normal, log-logistic, Weibull and Gamma densities. 
    \item[(b)] Condition \eqref{eq:C12} is satisfied for the families of log-normal, log-logistic, Weibull and Gamma densities. 
    \item[(c)] Conditions \eqref{eq:star1}, \eqref{eq:star2} and \eqref{eq:star3} are satisfied for the Frank, Joe, Clayton(90), Clayton(180) and Clayton(270) copulas, independently of the marginal distributions. For the Gumbel and Gaussian copulas we need, respectively, the following additional conditions:
        \begin{equation}
             \lim_{y\rightarrow 0} \frac{\log\{F_{C, \theta_C}(y)\}}{\log\{pF_{U, \theta_U}(y)\}} > 0 \quad \forall (\theta,\theta_U,\theta_C)^T \in \Theta \times \Theta_U \times \Theta_C,
        \label{hypGumbel2}
        \end{equation}
        and
        \begin{equation}
             \lim_{y\rightarrow 0} \left[\Phi^{-1}\left\{F_{C, \theta_C}(y)\right\}-\theta\Phi^{-1}\left\{pF_{U, \theta_U}(y)\right\} \right]= -\infty \quad \forall (\theta,\theta_U,\theta_C,p)^T \in A, \label{hypGaussian}
        \end{equation}
        where the set $\Theta$ in the case of the Gaussian copula is either a subset of $[-1,0)$ or $[0,1]$. 
\end{itemize}
\end{Theorem}

\smallskip 

Note that the limit in (\ref{hypGaussian}) is always equal to $-\infty$ for $\theta \le 0$.
The combination of these two theorems shows that our model is identified for the Frank, Gumbel, Joe, Clayton(90), Clayton(180), Clayton(270) and Gaussian copulas combined with numerous densities for the marginals. Since the proof of Theorem \ref{IdentCop} requires lengthy calculations that are moreover different from one copula family to another, some parts of the proof are given in the Appendix whereas others can be found in Appendix A of the Supplementary Material. They can be used as a starting point for showing the identifiability conditions for other copulas and marginal distributions, not given here.

Note that we did not mention the (non-rotated) Clayton copula in the previous theorem. The reason is simply that the proof does not work for the Clayton copula. In fact, conditions \eqref{eq:star1} and \eqref{eq:star2} never hold together for this copula. However, this does not necessarily mean that the model is not identifiable when a Clayton copula is used, since the conditions are sufficient but not necessary (see also Theorem 4 in Czado and Van Keilegom \cite{CzadoVanKeilegom} for the case without cure fraction).

\section{Parameter estimation}\label{sect4}

In this section we will present the estimation of the parameters of our model using a maximum likelihood approach. We will also comment on the consistency and asymptotic normality of the parameter estimators. Assume that we have an i.i.d.\ sample $\mathcal{D}=(Y_i,\Delta_i)_{i=1,\ldots,n}$. The likelihood of our model is given by
\begin{align*}
   L(\alpha;\mathcal{D}) & = \prod_{i=1}^n \left(pf_{U, \theta_U}(Y_i)\left[1-h_{C|T, \theta}\left\{F_{C, \theta_C}(Y_i)|pF_{U, \theta_U}(Y_i)\right\}\right]\right)^{\Delta_i} \\
   & \quad \times \prod_{i=1}^n \left(f_{C, \theta_C}(Y_i)\left[1-h_{T|C, \theta}\left\{pF_{U, \theta_U}(Y_i)|F_{C, \theta_C}(Y_i)\right\}\right]\right)^{1-\Delta_i}. 
\end{align*}

The log-likelihood is then
\begin{align*}
   \ell(\alpha;\mathcal{D}) & = \sum_{i=1}^n \Delta_i\log\left(pf_{U, \theta_U}(Y_i)\left[1-h_{C|T, \theta}\left\{F_{C, \theta_C}(Y_i)|pF_{U, \theta_U}(Y_i)\right\}\right]\right) \\
   & \quad + \sum_{i=1}^n (1-\Delta_i) \log\left(f_{C, \theta_C}(Y_i)\left[1-h_{T|C, \theta}\left\{pF_{U, \theta_U}(Y_i)|F_{C, \theta_C}(Y_i)\right\}\right]\right),
\end{align*}
and we can define the estimators of the parameters by
$$\hat{\alpha} = \left(\hat{\theta}, \hat{\theta}_U, \hat{\theta}_C, \hat{p}\right)^\top = \argmax_{\alpha \in A}\ell(\alpha;\mathcal{D}). $$
Note that the estimated parameters are consistent, asymptotically normal, and even efficient if conditions 1 to 4 of Theorem 8.17 and 1 to 3 of Theorem 8.18 in Mittelhammer \cite{Mittelhammer} are satisfied. These conditions are about properties of limits, first and second derivatives of the log-likelihood

\section{Simulation study} \label{sect5}

In this section we will investigate the finite sample performance of our estimation procedure for several copulas and for Weibull marginal distributions for $U$ and $C$. Although our theoretical results require that the support of $U$ is finite, we will work both with truncated and non-truncated Weibull distributions for $U$, and we will show that the performance is good in both cases.

We will investigate the quality of the estimators of the most important parameters in our model, namely the parameters of the distribution of $U$, the incidence $p$, and Kendall's tau $\tau_K$, which is easier to interpret than the copula parameter $\theta$, since it lies between -1 and 1 and it measures the strength of the dependence between $T$ and $C$. Knowing $\theta$ or knowing $\tau_K$ is equivalent, and it can be shown that $\tau_K$ is given by
$$ \tau_K = 4 \int_0^1\int_0^1 \C(u, v) c(u, v) dudv -1, $$
where $c(u,v)$ is the copula density.

To maximise the likelihood, we need to choose appropriate starting values. They are chosen in the following way. For the parameters of the Weibull distributions for $U$ (respectively $C$), we select the maximum likelihood estimators under the Weibull model using only the uncensored (respectively censored) observations, to which we add 9 perturbations by multiplying these estimators by 9 random values $U_1,\ldots,U_9$ from a $\mathcal{U}(0.5, 1.5)$ distribution. For the incidence, we take as starting value the proportion of subjects whose survival time (censored or uncensored) is smaller than or equal to the largest uncensored survival time, and we add again 9 perturbations which are this value raised to the power $U_1,\ldots,U_9$ (to ensure that the perturbed proportions are within the interval $[0,1]$ by construction). Finally, for Kendall's tau, we use a grid $0.1, 0.2, \ldots, 0.9$ (and also negative values in cases where it makes sense). Combining these values for Kendall's tau with the previously obtained perturbed parameters, we end up with at least 90 vectors of starting values.

We will start with the case where the model is correctly specified, there is a cure fraction, and the survival and censoring times are dependent. We will then examine what happens if the copula is misspecified, and finally we will use our dependent censoring model for data that are subject to independent censoring, in order to see if the model is able to detect it.

It is interesting to note that for each of the simulations described below, we run $1000$ repetitions. We then remove the optimisation results corresponding to $1\%$ of the smallest and $1\%$ of the largest estimated values of Kendall's tau (which is the most difficult parameter to estimate). We then average the remaining $980$ results for each of the model parameters to obtain our final estimators.

For the case of non-truncated distributions, we generate data from model (\ref{eq:model})-(\ref{copul}) in the following way. For the distribution of $U$ we consider a Weibull distribution with scale and shape parameters equal to 0.5 and 1, respectively, while the distribution of $C$ is a Weibull with scale and shape parameters both equal to 1. For each of the copulas considered, different strengths of the dependence are used, with a Kendall's tau of $0.2$, $0.5$, $0.8$ for the Gaussian, Joe, Gumbel and Clayton(180) copulas, to which we add the value $-0.5$ for the Frank copula that allows for positive and negative dependence and, in case of the Clayton(90) and Clayton(270) copulas, which model negative dependence, we take a Kendall's tau of $-0.2$, $-0.5$ and $-0.8$. The incidence in the model is $p=0.8, 0.6$ or $0.4$, and the sample sizes considered are $n=200, 500, 1000$ and $2500$. We restrict ourselves here to showing the results for $n=200$ and 1000, and for $p=0.8$. All results can be found in the Supplementary Material. Table \ref{tblC90} shows the empirical bias, standard deviation and root mean squared error (RMSE) as well as the standard deviation estimated via the inverse Fisher matrix for several copula families that are all correctly specified, while Table \ref{tblGlF} shows the results when the data are generated from the independence and the Gumbel copula, and the model is estimated based on a Joe and Frank copula, respectively.

\begin{table}[!h]
\centering
\fontsize{7.2}{7.2}\selectfont
\begin{tabular}{l|c|l|rrrr|rrrr}
\multicolumn{3}{c|}{} & \multicolumn{4}{c|}{$n=200$} & \multicolumn{4}{c}{$n=1000$}  \\
Copula & $\tau_{K}$ & & $\tau_{K}$ & $p$ & $\theta_{U1}$ & $\theta_{U2}$ & $\tau_{K}$ & $p$ & $\theta_{U1}$ & $\theta_{U2}$ \\ 
  \hline
Clayton(90) & -0.20 & Bias & -0.02 & -0.01 & -0.01 & 0.02 & 0.02 & 0.01 & -0.00 & 0.01 \\ 
& & SD & 0.22 & 0.10 & 0.08 & 0.10 & 0.13 & 0.06 & 0.03 & 0.05 \\ 
& & $\widehat{\mbox{SD}}$ & 0.20 & 0.10 & 0.09 & 0.10 & 0.17 & 0.07 & 0.04 & 0.05 \\ 
& & RMSE & 0.23 & 0.10 & 0.08 & 0.10 & 0.13 & 0.05 & 0.03 & 0.04 \\ 
  \cline{2-11}
& -0.50 & Bias & 0.08 & 0.06 & 0.02 & 0.04 & 0.02 & 0.02 & 0.01 & 0.01 \\ 
& & SD & 0.21 & 0.12 & 0.12 & 0.12 & 0.09 & 0.06 & 0.06 & 0.05 \\ 
& & $\widehat{\mbox{SD}}$ & 0.15 & 0.13 & 0.16 & 0.13 & 0.09 & 0.06 & 0.06 & 0.05 \\ 
& & RMSE & 0.23 & 0.13 & 0.13 & 0.12 & 0.09 & 0.06 & 0.05 & 0.04 \\ 
  \cline{2-11}
& -0.80 & Bias & -0.00 & 0.03 & 0.04 & 0.02 & 0.00 & 0.02 & 0.03 & 0.00 \\ 
& & SD & 0.06 & 0.14 & 0.16 & 0.12 & 0.03 & 0.09 & 0.10 & 0.06 \\ 
& & $\widehat{\mbox{SD}}$ & 0.07 & 0.24 & 0.28 & 0.13 & 0.02 & 0.10 & 0.12 & 0.05 \\ 
& & RMSE & 0.06 & 0.14 & 0.17 & 0.13 & 0.03 & 0.09 & 0.11 & 0.05 \\ 
  \hline
Frank & 0.20 & Bias & -0.07 & -0.02 & -0.01 & 0.00 & -0.01 & -0.00 & 0.00 & -0.01 \\ 
& & SD & 0.31 & 0.08 & 0.08 & 0.09 & 0.11 & 0.03 & 0.03 & 0.04 \\ 
& & $\widehat{\mbox{SD}}$ & 0.22 & 0.07 & 0.08 & 0.09 & 0.11 & 0.03 & 0.03 & 0.04 \\ 
& & RMSE & 0.32 & 0.09 & 0.08 & 0.09 & 0.11 & 0.03 & 0.03 & 0.04 \\ 
  \cline{2-11}
& 0.50 & Bias & -0.26 & -0.04 & -0.00 & -0.00 & -0.03 & -0.00 & 0.01 & -0.01 \\ 
& & SD & 0.52 & 0.11 & 0.07 & 0.09 & 0.20 & 0.04 & 0.03 & 0.04 \\ 
& & $\widehat{\mbox{SD}}$ & 0.17 & 0.05 & 0.06 & 0.08 & 0.09 & 0.02 & 0.03 & 0.04 \\ 
& & RMSE & 0.58 & 0.11 & 0.07 & 0.09 & 0.20 & 0.04 & 0.03 & 0.04 \\ 
  \cline{2-11}
& 0.80 & Bias & -0.05 & 0.00 & 0.01 & 0.01 & 0.00 & 0.01 & 0.01 & -0.01 \\ 
& & SD & 0.23 & 0.05 & 0.08 & 0.09 & 0.04 & 0.02 & 0.03 & 0.04 \\ 
& & $\widehat{\mbox{SD}}$ & 0.08 & 0.03 & 0.05 & 0.07 & 0.02 & 0.01 & 0.02 & 0.03 \\ 
& & RMSE & 0.24 & 0.05 & 0.08 & 0.09 & 0.04 & 0.00 & 0.03 & 0.04 \\ 
  \cline{2-11}
& -0.50 & Bias & 0.33 & 0.06 & -0.03 & 0.08 & 0.06 & 0.02 & -0.00 & 0.01 \\ 
& & SD & 0.49 & 0.13 & 0.12 & 0.13 & 0.22 & 0.08 & 0.07 & 0.06 \\ 
& & $\widehat{\mbox{SD}}$ & 0.20 & 0.12 & 0.13 & 0.10 & 0.08 & 0.07 & 0.06 & 0.05 \\ 
& & RMSE & 0.59 & 0.14 & 0.13 & 0.15 & 0.23 & 0.08 & 0.07 & 0.06 \\ 
  \hline
Gaussian & 0.20 & Bias & -0.04 & -0.02 & -0.01 & -0.01 & -0.08 & -0.02 & 0.00 & -0.02 \\ 
& & SD & 0.42 & 0.11 & 0.07 & 0.09 & 0.27 & 0.07 & 0.03 & 0.05 \\ 
& & $\widehat{\mbox{SD}}$ & 0.41 & 0.14 & 0.10 & 0.16 & 0.19 & 0.05 & 0.03 & 0.04 \\ 
& & RMSE & 0.42 & 0.11 & 0.07 & 0.09 & 0.28 & 0.07 & 0.03 & 0.05 \\ 
  \cline{2-11}
& 0.50 & Bias & -0.17 & -0.03 & 0.00 & -0.01 & -0.07 & -0.01 & 0.01 & -0.01 \\ 
& & SD & 0.46 & 0.10 & 0.07 & 0.10 & 0.21 & 0.05 & 0.03 & 0.04 \\ 
& & $\widehat{\mbox{SD}}$  & 0.26 & 0.09 & 0.09 & 0.10 & 0.12 & 0.03 & 0.03 & 0.04 \\ 
& & RMSE & 0.49 & 0.11 & 0.07 & 0.09 & 0.23 & 0.04 & 0.03 & 0.04 \\ 
  \cline{2-11}
& 0.80 & Bias & -0.16 & -0.02 & -0.00 & 0.02 & -0.08 & -0.01 & -0.00 & 0.01 \\ 
& & SD & 0.40 & 0.06 & 0.06 & 0.09 & 0.14 & 0.02 & 0.02 & 0.04 \\ 
& & $\widehat{\mbox{SD}}$  & 0.14 & 0.05 & 0.07 & 0.08 & 0.06 & 0.02 & 0.03 & 0.04 \\ 
& & RMSE & 0.43 & 0.06 & 0.06 & 0.09 & 0.16 & 0.03 & 0.03 & 0.03 \\ 
  \cline{2-11}
& -0.50 & Bias & 0.35 & 0.07 & -0.03 & 0.05 & 0.14 & 0.03 & -0.01 & 0.02 \\ 
& & SD & 0.59 & 0.14 & 0.14 & 0.12 & 0.40 & 0.09 & 0.08 & 0.06 \\ 
& & $\widehat{\mbox{SD}}$ & 0.46 & 0.08 & 0.16 & 0.14 & 0.14 & 0.06 & 0.05 & 0.05 \\ 
& & RMSE & 0.69 & 0.15 & 0.14 & 0.13 & 0.43 & 0.09 & 0.08 & 0.07 \\
\end{tabular}
\smallskip
\caption{Bias, standard deviation (SD), estimated SD and RMSE of the estimators of $\tau_K$, $p=0.8$, and the scale ($\theta_{U1}$) and shape ($\theta_{U2}$) of $U$, for several copulas and values of $\tau_K$ and $n$, and in the case of non-truncated Weibull distributions.}
\label{tblC90}
\end{table}

\begin{table}[!h]
\centering
\fontsize{7.5}{7.5}\selectfont
\begin{tabular}{l|c|l|rrrr|rrrr}
Copula & \multicolumn{2}{c|}{} & \multicolumn{4}{c|}{$n=200$} & \multicolumn{4}{c}{$n=1000$}  \\
data gen./ & \multirow{2}{*}{$\tau_{K}$} & & \multirow{2}{*}{$\tau_{K}$} & \multirow{2}{*}{$p$} & \multirow{2}{*}{$\theta_{U1}$} & \multirow{2}{*}{$\theta_{U2}$} & \multirow{2}{*}{$\tau_{K}$} & \multirow{2}{*}{$p$} & \multirow{2}{*}{$\theta_{U1}$} & \multirow{2}{*}{$\theta_{U2}$} \\
estim. & & & & & & & & & & \\ 
  \hline
 Indep./ & 0.00 & Bias & 0.17 & 0.07 & 0.02 & 0.02 & 0.05 & 0.03 & 0.02 & -0.00 \\ 
 Joe & & SD & 0.19 & 0.08 & 0.08 & 0.09 & 0.07 & 0.04 & 0.04 & 0.04 \\ 
& & $\widehat{\mbox{SD}}$ & 0.14 & 0.07 & 0.08 & 0.10 & 0.05 & 0.04 & 0.04 & 0.04 \\ 
& & RMSE & 0.25 & 0.10 & 0.08 & 0.09 & 0.08 & 0.05 & 0.04 & 0.04 \\ 
  \hline
Gumbel/ & 0.20 & Bias & -0.10 & -0.05 & -0.02 & 0.00 & -0.12 & -0.04 & -0.01 & -0.01 \\ 
Frank & & SD & 0.20 & 0.06 & 0.07 & 0.10 & 0.09 & 0.03 & 0.03 & 0.05 \\ 
& & $\widehat{\mbox{SD}}$  & 0.21 & 0.07 & 0.07 & 0.08 & 0.10 & 0.03 & 0.03 & 0.04 \\ 
& & RMSE & 0.22 & 0.08 & 0.08 & 0.10 & 0.15 & 0.05 & 0.03 & 0.04 \\ 
  \cline{2-11}
& 0.50 & Bias & -0.13 & -0.05 & -0.02 & 0.02 & -0.13 & -0.04 & -0.02 & 0.01 \\ 
& & SD & 0.17 & 0.05 & 0.06 & 0.09 & 0.07 & 0.02 & 0.03 & 0.04 \\ 
& & $\widehat{\mbox{SD}}$  & 0.18 & 0.05 & 0.06 & 0.08 & 0.08 & 0.02 & 0.02 & 0.04 \\ 
& & RMSE & 0.22 & 0.07 & 0.07 & 0.09 & 0.15 & 0.04 & 0.03 & 0.04 \\ 
  \cline{2-11}
& 0.80 & Bias & -0.14 & -0.03 & -0.01 & 0.06 & -0.13 & -0.04 & -0.02 & 0.05 \\ 
& & SD & 0.12 & 0.04 & 0.07 & 0.10 & 0.06 & 0.02 & 0.02 & 0.05 \\ 
& & $\widehat{\mbox{SD}}$ & 0.11 & 0.04 & 0.05 & 0.08 & 0.06 & 0.02 & 0.02 & 0.04 \\ 
& & RMSE & 0.18 & 0.05 & 0.07 & 0.12 & 0.14 & 0.04 & 0.03 & 0.07 \\
\end{tabular}
\smallskip
\caption{Bias, standard deviation (SD), estimated SD and RMSE of the estimators of $\tau_K$, $p=0.8$, and the scale ($\theta_{U1}$) and shape ($\theta_{U2}$) of $U$, for several values of $\tau_K$ and $n$, and in the case of non-truncated Weibull distributions. The data are generated from an independence and a Gumbel copula, while the fitted model is based on a Joe and a Frank copula, respectively.}
\label{tblGlF}
\end{table}

We can see that globally the bias and RMSE are very satisfactory, even when the copula is misspecified or when there is no dependence between the survival and censoring times. This seems to indicate that the model is also identifiable in the case of non-truncated distributions. Note that the results for the parameters of $F_C$ are not shown in the tables due to space limitations, but they can be found in the Supplementary Material. The tables also show that in most cases the empirical standard deviations and the estimated standard deviations are close to each other, although for $\tau_K$ there are sometimes some discrepancies between these two standard deviations, probably caused by the fact that this is the most difficult parameter to estimate, given that we never observe both $T$ and $C$ for a given subject. Finally note that the results for the Gaussian copula are for the case where the parameters satisfy the identification condition (\ref{hypGaussian}). It should be noted, however, that we have carried out simulations with parameters that do not satisfy this condition, and we also obtain satisfactory results.

For the case of a truncated Weibull model for $U$, we keep all other model settings as for the non-truncated case and we consider $p = 0.6$, in order to see how the model performs with a fairly high cure fraction, similar to that of the real (Veridex) data analysed in this paper. The results are given in Tables \ref{tblC90Tr}-\ref{tblGlFTr} for a sample size of $1000$. We truncate $5\%$ of the upper tail. We can see that for this sample size the bias and the RMSE are generally small, except in certain cases (certain combinations of copulas and dependencies, e.g.\ the Clayton(90) copula with a Kendall's tau of $-0.8$), where the bias is less satisfactory but the RMSE is still small.

\begin{table}[!h]
\centering
\fontsize{7.5}{7.5}\selectfont
\begin{tabular}{l|c|l|rrrr}
Copula & $\tau_{K}$ & & $\tau_{K}$ & $p$ & $\theta_{U1}$ & $\theta_{U2}$ \\ 
  \hline
Clayton(90) & -0.20 & Bias & 0.02 & 0.01 & 0.01 & 0.00 \\ 
& & SD & 0.13 & 0.05 & 0.05 & 0.05 \\ 
& & $\widehat{\mbox{SD}}$ & 0.16 & 0.07 & 0.06 & 0.06 \\ 
& & RMSE & 0.13 & 0.05 & 0.05 & 0.05 \\ 
  \cline{2-7}
& -0.50 & Bias & 0.04 & 0.02 & 0.03 & 0.00 \\ 
& & SD & 0.12 & 0.06 & 0.08 & 0.05 \\ 
& & $\widehat{\mbox{SD}}$ & 0.12 & 0.06 & 0.08 & 0.05 \\ 
& & RMSE & 0.13 & 0.06 & 0.08 & 0.05 \\ 
  \cline{2-7}
& -0.80 & Bias & 0.16 & 0.09 & 0.30 & -0.07 \\ 
& & SD & 0.00 & 0.00 & 0.00 & 0.00 \\ 
& & $\widehat{\mbox{SD}}$ & 0.06 & 0.05 & 0.27 & 0.06 \\ 
& & RMSE & 0.16 & 0.09 & 0.30 & 0.07 \\ 
  \hline
Frank & 0.20 & Bias & -0.00 & -0.00 & 0.02 & -0.01 \\ 
& & SD & 0.11 & 0.04 & 0.05 & 0.05 \\ 
& & $\widehat{\mbox{SD}}$ & 0.12 & 0.04 & 0.06 & 0.05 \\ 
& & RMSE & 0.11 & 0.03 & 0.05 & 0.05 \\ 
  \cline{2-7}
& 0.50 & Bias & -0.14 & -0.03 & 0.01 & -0.02 \\ 
& & SD & 0.39 & 0.08 & 0.07 & 0.05 \\ 
& & $\widehat{\mbox{SD}}$ & 0.10 & 0.04 & 0.05 & 0.05 \\ 
& & RMSE & 0.41 & 0.08 & 0.07 & 0.05 \\ 
  \cline{2-7}
& 0.80 & Bias & -0.01 & -0.01 & 0.02 & -0.00 \\ 
& & SD & 0.06 & 0.04 & 0.08 & 0.05 \\ 
& & $\widehat{\mbox{SD}}$ & 0.06 & 0.03 & 0.08 & 0.05 \\ 
& & RMSE & 0.06 & 0.04 & 0.08 & 0.05 \\ 
  \cline{2-7}
& -0.50 & Bias & 0.05 & 0.02 & 0.04 & -0.00 \\ 
& & SD & 0.19 & 0.07 & 0.10 & 0.06 \\ 
& & $\widehat{\mbox{SD}}$ & 0.09 & 0.05 & 0.10 & 0.05 \\ 
& & RMSE & 0.19 & 0.07 & 0.10 & 0.05 \\ 
  \hline
Gaussian & 0.20 & Bias & -0.06 & -0.01 & 0.00 & -0.01 \\ 
& & SD & 0.27 & 0.07 & 0.05 & 0.05 \\ 
& & $\widehat{\mbox{SD}}$ & 0.20 & 0.06 & 0.06 & 0.05 \\ 
& & RMSE & 0.27 & 0.08 & 0.05 & 0.05 \\ 
  \cline{2-7}
& 0.50 & Bias & -0.13 & -0.02 & 0.01 & -0.01 \\ 
& & SD & 0.39 & 0.09 & 0.07 & 0.05 \\ 
& & $\widehat{\mbox{SD}}$  & 0.15 & 0.05 & 0.06 & 0.05 \\ 
& & RMSE &  0.41 & 0.09 & 0.07 & 0.05 \\ 
  \cline{2-7}
& 0.80 & Bias & -0.12 & -0.03 & 0.00 & 0.01 \\ 
& & SD & 0.34 & 0.07 & 0.09 & 0.06 \\ 
& & $\widehat{\mbox{SD}}$  & 0.10 & 0.05 & 0.08 & 0.05 \\ 
& & RMSE & 0.36 & 0.08 & 0.09 & 0.05 \\ 
  \cline{2-7}
& -0.50 & Bias & 0.21 & 0.06 & 0.02 & 0.01 \\ 
& & SD & 0.45 & 0.13 & 0.09 & 0.06 \\ 
& & $\widehat{\mbox{SD}}$ & 0.15 & 0.06 & 0.09 & 0.05 \\ 
& & RMSE & 0.50 & 0.15 & 0.09 & 0.06 \\
\end{tabular}
\smallskip
\caption{Bias, standard deviation (SD), estimated SD and RMSE of the estimators of $\tau_K$, $p=0.6$, and the scale ($\theta_{U1}$) and shape ($\theta_{U2}$) of $U$, for several copulas and values of $\tau_K$ and $n=1000$, and in the case of a truncated Weibull distribution for the variable $U$.}
\label{tblC90Tr}
\end{table}

It is interesting to note that what might explain, at least in part, the fact that some combinations of copulas and levels of dependence give less good results is the censoring rate. Table \ref{PerctgCensoring} shows the percentages of censoring for the models considered in Tables \ref{tblC90}-\ref{tblGlFTr}, while Appendix B of the Supplementary Material contains similar tables for all the models. For example, in the table corresponding to the non-truncated distributions with $p=0.4$ (see Appendix B), we observe that some cases with less good results, such as Gumbel and Clayton(180) with $\tau_K=0.8$, correspond to very high censoring rates, which could explain the less good performance of the model in these situations. The same observation applies to the Gaussian copula, which sometimes gives poorer results than other copulas, but also has higher censoring rates. Another hypothesis to explain the less satisfactory results that are sometimes obtained for small sample sizes is the fact that the maximum likelihood estimator may not always exist in the case of truncated distributions, see e.g.\ Mittal and Dahiya \cite{MittalDahiya} and Ghosh \cite{Ghosh}. However, this problem is typical for small samples and disappears when $n$ increases.

\begin{table}[!h]
\centering
\fontsize{7.5}{7.5}\selectfont
\begin{tabular}{l|c|l|rrrr}
Copula & \multicolumn{6}{c}{} \\
data gen./ & \multirow{2}{*}{$\tau_{K}$} & & \multirow{2}{*}{$\tau_{K}$} & \multirow{2}{*}{$p$} & \multirow{2}{*}{$\theta_{U1}$} & \multirow{2}{*}{$\theta_{U2}$} \\
estim. & & \\
  \hline
 Indep./ & 0.00 & Bias & 0.06 & 0.02 & 0.04 & -0.00 \\ 
 Joe & & SD & 0.08 & 0.04 & 0.07 & 0.05 \\ 
& & $\widehat{\mbox{SD}}$ & 0.07 & 0.04 & 0.07 & 0.05 \\ 
& & RMSE & 0.29 & 0.12 & 2.19 & 0.11 \\ 
  \hline
Gumbel/ & 0.20 & Bias & -0.10 & -0.03 & -0.02 & -0.00 \\ 
Frank & & SD & 0.09 & 0.03 & 0.04 & 0.05 \\ 
& & $\widehat{\mbox{SD}}$  & 0.11 & 0.04 & 0.05 & 0.05 \\ 
& & RMSE & 0.24 & 0.08 & 0.25 & 0.11 \\ 
  \cline{2-7}
& 0.50 & Bias & -0.14 & -0.04 & -0.03 & 0.02 \\ 
& & SD & 0.09 & 0.03 & 0.04 & 0.05 \\ 
& & $\widehat{\mbox{SD}}$  & 0.11 & 0.04 & 0.04 & 0.05 \\ 
& & RMSE & 0.27 & 0.09 & 1.22 & 0.12 \\
  \cline{2-7}
& 0.80 & Bias & -0.17 & -0.07 & -0.04 & 0.06 \\ 
& & SD & 0.12 & 0.04 & 0.06 & 0.06 \\ 
& & $\widehat{\mbox{SD}}$ & 0.10 & 0.04 & 0.06 & 0.06 \\ 
& & RMSE & 0.42 & 0.13 & 4.07 & 0.14 \\
\end{tabular}
\smallskip
\caption{Bias, standard deviation (SD), estimated SD and RMSE of the estimators of $\tau_K$, $p=0.6$, and the scale ($\theta_{U1}$) and shape ($\theta_{U2}$) of $U$, for several values of $\tau_K$ and $n=1000$, and in the case of a truncated Weibull distribution for the variable $U$. The data are generated from an independence and a Gumbel copula, while the fitted model is based on a Joe and a Frank copula, respectively.}
\label{tblGlFTr}
\end{table}

\begin{table}[!h]
\centering
\fontsize{8}{8}\selectfont
\begin{tabular}{l|l|ccccccc}
\multirow{2}{*}{Model} & \multirow{2}{*}{Copula} & \multicolumn{7}{c}{$\tau_K$} \\
 & & 0.00 & 0.20 & 0.50 & 0.80 & -0.20 & -0.50 & -0.80 \\
\hline
\multirow{5}{*}{(1)} & Clayton(90) & & & & & 43.60 & 42.78 & 43.06 \\
& Frank & & 47.30 & 46.86 & 38.00 & & 45.48 & \\
& Gaussian & & 71.58 & 76.58 & 83.46 & & 60.98 & \\
& Indep. & 46.78 & & & & & & \\
& Gumbel & & 46.78 & 46.64 & 40.28 & & & \\
\hline
\multirow{5}{*}{(2)} & Clayton(90) & & & & & 55.14 & 51.24 & 50.54 \\
& Frank & & 60.68 & 65.10 & 69.00 & & 54.12 & \\
& Gaussian & & 81.08 & 87.66 & 96.84 & & 65.30 &\\
& Indep. & 58.04 & & & & & & \\
& Gumbel & & 59.66 & 63.58 & 67.44 & & & 
\end{tabular}
\smallskip
\caption{Censoring rates for several combinations of copulas and values of $\tau_{K}$ in the case of non-truncated Weibull distributions and an incidence level of $p=0.8$ (model (1)) and in the case of a truncated Weibull distribution for the variable $U$ and an incidence level of $p=0.6$ (model (2)).}
\label{PerctgCensoring}
\end{table}
  
\section{Illustration on Veridex data} \label{sect6}

\begin{figure}[!h]
    \centering
    \includegraphics[scale=0.6]{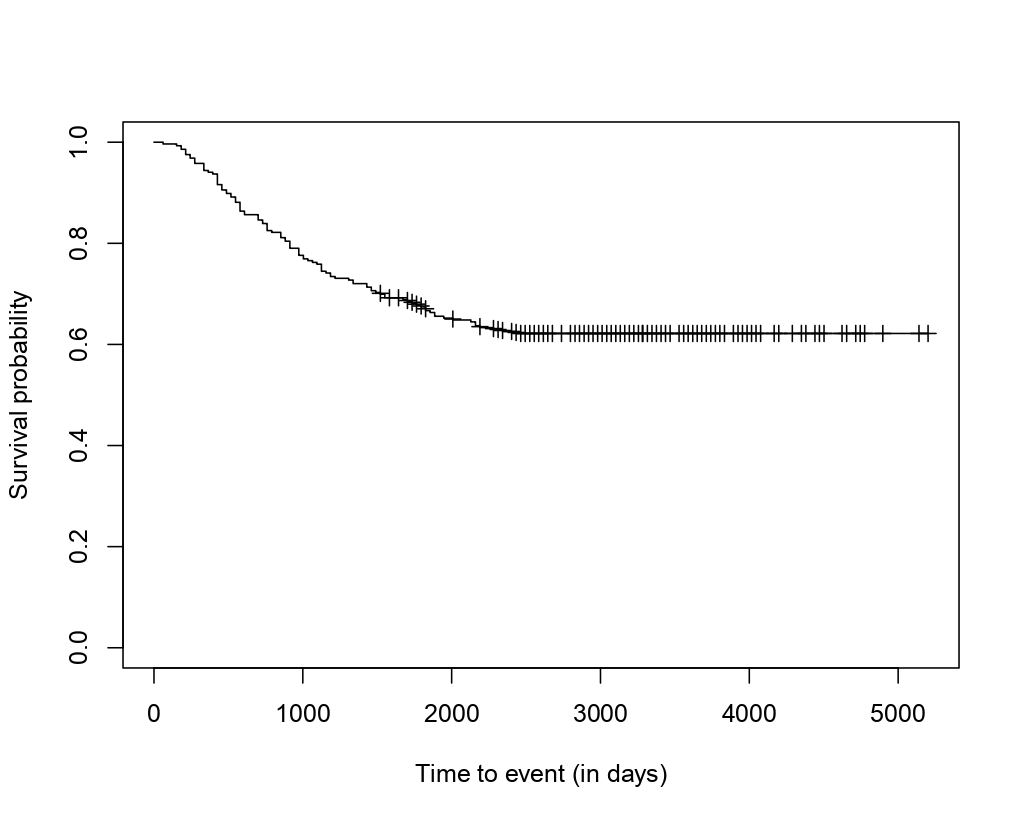}
    \caption{Kaplan-Meier curve for the Veridex data}
    \label{fig:VeridexKM}
\end{figure}

We now illustrate our model and estimation procedure using data (also known as the Veridex data) on 286 lymph-node-negative breast cancer patients who were treated between 1980 and 1995. We refer to Wang et al. \cite{WangAl} for more information on this dataset. The event of interest is the development of distant metastasis and the time to this event is expressed in days. Some patients may remain metastasis-free, and so it is logical to expect a cure rate in these data. Furthermore, Figure \ref{fig:VeridexKM} clearly shows a long plateau with many censored observations, which is a strong indication of the presence of cured individuals (see Amico and Van Keilegom \cite{AmicoVK}).

The data are also possibly subject to dependent censoring. Indeed, as is explained in Wang et al. \cite{WangAl}, 83 ($29\%$) patients died after a previous relapse, and we can suspect that the time to death and the time to distant metastasis are related, since they both depend on how the cancer evolves over time.

For all these reasons, it would be very interesting to apply our model to these data. After changing the unit of the event time to months (instead of days), brief descriptive analyses of the data were carried out and our model and estimation procedure were applied for different combinations of copulas and marginal distributions. 

The results for truncated distributions for $U$ are given in Table \ref{tblVeridexTr}, and those for non-truncated distributions are in Appendix B of the Supplementary Material, where truncation is done after the last uncensored observation. Each time we take the survival and censoring distributions from the same family (for simplicity, to limit the number of models to fit). The tables contain the estimated values of $\tau_K, p$ and the parameters of the laws of $U$ and $C$, together with their estimated error based on the inverse Fisher matrix. We also include in these tables the estimated median of $U$, the value of the negative log-likelihood and the AIC given by $\text{AIC} = -2\ell(\alpha;{\cal D}) + 2k$, where $k$ is the number of parameters in the model. The tables show that truncation does not seem to have a major impact on the estimated parameters for the selected model (in the sense that the results with and without truncation are similar). This seems to indicate that the truncation has no real effect on the results and that our model is stable.

A closer look at the AIC for the different models shows that the model that best fits these data is a Joe copula model with a truncated log-normal margin for $U$ and a non-truncated log-normal margin for $C$. For the latter model, we observe that the estimated dependence is quite high, with a Kendall's tau of $0.62$, confirming the usefulness of our model for these data. We also observe that the estimated incidence is 0.39, which is very close to what the Kaplan-Meier curve gives us (0.38), and that the estimated incidence is quite stable across all fitted models. Note that the Joe copula has a positive upper-tail dependence, which confirms that large survival times are associated with large censoring times, as was already suggested by Figure \ref{fig:VeridexKM}.

Next, in order to confirm the presence of dependent censoring in these data, we perform a likelihood ratio test with a type I error of $5\%$. We do this by comparing the model containing the Joe copula with the model assuming independent censoring, where in both models we use a truncated log-normal margin for $U$ and a non-truncated log-normal margin for $C$. The test statistic is
$$\lambda_{LR}=-2\left(\text{logLik}_{Ind} - \text{logLik}_{Joe}\right) = -2(-1474.1+1469.5)=9.2,$$
which should be compared with the quantile $\chi_{1, 0.95}^2=3.84$. Since $\lambda_{LR} > \chi_{1, 0.95}^2$, we conclude that the dependent censoring model is preferable to the independence model.

\begin{table}[!h]
\centering
\fontsize{6.7}{6.7}\selectfont
\begin{tabular}{l|rrrrrrrrr}
  Copula & \multirow{3}{*}{$\tau_K$} & \multirow{3}{*}{$p$} & \multirow{3}{*}{$\theta_{U1}$} & \multirow{3}{*}{$\theta_{U2}$} & \multirow{3}{*}{$\theta_{C1}$} & \multirow{3}{*}{$\theta_{C2}$} & \multirow{3}{*}{Med$_U$} & \multirow{3}{*}{-logLik} & \multirow{3}{*}{AIC} \\
   & & & & & & & & & \\
  Margins & & & & & & & & & \\
  \hline
 Independ. & & & & & & & & & \\
 & & & & & & & & & \\
 Weibull & & 0.38 & 39.72 & 1.52 & 115.08 & 4.61 & 29.64 & 1477.6 & 2965.1 \\
  & & (0.03) & (3.87) & (0.15) & (1.95) & (0.25) & & & \\ 
 Gamma & & 0.38 & 18.86 & 1.98 & 6.03 & 17.42 & 29.07 & 1472.6 & 2955.1 \\ 
  & & (0.03) & (4.44) & (0.31) & (0.63) & (1.78) & & & \\ 
 Log-normal & & 0.38 & 0.95 & 3.57 & 0.25 & 4.63 & 28.11 & 1474.1 & 2958.2 \\ 
  & & (0.03) & (0.12) & (0.19) & (0.01) & (0.02) & & & \\
 \hline
 Frank & & & & & & & & & \\ 
  & & & & & & & & & \\
 Weibull & -0.36 & 0.38 & 39.16 & 1.53 & 123.55 & 4.81 & 29.38 & 1474.5 & 2961.0 \\
 & (0.10) & (0.03) & (3.65) & (0.15) & (3.39) & (0.29) & & & \\ 
 Gamma & 0.43 & 0.39 & 20.34 & 1.92 & 6.92 & 13.84 & 29.82 & 1470.8 & 2953.6 \\ 
 & (0.13) & (0.03) & (5.21) & (0.31) & (1.04) & (2.39) & & & \\ 
 Log-normal & 0.43 & 0.39 & 0.98 & 3.65 & 0.27 & 4.53 & 29.00 & 1470.7 & 2953.3 \\ 
 & (0.10) & (0.03) & (0.13) & (0.23) & (0.02) & (0.03) & & & \\ 
 \hline
 Gumbel & & & & & & & & & \\
 & & & & & & & & & \\
 Weibull & 0.73 & 0.40 & 43.73 & 1.46 & 99.39 & 3.16 & 31.14 & 1473.0 & 2958.0 \\
 & (0.05) & (0.03) & (5.97) & (0.15) & (2.45) & (0.23) & & & \\ 
 Gamma & 0.51 & 0.39 & 20.43 & 1.92 & 7.16 & 13.26 & 29.86 & 1471.6 & 2955.3 \\ 
 & (0.15) & (0.03) & (5.28) & (0.31) & (1.35) & (2.87) & & & \\ 
 Log-normal & 0.48 & 0.39 & 0.98 & 3.65 & 0.27 & 4.52 & 28.94 & 1472.2 & 2956.5 \\ 
 & (0.12) & (0.03) & (0.13) & (0.22) & (0.02) & (0.03) & & & \\ 
 \hline
 {\it Joe} & & & & & & & & & \\ 
 & & & & & & & & & \\
 Weibull & 0.77 & 0.39 & 43.61 & 1.46 & 99.55 & 3.21 & 30.98 & 1475.2 & 2962.4 \\
 & (0.05) & (0.03) & (5.89) & (0.15) & (2.28) & (0.22) & & & \\ 
 Gamma & 0.63 & 0.39 & 21.20 & 1.89 & 7.33 & 12.75 & 30.21 & 1470.4 & 2952.8 \\ 
 & (0.08) & (0.03) & (5.63) & (0.31) & (1.09) & (2.08) & & & \\ 
 \it{Log-normal} & \it{0.62} & \it{0.39} & \it{1.00} & \it{3.69} & \it{0.27} & \it{4.51} & \it{29.36} & \it{1469.5} & \it{2950.9} \\ 
 & (0.07) & (0.03) & (0.14) & (0.24) & (0.02) & (0.03) & & & \\ 
 \hline
 Clayton(90) & & & & & & & & & \\
 & & & & & & & & & \\
 Weibull & -0.53 & 0.38 & 38.93 & 1.54 & 125.79 & 4.91 & 29.29 & 1473.0 & 2958.0 \\
 & (0.08) & (0.03) & (3.56) & (0.15) & (2.90) & (0.31) & & & \\ 
 Gamma & -0.63 & 0.37 & 18.14 & 2.01 & 9.10 & 13.25 & 28.70 & 1472.4 & 2956.8 \\ 
 & (0.09) & (0.03) & (4.11) & (0.31) & (1.84) & (2.30) & & & \\ 
 Log-normal & -0.72 & 0.38 & 0.93 & 3.53 & 0.32 & 4.79 & 27.70 & 1472.7 & 2957.5 \\ 
 & (0.06) & (0.03) & (0.12) & (0.17) & (0.03) & (0.03) & & & \\ 
 \hline
 Clayton(180) & & & & & & & & & \\
 & & & & & & & & & \\
 Weibull & 0.77 & 0.39 & 43.65 & 1.46 & 99.52 & 3.21 & 30.99 & 1475.1 & 2962.2 \\
 & (0.05) & (0.03) & (5.91) & (0.15) & (2.29) & (0.22) & & & \\ 
 Gamma & 0.63 & 0.39 & 21.21 & 1.89 & 7.30 & 12.80 & 30.21 & 1470.5 & 2953.0 \\ 
 & (0.09) & (0.03) & (5.63) & (0.31) & (1.12) & (2.15) & & & \\ 
 Log-normal & 0.61 & 0.39 & 1.00 & 3.69 & 0.27 & 4.51 & 29.38 & 1469.7 & 2951.4 \\ 
 & (0.08) & (0.03) & (0.14) & (0.24) & (0.02) & (0.03) & & & \\ 
 \hline
 Clayton(270) & & & & & & & & & \\
 & & & & & & & & & \\
 Weibull & -0.02 & 0.38 & 39.72 & 1.52 & 115.74 & 4.61 & 29.63 & 1477.6 & 2967.1 \\
 & (0.14) & (0.03) & (3.88) & (0.15) & (4.67) & (0.25) & & & \\ 
 Gamma & -0.30 & 0.38 & 18.56 & 1.99 & 8.19 & 13.98 & 28.92 & 1471.9 & 2955.8 \\ 
 & (0.19) & (0.03) & (4.30) & (0.31) & (2.05) & (2.75) & & & \\ 
 Log-normal & -0.49 & 0.38 & 0.93 & 3.54 & 0.31 & 4.77 & 27.81 & 1471.4 & 2954.7 \\ 
 & (0.13) & (0.03) & (0.12) & (0.18) & (0.03) & (0.05) & & & \\ 
 \hline
 Gaussian & & & & & & & & & \\ 
 & & & & & & & & & \\
 Weibull & 0.69 & 0.40 & 43.68 & 1.47 & 99.52 & 3.16 & 31.19 & 1472.9 & 2957.9 \\ 
 & (0.06) & (0.03) & (5.92) & (0.15) & (2.60) & (0.25) & & & \\ 
 Gamma & -0.57 & 0.37 & 18.13 & 2.01 & 9.78 & 12.41 & 28.71 & 1471.6 & 2955.2 \\ 
 & (0.12) & (0.03) & (4.10) & (0.31) & (2.09) & (2.23) & & & \\ 
 Log-normal & -0.67 & 0.38 & 0.93 & 3.53 & 0.33 & 4.79 & 27.72 & 1470.9 & 2953.8 \\ 
 & (0.08) & (0.03) & (0.12) & (0.17) & (0.03) & (0.03) & & & \\
\end{tabular}
\smallskip
\caption{Results for the Veridex data in the case of truncated distributions for $U$. For each combination of copulas and margins, the first row gives the estimated value, while the second row gives the standard error based on Fisher's information.}
\label{tblVeridexTr}
\end{table}

Now that the model has been selected, we can estimate, for example, the median survival time of the uncured individuals, or any other quantity that might be of interest in practice. For the selected model we obtain a median of $29.36$ months, which is $1.25$ months longer than for the independence copula. We have numerically verified, by bootstrapping, whether the difference between these two medians is statistically significant, at a significance level of $5\%$. To do this, we drew 1000 bootstrap samples from the original sample with replacement, and calculated each time the difference between the median survival times for uncured individuals in the case of the independence and the Joe copula. We then computed a $95\%$ confidence interval for this difference using the formula $-1.25 \pm z_{0.975} \hat\sigma_D$, where $\hat\sigma_D$ is the estimated standard deviation of the difference of the two median estimators, based on the 1000 bootstrap samples. We obtained $(-2.09,-0.41)$. Since this interval does not include 0, we can conclude that the two estimators are significantly different, and so accounting for dependent censoring in the data can make a difference in practice for medically relevant questions.

\newpage
\section{Discussion and future research} \label{sect7}

We believe that this work is an important step forward in the field of cure models with dependent censoring, as it allows data to be analysed without making any prior assumptions about the cure rate or the strength of the dependence. Moreover, it provides estimates of these quantities, which could be of great interest in the biomedical field, for example.

In the future it would be useful to extend the model by allowing for covariates. This can be done by adding covariates to the marginal distributions of $U$ and $C$, to the cure rate, and/or by letting the copula parameter depend on covariates. It would also be of interest to extend the paper to semiparametric margins for $U$. This has been done in the absence of a cure fraction by Deresa and Van Keilegom \cite{DeresaVK_2023}, who worked with a semiparametric Cox model under dependent censoring. It is unclear for the moment under which conditions the model would be identified if the Cox model would be replaced by e.g.\ a logistic/Cox mixture cure model. Some of these extensions are currently in preparation.

\section*{Acknowledgements}

Computational resources have been provided by the supercomputing facilities of the Universit\'e catholique de Louvain (CISM/UCL) and the `Consortium des Equipements de Calcul Intensif en F\'ed\'eration Wallonie Bruxelles (CECI), funded by the `Fond de la Recherche Scientifique de Belgique' (F.R.S.-FNRS) under convention 2.5020.11 and by the Walloon Region.

\section*{Declarations}

\subsection*{Funding}

Morine Delhelle and Ingrid Van Keilegom acknowledge the support of the ARC project (Projet d'Actions de Recherche Concert\'ees) `Imperfect data : From mathematical foundations to applications in life sciences' of the `Communaut\'e fran\c{c}aise de Belgique', granted by the `Acad\'emie universitaire Louvain' (2020-2025).

\subsection*{Conflict of interest}

The authors declare that they have no conflict of interest. 

\section*{Supplementary Material}

The Supplementary Material consists of three parts: Supplement A contains the proof of Theorem \ref{IdentCop} for densities and copulas not covered in this document, Supplement B presents additional results of Veridex data analysis and Supplement C shows the results of additional simulations.

\section*{Appendix : Proofs}

\begin{proof}[Theorem \ref{IdentMod}]
Suppose that (\ref{eq:identification}) holds for all $y$ and for $\delta\in\{0, 1\}$. If $\delta=0$ we have
\begin{align}
    & f_{C, \theta_C}(y)\big[1-h_{T|C, \theta}\{pF_{U, \theta_U}(y)|F_{C, \theta_C}(y)\}\big] \nonumber \\
    & \quad\quad = f_{C, \tilde\theta_C}(y)\big[1-h_{T|C, \tilde\theta}\{\tilde pF_{U, \tilde\theta_U}(y)|F_{C, \tilde\theta_C}(y)\}\big], \label{eq:delta0}
\end{align}
for all $y$, and hence it follows from \eqref{eq:star1} that (where $a=0$ or $a=\infty$)
$$ \lim_{y\rightarrow a} \frac{f_{C, \theta_C}(y)}{f_{C, \tilde\theta_C}(y)}=1.$$
Thanks to equation \eqref{eq:C12} we obtain $\theta_C=\tilde\theta_C$. Hence, equation \eqref{eq:delta0} becomes 
\begin{equation}
	h_{T|C, \theta}\{pF_{U, \theta_U}(y)|F_{C, \theta_C}(y)\}=h_{T|C, \tilde\theta}\{\tilde pF_{U, \tilde\theta_U}(y)|F_{C, \theta_C}(y)\}. \label{eq:delta0bis}
\end{equation}
In particular, if $y > \tau=\inf\{y : F_U(y)=1\}$, we have
\begin{equation*}
	h_{T|C, \theta}\{p|F_{C, \theta_C}(y)\}=h_{T|C, \tilde\theta}\{\tilde p|F_{C, \theta_C}(y)\}, 
\end{equation*}
which, by \eqref{eq:star3}, implies that $p=\tilde p$ and $\theta=\tilde\theta$.

Next, we consider the case $\delta=1$ for which we have the relation
\begin{align}
    & pf_{U, \theta_U}(y)\big[1-h_{C|T, \theta}\{F_{C, \theta_C}(y)|pF_{U, \theta_U}(y)\}\big] \nonumber \\
    & \quad\quad = pf_{U, \tilde\theta_U}(y)\big[1-h_{C|T, \theta}\{F_{C, \theta_C}(y)| pF_{U, \tilde\theta_U}(y)\}\big]. \label{eq:delta1}
\end{align}
It follows from \eqref{eq:star2} that
$$ \lim_{y\rightarrow 0} \frac{f_{U, \theta_U}(y)}{f_{U, \tilde\theta_U}(y)}=1, $$
and thanks to equation \eqref{eq:C11} we obtain $\theta_U=\tilde\theta_U$, which shows the result.
\end{proof}

\begin{proof}[Theorem \ref{IdentCop}(a)]
We will show that condition \eqref{eq:C11} holds for the truncated log-logistic and Weibull densities. The proofs for the truncated log-normal and Gamma densities are similar, and can be found in Appendix A of the Supplementary Material.

We start with the truncated log-logistic density, which is given by 
$$ f_{U,\alpha,\beta}(t) = \frac{\I(0 \le t \le \tau)}{\frac{\tau^{\beta}}{\alpha^{\beta}+\tau^{\beta}}}\cdot\frac{\frac{\beta}{\alpha}\cdot\left(\frac{t}{\alpha}\right)^{\beta-1}}{\left\{1+\left(\frac{t}{\alpha}\right)^{\beta}\right\}^2}, $$
and hence we have that
\begin{align}
\lim_{t \rightarrow 0}\frac{f_{U,\alpha_1,\beta_1}(t)}{f_{U,\alpha_2,\beta_2}(t)} 
& = \lim_{t \rightarrow 0} \frac{\frac{\beta_1}{\alpha_1}\cdot\left(\frac{t}{\alpha_1}\right)^{\beta_1-1}}{\left\{1+\left(\frac{t}{\alpha_1}\right)^{\beta_1}\right\}^2} \cdot
\frac{\left\{1+\left(\frac{t}{\alpha_2}\right)^{\beta_2}\right\}^2}{\frac{\beta_2}{\alpha_2}\cdot\left(\frac{t}{\alpha_2}\right)^{\beta_2-1}} \cdot
\frac{\tau^{\beta_2}}{\alpha_2^{\beta_2}+\tau^{\beta_2}} \cdot
\frac{\alpha_1^{\beta_1}+\tau^{\beta_1}}{\tau^{\beta_1}} \nonumber\\
& = \frac{\beta_1}{\beta_2} \cdot \frac{\alpha_2^{\beta_2}}{\alpha_1^{\beta_1}} \cdot \frac{\tau^{\beta_2}}{\tau^{\beta_1}} \cdot \frac{\alpha_1^{\beta_1}+\tau^{\beta_1}}{\alpha_2^{\beta_2}+\tau^{\beta_2}}\cdot\lim_{t \rightarrow 0} t^{\beta_1-\beta_2}, \label{lim1}
\end{align}
since $\lim_{t \rightarrow 0} \left\{1+\left(t/\alpha\right)^{\beta}\right\} = 1$. The limit in (\ref{lim1}) can only be equal to one if $ \beta_1=\beta_2$, which we denote now simply by $\beta$. 
We then have
\begin{align}
    \lim_{t \rightarrow 0}\frac{f_{U,\alpha_1,\beta_1}(t)}{f_{U,\alpha_2,\beta_2}(t)} & = \frac{\alpha_2^{\beta}}{\alpha_1^{\beta}}\cdot\frac{\alpha_1^{\beta}+\tau^{\beta}}{\alpha_2^{\beta}+\tau^{\beta}}. \nonumber
\end{align}
It is easily seen that this limit is equal to one if and only if $\alpha_1=\alpha_2$. 

We next consider the truncated Weibull density, which is given by 
$$ f_{U,k,\lambda}(t) = \frac{k}{\lambda^k}\cdot\frac{t^{k-1} \exp\left\{-\left(\frac{t}{\lambda}\right)^k\right\}}{1-\exp\left\{-\left(\frac{\tau}{\lambda}\right)^k\right\}}\cdot \I(0 \le t \le \tau), $$
and hence,
\begin{align}
    \lim_{t \rightarrow 0}\frac{f_{U,k_1,\lambda_1}(t)}{f_{U, k_2,\lambda_2}(t)}     
    & = \frac{k_1\lambda_2^{k_2}}{k_2\lambda_1^{k_1}}\cdot \frac{1-\exp\left\{-\left(\frac{\tau}{\lambda_2}\right)^{k_2}\right\}}{1-\exp\left\{-\left(\frac{\tau}{\lambda_1}\right)^{k_1}\right\}} \cdot \lim_{t \rightarrow 0} t^{k_1-k_2}, \nonumber  
\end{align}
since $\lim_{t \rightarrow 0} \exp\{-(t/\lambda)^k\}=1$. 

The latter can only be equal to one if $k_1=k_2=k$ (say). It follows that
\begin{align}
\frac{\lambda_2^{k}}{\lambda_1^{k}}\cdot \frac{1-\exp\left\{-\left(\frac{\tau}{\lambda_2}\right)^k\right\}}{1-\exp\left\{-\left(\frac{\tau}{\lambda_1}\right)^k\right\}} =1.\label{lim2}
\end{align}
The derivative with respect to $\lambda$ of the function $g(\lambda) = \lambda^k[1-\exp\{-(\tau/\lambda)^k\}]$ equals 
$$g'(\lambda) = k\lambda^{k-1}\left[1-\exp\left\{-\left(\frac{\tau}{\lambda}\right)^k\right\}\cdot\left\{1+\left(\frac{\tau}{\lambda}\right)^k\right\}\right].$$
This derivative is strictly positive since $\e^{-x}(1+x)<1$ for $x >0$. Hence, the function $g(\lambda)$ is strictly increasing and equation (\ref{lim2}) implies that $\lambda_1=\lambda_2$.
\end{proof}

\begin{proof}[Theorem \ref{IdentCop}(b)]
Condition \eqref{eq:C12} has been shown in Theorem 2 in Czado and Van Keilegom \cite{CzadoVanKeilegom} for the log-normal, log-logistic and Weibull densities. The case of the Gamma density can be treated in a similar way and can be found in Appendix A of the Supplementary Material.
\end{proof}

\begin{proof}[Theorem \ref{IdentCop}(c)]
We will show that conditions \eqref{eq:star1}, \eqref{eq:star2} and \eqref{eq:star3} are satisfied for the Frank and Gumbel copulas. The verification of these conditions for the Joe, Gaussian, and Clayton($\alpha$) copulas with $\alpha=90, 180$ or 270 is given in Appendix A of the Supplementary Material.

The Frank copula function is given by 
\begin{equation*}
C_{\theta}(u,v)=-\frac{1}{\theta}\cdot\log\left\{ 1+\frac{\left(\e^{-\theta u}-1\right)\left(\e^{-\theta v}-1\right)}{\e^{-\theta}-1}\right\}, \label{eq:frank}
\end{equation*} 
for $\theta\in(-\infty,\infty)\backslash\{0\}$. Hence, it is easily seen that
\begin{equation*}
h_{T|C, \theta}(u|v)=\frac{\partial}{\partial v}C_{\theta}(u,v)=\frac{\e^{-\theta v}\left(\e^{-\theta u}-1\right)}{\e^{-\theta}-1+\left(\e^{-\theta u}-1\right)\left(\e^{-\theta v}-1\right)}, \label{eq:frankhTC}
\end{equation*}
and similarly for $h_{C|T, \theta}(v|u)$. 

The verification of \eqref{eq:star1} and \eqref{eq:star2} is obvious, since it is clear that
\begin{equation*}
\lim_{y\rightarrow 0} \{\text{e}^{-\theta F_{C, \theta_C}(y)}-1\} = \lim_{y\rightarrow 0} \{\text{e}^{-\theta pF_{U, \theta_U}(y)}-1\} = 0 ,
\end{equation*}
and hence
\begin{equation*}
\lim_{y\rightarrow 0} h_{T|C,\alpha}(y) = 0 =  \lim_{y\rightarrow 0} h_{C|T,\alpha}(y) ,
\end{equation*}
for all $\theta, \theta_C, \theta_U, p$.

To show \eqref{eq:star3}, we define the function $g_{\theta,p}(y) = 1/h_{T|C, \theta}\{p|F_{C}(y)\}$. We need to show that if $g_{\theta,p}(y) = g_{\tilde\theta,\tilde p}(y)$ for all $y>\tau$, then $p=\tilde p$ and $\theta=\tilde\theta$. Some simple algebra shows that
$$ g_{\theta,p}(y) = \frac{\e^{\theta F_{C}(y)} \left(\e^{-\theta}-1\right) + \left(\e^{-\theta p}-1\right) \left\{1-\e^{\theta F_{C}(y)}\right\}}{\e^{-\theta p}-1}. $$
Then, $g'_{\theta,p}(y) = (\text{d}/\text{dy}) g_{\theta,p}(y)$ equals
$$g'_{\theta,p}(y) =  \frac{\theta f_{C}(y) \e^{\theta F_{C}(y)} \left(\e^{-\theta}-\e^{-\theta p}\right)}{\e^{-\theta p}-1}.$$
It follows that for any $y_1,y_2 > \tau$, 
$$ \frac{g'_{\theta,p}(y_1)}{g'_{\theta,p}(y_2)} = \frac{f_{C}(y_1)}{f_{C}(y_2)}\cdot \text{e}^{\theta\left\{F_C(y_1)-F_C(y_2)\right\}}. $$
Now, since $g_{\theta,p}(y) = g_{\tilde\theta,\tilde p}(y)$ for all $y>\tau$, we also have that 
$$ \frac{g'_{\theta,p}(y_1)}{g'_{\theta,p}(y_2)} = \frac{g'_{\tilde\theta,\tilde p}(y_1)}{g'_{\tilde \theta,\tilde p}(y_2)}, $$ 
and hence $\e^{\theta \{F_C(y_1)-F_C(y_2)\}} = \e^{\tilde \theta \{F_C(y_1)-F_C(y_2)\}}$, which implies that $\theta=\tilde \theta$. Some straightforward algebra now shows that $p=\tilde p$. 

Next, we consider the Gumbel copula, which is given by 
\begin{equation*}
C_{\theta}(u,v) = \exp\left[-\left\{(-\log u)^{\theta} + (-\log v)^{\theta}\right\}^\frac{1}{\theta}\right], \label{eq:gumbel}
\end{equation*} 
for $\theta\in[1,\infty)$, and hence
\begin{equation*}
h_{T|C, \theta}(u|v) = \frac{\partial}{\partial v}C_{\theta}(u,v) 
= \frac{(-\log v)^{\theta-1}\cdot\exp\left[-\left\{(-\log u)^\theta + (-\log v)^\theta\right\}^\frac{1}{\theta}\right]}{v\left\{(-\log u)^\theta + (-\log v)^\theta\right\}^\frac{\theta-1}{\theta}}, \label{eq:gumbelhTC}
\end{equation*}
and similarly for $h_{C|T, \theta}(v|u)$. For the verification of \eqref{eq:star1}, we consider the limit when $y$ tends to infinity. Since 
$$ \lim_{y \rightarrow \infty} \left(\left[-\log\left\{pF_{U, \theta_U}(y)\right\}\right]^\theta + \left[-\log\left\{F_{C, \theta_C}(y)\right\}\right]^\theta\right) = (-\log p)^\theta, $$ 
and $\lim_{y \rightarrow \infty} \log\{F_{C, \theta_C}(y)\}=0$, it follows that $\lim_{y\rightarrow \infty} h_{T|C,\alpha}(y) = 0$. For \eqref{eq:star2}, we have
\begin{align*}
    \lim_{y\rightarrow 0} h_{C|T,\alpha}(y) & = \lim_{y\rightarrow 0} \frac{u_y^{\theta-1}\cdot\exp\left\{-(u_y^\theta + v_y^\theta)^\frac{1}{\theta}\right\}}{\exp(-u_y)\cdot\left(u_y^\theta + v_y^\theta\right)^\frac{\theta-1}{\theta}} \\
    &  = \lim_{y\rightarrow 0} \frac{\exp\left(\left[-\left\{1 + \left(v_y/u_y\right)^\theta\right\}^\frac{1}{\theta}+1\right]u_y\right)}{\left\{1 + \left(v_y/u_y\right)^\theta\right\}^\frac{\theta-1}{\theta}},
\end{align*}
where $u_y = -\log\{pF_{U, \theta_U}(y)\}$ and $v_y = -\log\{F_{C, \theta_C}(y)\}$. If $\lim_{y\rightarrow 0} v_y/u_y > 0$, then the above limit equals 0, and so \eqref{eq:star2} is satisfied.

Finally, for \eqref{eq:star3} note that $h_{T|C, \theta}\{p|F_C(y)\}$ can be written as 
$$ h_{T|C, \theta}\{p|F_C(y)\} = \frac{\left\{-\log F_{C}(y)\right\}^{\theta-1}}{F_{C}(y)}\cdot\frac{\e^{-g_{\theta,p}(y)^\frac{1}{\theta}}}{g_{\theta,p}(y)^\frac{\theta-1}{\theta}}, $$
where $g_{\theta,p}(y) = (-\log p)^\theta + \{-\log F_C(y)\}^\theta$. Since for all $y>\tau$
$$h_{T|C, \theta}\{p|F_C(y)\} = h_{T|C, \tilde\theta}\{\tilde p|F_C(y)\},$$
we have that
\begin{equation}
\left\{-\log F_{C}(y)\right\}^{\theta-\tilde\theta}\cdot\frac{\e^{-g_{\theta,p}(y)^\frac{1}{\theta}}}{g_{\theta,p}(y)^\frac{\theta-1}{\theta}} = \frac{\e^{-g_{\tilde\theta,\tilde p}(y)^\frac{1}{\tilde \theta}}}{g_{\tilde \theta,\tilde p}(y)^\frac{\tilde \theta-1}{\tilde \theta}}. \label{gum}
\end{equation}
Now suppose $\theta > \tilde \theta$ (the case $\theta < \tilde \theta$ can be considered similarly). Then, taking the limit to infinity on both sides of (\ref{gum}) gives $0 = \tilde p/(-\log \tilde p)^{\tilde \theta-1}$, since $\lim_{y \rightarrow \infty} g_{\theta,p}(y) = (-\log p)^\theta$, which is a contradiction. This shows that $\theta=\tilde \theta$. It now follows from (\ref{gum}) that $p=\tilde p$, since $\e^{-g_{\theta,p}(y)^\frac{1}{\theta}} / g_{\theta,p}(y)^\frac{\theta-1}{\theta}$ is an increasing function of $p$ for fixed $\theta$ and $y$.
\end{proof}


\bibliographystyle{dcu}
\bibliography{References}

\end{document}


\thispagestyle{empty}
\begin{center}	
\Large{\bf Supplementary Material for} \\
\Large{\bf `Copula based dependent censoring in cure models'} \\
\bigskip
\large{Morine Delhelle$\,^a$ and Ingrid Van Keilegom$\,^{a,b}$} \\
\bigskip
\small{$^a$ Institute of Statistics, Biostatistics and Actuarial Sciences (ISBA),\\
UCLouvain, Voie du Roman Pays 20, bte L1.04.01, Louvain-la-Neuve, 1348, Belgium\\
$^b$ Research Centre for Operations Research and Statistics (ORSTAT),\\
KU Leuven, Naamsestraat 69 - bus 3555, Leuven, 3000, Belgium\\}
\bigskip
E-mails : morine.delhelle@uclouvain.be and ingrid.vankeilegom@kuleuven.be
\medskip
\end{center}

\tableofcontents
\newpage

\addcontentsline{toc}{section}{\large Supplement A: Proof of Theorem 3.2 (continued)}

\noindent
{\Large {\bf Supplement A: Proof of Theorem 3.2 (continued)}}
\medskip

\noindent
{\bf Proof of Theorem 3.2(a).} 
We give here the proof of equation (4) for the truncated log-normal and Gamma densities.

We start with the truncated log-normal density, which is given by
$$f_{U,\mu,\sigma}(t) = \frac{\I(0 \le t \le \tau)}{t\sigma\sqrt{2\pi}}\cdot\frac{\exp\left[-\frac{1}{2}\left\{\frac{\log(t)-\mu}{\sigma}\right\}^2\right]}{\frac{1}{2}\left[1+\erf\left\{\frac{\log(\tau)-\mu}{\sigma\sqrt{2}}\right\}\right]},$$
where $\erf(y)=(2/\sqrt{\pi}) \int_0^y\e^{-t^2}\text{dt}$, and hence we have that
\begin{align}
    & \lim_{t \rightarrow 0}\frac{f_{U,\mu_1,\sigma_1}(t)}{f_{U,\mu_2,\sigma_2}(t)} \nonumber \\
    & \quad\quad\quad = \lim_{t \rightarrow 0}\frac{\frac{1}{t\sigma_1\sqrt{2\pi}}\cdot \exp\left[-\frac{1}{2}\left\{\frac{\log(t)-\mu_1}{\sigma_1}\right\}^2\right]}{\frac{1}{t\sigma_2\sqrt{2\pi}}\cdot \exp\left[-\frac{1}{2}\left\{\frac{\log(t)-\mu_2}{\sigma_2}\right\}^2\right]}\cdot\frac{\frac{1}{2}\left[1+\erf\left\{\frac{\log(\tau)-\mu_2}{\sigma_2\sqrt{2}}\right\}\right]}{\frac{1}{2}\left[1+\erf\left\{\frac{\log(\tau)-\mu_1}{\sigma_1\sqrt{2}}\right\}\right]} \nonumber \\
    & \quad\quad\quad = \frac{\sigma_2}{\sigma_1}\cdot\frac{\left[1+\erf\left\{\frac{\log(\tau)-\mu_2}{\sigma_2\sqrt{2}}\right\}\right]}{\left[1+\erf\left\{\frac{\log(\tau)-\mu_1}{\sigma_1\sqrt{2}}\right\} \right]}\cdot \lim_{t \rightarrow 0}\exp\left[\frac{1}{2}\left\{\frac{\log(t)-\mu_2}{\sigma_2}\right\}^2-\frac{1}{2}\left\{\frac{\log(t)-\mu_1}{\sigma_1}\right\}^2\right] \nonumber \\
    & \quad\quad\quad = \frac{\sigma_2}{\sigma_1}\cdot\frac{\left[1+\erf\left\{\frac{\log(\tau)-\mu_2}{\sigma_2\sqrt{2}}\right\}\right]}{\left[1+\erf\left\{\frac{\log(\tau)-\mu_1}{\sigma_1\sqrt{2}}\right\} \right]} \nonumber \\
    & \quad\quad\quad\quad\quad \times \lim_{t \rightarrow 0} \exp\left\{\log(t)^2\left(\frac{1}{2\sigma_2^2}-\frac{1}{2\sigma_1^2}\right)-\log(t)\left(\frac{\mu_2}{\sigma_2^2}-\frac{\mu_1}{\sigma_1^2}\right)+\frac{1}{2}\left(\frac{\mu_2^2}{\sigma_2^2}-\frac{\mu_1^2}{\sigma_1^2}\right)\right\}. \label{limLN}
\end{align}
Since $\lim_{t \rightarrow 0}\log(t)=-\infty$, the limit in (\ref{limLN}) can only be equal to one if $\frac{1}{2\sigma_2^2}=\frac{1}{2\sigma_1^2}$ and $\frac{\mu_2}{\sigma_2^2}=\frac{\mu_1}{\sigma_1^2}.$ The first equality leads to $\sigma_1=\sigma_2$ (since $\sigma_1$, $\sigma_2 > 0$ by definition) which injected in the second one, gives $\mu_1=\mu_2$.

Let $\gamma(s,y)=\int_{0}^{y}t^{s-1}\e^{-t}\text{dt}$ be the lower incomplete gamma function. We next consider the truncated Gamma density, which is given by
$$f_{U,k,\theta}(t) = \frac{\I(0 \le t \le \tau)}{\Gamma(k)\theta_1^{k}}\cdot\frac{t^{k-1}\cdot \e^{-\frac{t}{\theta}}}{\frac{1}{\Gamma(k)}\cdot\gamma(k, \frac{\tau}{\theta})},$$
and hence,
\begin{align}
    \lim_{t \rightarrow 0}\frac{f_{U,k_1,\theta_1}(t)}{f_{U, k_2,\theta_2}(t)} & =  \lim_{t \rightarrow 0}\frac{\frac{1}{\Gamma(k_1)\theta_1^{k_1}}\cdot t^{k_1-1}\cdot \e^{-\frac{t}{\theta_1}}}{\frac{1}{\Gamma(k_2)\theta_2^{k_2}}\cdot t^{k_2-1}\cdot \e^{-\frac{t}{\theta_2}}}\cdot\frac{\frac{1}{\Gamma(k_2)}\cdot\gamma(k_2, \frac{\tau}{\theta_2})}{\frac{1}{\Gamma(k_1)}\cdot\gamma(k_1, \frac{\tau}{\theta_1})} \nonumber\\
    & =  \frac{\theta_2^{k_2}\gamma(k_2, \frac{\tau}{\theta_2})}{\theta_1^{k_1}\gamma(k_1, \frac{\tau}{\theta_1})}\cdot\lim_{t \rightarrow 0}t^{k_1-k_2}\cdot \exp\left\{t\left(\frac{1}{\theta_2}-\frac{1}{\theta_1}\right)\right\}. \label{limGt}
\end{align}
The latter can only be equal to one if $k_1=k_2=k$ (say). Moreover, since $\lim_{t \rightarrow 0}\exp\big\{{t\big(\frac{1}{\theta_2}-\frac{1}{\theta_1}\big)}\big\}=1$, the limit in (\ref{limGt}) is equal to one if $\theta_2^{k}\cdot\gamma(k, \frac{\tau}{\theta_2})=\theta_1^{k}\cdot\gamma(k, \frac{\tau}{\theta_1})$.
Computing the derivative with respect to $\theta$ of the function $\theta^k\gamma\left(k, \frac{\tau}{\theta}\right)$ gives us $k\theta^{k-1}\int_0^{\frac{\tau}{\theta}}t^{k-1}\cdot \e^{-t}\text{dt}-\frac{\tau^k}{\theta}\e^{-\frac{\tau}{\theta}}$.
We can verify numerically that this derivative is strictly positive and the function $\theta^k\gamma\left(k, \frac{\tau}{\theta}\right)$ is strictly increasing. The above equality implies that $\theta_1=\theta_2$. \hfill $\Box$

\medskip

\noindent
{\bf Proof of Theorem 3.2(b).}
We will show that condition (5) is verified for the Gamma density which is given by
$$f_{C,k,\theta}(t) = \frac{1}{\Gamma(k)\theta^{k}}\cdot t^{k-1}\cdot \text{e}^{-\frac{t}{\theta}}.$$

For the limit as $t$ goes to zero the proof is even simpler than the one for the truncated Gamma density above. Indeed, since $\lim_{t\rightarrow 0}\e^{-\frac{t}{\theta}}=1$ the limit is
\begin{align}
    \lim_{t \rightarrow 0}\frac{f_{C,k_1,\theta_1}(t)}{f_{C, k_2,\theta_2}(t)} & =  \frac{\theta_2^{k_2}}{\theta_1^{k_1}}\cdot\frac{\Gamma(k_2)}{\Gamma(k_1)}\cdot\lim_{t \rightarrow 0}t^{k_1-k_2}. \nonumber
\end{align}
The latter can only be equal to one if $k_1=k_2=k$ (say). It follows that
\begin{align*}
    \frac{\theta_2^{k}}{\theta_1^{k}}\cdot\frac{\Gamma(k)}{\Gamma(k)}=1 \Leftrightarrow \left(\frac{\theta_2}{\theta_1}\right)^k=1\Leftrightarrow \theta_1=\theta_2.
\end{align*}

We next consider the case in which $t$ goes to infinity:
\begin{align}
    \lim_{t \rightarrow \infty}\frac{f_{C,k_1,\theta_1}(t)}{f_{C, k_2,\theta_2}(t)}  & =  \frac{\theta_2^{k_2}}{\theta_1^{k_1}}\cdot\frac{\Gamma(k_2)}{\Gamma(k_1)}\cdot\lim_{t \rightarrow \infty}t^{k_1-k_2}\cdot \e^{t\left(\frac{1}{\theta_2}-\frac{1}{\theta_1}\right)}. \label{limGinf}
\end{align}
The limit in (\ref{limGinf}) can only be equal to one if $k_1=k_2$ and $1/\theta_1=1/\theta_2$, which is obviously equivalent to $\theta_1=\theta_2$. \hfill $\Box$

\medskip

\noindent
{\bf Proof of Theorem 3.2(c).} 
We will show that conditions (6), (7) and (8) are satisfied for the Joe, Gaussian and rotated Clayton copulas.

The Joe copula is in the family of Archimedean copulas and is given by
\begin{equation}
\C_{\theta}(u,v)=1-\left\{ \left( 1-u \right)^{\theta}+\left( 1-v \right)^{\theta}-\left( 1-u \right)^{\theta}\left( 1-v \right)^{\theta} \right\}^{\frac{1}{\theta}}, \label{eq:joe}
\end{equation}
for $\theta\in[1, \infty)$.
Hence it is easily seen that
\begin{equation}
h_{T|C, \theta}(u|v)=\frac{\partial}{\partial v}\C_{\theta}(u,v)=\frac{\left( 1-v \right)^{\theta-1}\left\{ 1-\left( 1-u \right)^{\theta} \right\}}{\left\{ \left( 1-u \right)^{\theta}+\left( 1-v \right)^{\theta}-\left( 1-u \right)^{\theta}\left( 1-v \right)^{\theta} \right\}^{\frac{\theta-1}{\theta}}}, \label{eq:joehTC}
\end{equation}
and similarly for $h_{C|T, \theta}(v|u)$.

The verification of (6) and (7) is obvious, since it is clear that $\lim_{y \rightarrow 0}\{1-pF_{U, \theta_U}(y)^{\theta}\} = \lim_{y \rightarrow 0}\{1-F_{C, \theta_C}(y)^{\theta}\} = 1$, and hence
\begin{equation*}
\lim_{y\rightarrow 0} h_{T|C, \theta}\{pF_{U, \theta_U}(y)|F_{C, \theta_C}(y)\} = 0 =  \lim_{y\rightarrow 0} h_{C|T, \theta}\{F_{C, \theta_C}(y)|pF_{U, \theta_U}(y)\},
\end{equation*}
for all $\theta, \theta_C, \theta_U, p$.

To show (8), note that $h_{T|C, \theta}\{p|F_{C}(y)\}$ can be rewritten as
$$h_{T|C, \theta}\{p|F_{C}(y)\}=\frac{g_{\theta,p}\left\{ 1-F_{C}(y) \right\}^{\theta-1}}{\left( 1-g_{\theta,p}\left[ 1-\left\{1-F_{C}(y)\right\}^{\theta} \right] \right)^{\frac{\theta-1}{\theta}}},$$
where $g_{\theta,p} = 1-\left(1-p\right)^{\theta}$.

The derivative $h'_{T|C, \theta}\{p|F_{C}(y)\} = (\text{d}/\text{dy}) h_{T|C, \theta}\{p|F_{C}(y)\}$ equals
\begin{align*}
    & h'_{T|C, \theta}\{p|F_{C}(y)\} \\
    & \quad =  \frac{-f_{C}(y)g_{\theta,p}\left( \theta-1 \right)\left\{ 1-F_{C}(y) \right\}^{\theta-2}\left\{1-g_{\theta,p}\left\{ 1-F_{C}(y) \right\}^{\theta}\left( 1-g_{\theta,p}\left[ 1-\left\{1-F_{C}(y)\right\}^{\theta} \right] \right)^{-1}\right\}}{\left( 1-g_{\theta,p}\left[ 1-\left\{1-F_{C}(y)\right\}^{\theta} \right] \right)^{\frac{\theta-1}{\theta}}}.
\end{align*}
Considering the case $\tau<y<\infty$, we have that
\begin{align*}
    & \frac{h'_{T|C, \theta}\{p|F_{C}(y)\}}{h_{T|C, \theta}\{p|F_{C}(y)\}} \\
    & \quad\quad = -f_{C}(y)\left( \theta-1 \right)\left\{ 1-F_{C}(y) \right\}^{-1}\left\{1-g_{\theta,p}\left\{ 1-F_{C}(y) \right\}^{\theta}\left( 1-g_{\theta,p}\left[ 1-\left\{1-F_{C}(y)\right\}^{\theta} \right] \right)^{-1}\right\}.
\end{align*}
Since $h_{T|C, \theta}\{p|F_{C}(y)\}=h_{T|C, \tilde\theta}\{\tilde p|F_{C}(y)\}$ for all $y>\tau$, we have that
$$\lim_{y \rightarrow \infty}\frac{h'_{T|C, \theta}\{p|F_{C}(y)\}}{h_{T|C, \theta}\{p|F_{C}(y)\}}=\lim_{y \rightarrow \infty}\frac{h'_{T|C, \tilde\theta}\{\tilde p|F_{C}(y)\}}{h_{T|C, \tilde \theta}\{\tilde p|F_{C}(y)\}},$$
which, since $\lim_{y \rightarrow \infty}\left\{1-F_{C}(y)\right\}=0$, yields $\theta-1=\tilde\theta-1 \Leftrightarrow \theta=\tilde\theta$.

Injecting $\theta=\tilde\theta$ in $h'_{T|C, \theta}\{p|F_{C}(y)\}/h_{T|C, \theta}\{p|F_{C}(y)\}=h'_{T|C, \tilde\theta}\{\tilde p|F_{C}(y)\}/h_{T|C, \tilde \theta}\{\tilde p|F_{C}(y)\}$ and some straightforward calculations shows that $p=\tilde p$.

Next we consider the Gaussian copula which is in the family of elliptical copulas and which is given by
$$\C_{\theta}(u,v)=\Phi_{\theta}\left\{\Phi^{-1}(u),\Phi^{-1}(v)\right\},$$
with $\theta \in (-1,1)$, $\Phi^{-1}$ the inverse cumulative distribution function of a standard normal and $\Phi_{\theta}$ the cumulative distribution of a bivariate standard normal random vector with correlation $\theta$.

The calculations of the conditional distributions given below extend those given in \cite{CzadoVanKeilegom} to the case with a cure fraction. If the dependence between $T$ and $C$ is modelled by a Gaussian copula, then by definition
$$\Proba\left(T \leq t, C \leq c\right)=\C_{\theta}\{F_T(t),F_C(c)\}=\Phi_{\theta}\left[\Phi^{-1}\{F_T(t)\},\Phi^{-1}\{F_C(c)\}\right].$$
This leads to
$$\Proba\left[\Phi^{-1}\{F_T(t)\} \leq t,\Phi^{-1}\{F_C(c)\} \leq c\right] = \Proba\left[T \leq F_T^{-1}\{\Phi(t)\}, C \leq F_C^{-1}\{\Phi(c)\} \right] = \Phi_{\theta}\left(t, c\right).$$
Hence $\Phi^{-1}\{F_T(T)\} | \Phi^{-1}\{F_C(C)\} \sim \mathcal{N}\left(\theta\Phi^{-1}\{F_C(C)\}, 1-\theta^2 \right)$ and the conditional distribution is
\begin{align}
   h_{T|C, \theta}\left\{pF_U(y)|F_C(y)\right\} & = F_{T|C}(y|y) \nonumber\\
   & = \Proba\left[ \Phi^{-1}\{F_T(T)\} \leq \Phi^{-1}\{pF_U(y)\} | \Phi^{-1}\{F_C(C)\}=\Phi^{-1}\{F_C(y)\} \right] \nonumber\\
   & = \Phi\left[\frac{\Phi^{-1}\{pF_U(y)\}-\theta\Phi^{-1}\{F_C(y)\}}{\sqrt{1-\theta^2}}\right]. \nonumber
\end{align}
With a similar reasoning we obtain the other conditional distribution:
$$h_{C|T, \theta}\left\{F_C(y)|pF_U(y)\right\}=\Phi\left[\frac{\Phi^{-1}\{F_C(y)\}-\theta\Phi^{-1}\{pF_U(y)\}}{\sqrt{1-\theta^2}}\right].$$
For the verification of (6), note that $\Phi^{-1}(1)=\infty$ and $\Phi^{-1}(0)=-\infty$, and hence we can easily see that $\lim_{y \rightarrow \infty}h_{T|C, \theta}\left\{pF_U(y)|F_C(y)\right\}=0$ for $\theta > 0$ and $\lim_{y \rightarrow 0}h_{T|C, \theta}\left\{pF_U(y)|F_C(y)\right\}=0$ for $\theta\leq 0$. This shows that (6) is satisfied if $\Theta$ is an open interval including the true $\theta$ but not including zero.
Similarly, for the verification of condition (7), note that $\lim_{y \rightarrow 0}h_{C|T, \theta}\left\{F_C(y)|pF_U(y)\right\}=0$ if and only if $\lim_{y \rightarrow 0} \left[\Phi^{-1}\{F_C(y)\}-\theta\Phi^{-1}\{pF_U(y)\}\right] = -\infty$, which is always satisfied for $\theta \le 0$, whereas for $\theta>0$ the limit depends on the marginal distributions $F_U$ and $F_C$.

To show (8), since $\Phi$ is strictly increasing, we need to show that if $g_{\theta,p}(y) = g_{\tilde\theta,\tilde p}(y)$ for all $y>\tau$, then $p=\tilde p$ and $\theta=\tilde\theta$, where $g_{\theta,p}(y) = \left[\Phi^{-1}(p)-\theta\Phi^{-1}\{F_C(y)\}\right]/\sqrt{1-\theta^2}$.
Computing the derivative $g'_{\theta,p}(y) = (\text{d}/\text{dy}) g_{\theta,p}(y)$ and using that $g'_{\theta,p}(y) = g'_{\tilde\theta,\tilde p}(y)$, yields $\theta/\sqrt{1-\theta^2} = \tilde\theta/\sqrt{1-\tilde\theta^2}$, which is equivalent to $\theta=\tilde\theta$ since the function $x \rightarrow x/\sqrt{1-x^2}$ is strictly increasing. Taking this into account, we have $g_{\theta,p}(y) = g_{\theta,\tilde p}(y)$ which directly gives $p=\tilde p$.

We will end this section with the proofs in the case of three rotations of the Clayton copula. To do this, we first need a bit of theory about copula rotations. We will consider copulas $\C_{\beta}\{F_T(t), F_C(c)\}$ where $\beta$ is the angle of rotation of the copula. First of all recall the following equalities:
\begin{equation}
	\C_{90}(u, v)= v-\C(1-u, v), \quad c_{90}(u, v)=c(1-u, v); \label{eq:C_C90}
\end{equation}
\begin{equation}
	\C_{180}(u, v)= u+v-1+\C(1-u, 1-v), \quad c_{180}(u, v)=c(1-u, 1-v); \label{eq:C_C180}
\end{equation}
\begin{equation}
	\C_{270}(u, v)= u-\C(u, 1-v), \quad c_{270}(u, v)=c(u, 1-v). \label{eq:C_C270}
\end{equation}
Since $\C_{90}$, $\C_{180}$ and $\C_{270}$ are also copulas, we can use the formulas derived in Section 2 of the paper, and the formula of the likelihood given in Section 4.
Since the Clayton copula is given by $\C_{\theta}(u,v)=\left(u^{-\theta}+v^{-\theta}-1\right)^{-\frac{1}{\theta}}$ for $\theta\in(0,\infty)$, the Clayton(90) copula equals
\begin{equation*}
    \C_{90, \theta}(u,v)=v-\left\{\left(1-u\right)^{-\theta}+v^{-\theta}-1\right\}^{-\frac{1}{\theta}}\label{eq:Clayton(90)}
\end{equation*}
for $\theta\in(0,\infty)$. Hence it is easily seen that
\begin{equation*}
    h^{90}_{T|C, \theta}(u|v)=\frac{\partial}{\partial v}\C_{90, \theta}(u,v)=1-\left\{ 1-v^{\theta}+\left(\frac{v}{1-u}\right)^{\theta} \right\}^{-\frac{\theta+1}{\theta}}, \label{eq:Clayton(90)hTC}
\end{equation*}
and similarly for $h^{90}_{C|T, \theta}(v|u)$.

The verification of conditions (6) and (7) is obvious, since it is clear that $\lim_{y\rightarrow 0}F_{C, \theta_C}(y)=\lim_{y\rightarrow 0}F_{U, \theta_U}(y)=0$, and hence
\begin{equation*}
    \lim_{y\rightarrow 0} h^{90}_{T|C, \theta}\{pF_{U, \theta_U}(y)|F_{C, \theta_C}(y)\} = 0 = \lim_{y\rightarrow 0} h^{90}_{C|T, \theta}\{F_{C, \theta_C}(y)|pF_{U, \theta_U}(y)\},
\end{equation*}
for all $\theta, \theta_C, \theta_U, p$.

To show (8), we compute $h^{90'}_{T|C, \theta}\{p|F_{C}(y)\} = (d/dy) h^{90}_{T|C, \theta}\{p|F_{C}(y)\}$, which equals
\begin{equation*}
    -\left( \theta+1 \right)f_{C}(y)\left\{ F_{C}(y) \right\}^{\theta-1}\left\{ 1-\left(1-p\right)^{-\theta} \right\}\left[ 1-F_{C}(y)^{\theta}+\left\{\frac{F_{C}(y)}{1-p}\right\}^{\theta} \right]^{-\frac{2\theta+1}{\theta}}.
\end{equation*}
If $h^{90}_{T|C, \theta}\{p|F_{C}(y)\}=h^{90}_{T|C, \tilde\theta}\{\tilde p|F_{C}(y)\}$ for all $y>\tau$, a fortiori $h^{90'}_{T|C, \theta}\{p|F_{C}(y)\}=h^{90'}_{T|C, \tilde\theta}\{\tilde p|F_{C}(y)\}$. Simplifying this equality and then taking the limit as $y$ goes to infinity of both members gives 
$$ (\theta+1)(1-p)^{2\theta+1}\left\{ 1-\left(1-p\right)^{-\theta} \right\}=(\tilde\theta+1)(1-\tilde p)^{2\tilde\theta+1}\left\{ 1-\left(1-\tilde p\right)^{-\tilde\theta} \right\}. $$ 
Combining this with the fact that $\lim_{y \rightarrow \infty}h^{90}_{T|C, \theta}\{p|F_{C}(y)\}=\lim_{y \rightarrow \infty}h^{90}_{T|C, \tilde\theta}\{\tilde p|F_{C}(y)\}$ (i.e.\ \linebreak $(1-p)^{\theta+1} = (1-\tilde p)^{\tilde\theta+1}$) yields $(\theta+1)\big\{(1-p)^{\theta}-1\big\} = (\tilde\theta+1)\big\{(1-\tilde p)^{\tilde\theta}-1\big\}$. Then multiplying each side of the equality by $\left(1-p\right)\left(1-\tilde p\right)$, using the equality between limits once again, and doing some simple algebra leads to
$$ 1-\tilde p =\frac{\left(1-p\right)\left\{\tilde\theta\left(1-p\right)^{\theta}+\left(1- p\right)^{\theta}\right\}}{\theta\left(1-p\right)^{\theta}+\left(1-p\right)^{\theta}-\theta+\tilde\theta}.$$
Suppose now that $\tilde\theta>\theta$ (the case $\theta < \tilde \theta$ can be considered similarly). The above equality can be inserted into $(1-p)^{\theta+1} = (1-\tilde p)^{\tilde\theta+1}$ which leads to
$$\left(1-p\right)^{\theta+1} = \left(\frac{\left(1-p\right)\left(\tilde\theta\left(1-p\right)^{\theta}+\left(1- p\right)^{\theta}\right)}{\theta\left(1-p\right)^{\theta}+\left(1-p\right)^{\theta}-\theta+\tilde\theta}\right)^{\tilde\theta+1}$$
and some simple calculations yields
$$\left(1-p\right)^{\tilde\theta-\theta}=\frac{\left(\theta\left(1-p\right)^{\theta}+\left(1-p\right)^{\theta}-\theta+\tilde\theta\right)^{\tilde\theta+1}}{\left(\tilde\theta\left(1-p\right)^{\theta}+\left(1- p\right)^{\theta}\right)^{\tilde\theta+1}}$$
As $(1-p)^{\tilde\theta-\theta}<1$ we have $\big\{\theta(1-p)^{\theta}+(1-p)^{\theta}-\theta+\tilde\theta\big\}^{\tilde\theta+1} < \big\{\tilde\theta(1-p)^{\theta}+(1- p)^{\theta}\big\}^{\tilde\theta+1}$. Since the function $x \rightarrow x^{1/(\tilde\theta+1)}$ is strictly increasing, we obtain $1<\left(1-p\right)^{\theta}$ which is a contradiction. This shows that $\theta=\tilde \theta$. It now follows from $\lim_{y \rightarrow \infty}h^{90}_{T|C, \theta}\{p|F_{C}(y)\}=\lim_{y \rightarrow \infty}h^{90}_{T|C, \tilde\theta}\{\tilde p|F_{C}(y)\}$ that $p=\tilde p$.

Next, we consider the Clayton(180) copula:
\begin{equation*}
    \C_{180, \theta}(u,v)=u+v-1+\left\{\left(1-u\right)^{-\theta}+\left(1-v\right)^{-\theta}-1\right\}^{-\frac{1}{\theta}}, \label{eq:Clayton(180)}
\end{equation*}
for $\theta\in(0,\infty)$, and hence
\begin{equation*}
    h^{180}_{T|C, \theta}(u|v)=\frac{\partial}{\partial v}\C_{180, \theta}(u,v)=1-\left\{ 1-\left(1-v\right)^{\theta}+\left(\frac{1-v}{1-u}\right)^{\theta} \right\}^{-\frac{\theta+1}{\theta}}, \label{eq:Clayton(180)hTC}
\end{equation*}
and similarly for $h^{180}_{C|T, \theta}(v|u)$. The verification of (6) and (7) is similar to the case of Clayton(90). Indeed, $\lim_{y\rightarrow 0}F_{C, \theta_C}(y)=\lim_{y\rightarrow 0}F_{U, \theta_U}(y)=0$ and hence
\begin{equation*}
    \lim_{y\rightarrow 0} h^{180}_{T|C, \theta}\{pF_{U, \theta_U}(y)|F_{C, \theta_C}(y)\} = 0 = \lim_{y\rightarrow 0} h^{180}_{C|T, \theta}\{F_{C, \theta_C}(y)|pF_{U, \theta_U}(y)\},
\end{equation*}
for all $\theta, \theta_C, \theta_U, p$.

If $h^{180}_{T|C, \theta}\{p|F_{C}(y)\}=h^{180}_{T|C, \tilde\theta}\{\tilde p|F_{C}(y)\}$ for all $y > \tau$, a fortiori $h^{180'}_{T|C, \theta}\{p|F_{C}(y)\}=h^{180'}_{T|C, \tilde\theta}\{\tilde p|F_{C}(y)\}$. Let $g(y) = 1-F_{C}(y)$. Calculating the derivative and making some simplifications gives
$$
\begin{array}{l}
(\theta+1)g(y)^{\theta-1}\left\{ 1-\left( 1-p \right)^{-\theta} \right\}\left[ 1-g(y)^{\theta}+\left\{\frac{g(y)}{1-p}\right\}^{\theta} \right]^{-\frac{\left(2\theta+1\right)}{\theta}}\\
  \quad\quad\quad\quad\quad\quad\quad\quad\quad = (\tilde\theta+1)g(y)^{\tilde\theta-1}\left\{ 1-\left( 1-\tilde p \right)^{-\tilde\theta} \right\}\left[ 1-g(y)^{\tilde\theta}+\left\{\frac{g(y)}{1-\tilde p}\right\}^{\tilde\theta} \right]^{-\frac{\left(2\tilde\theta+1\right)}{\tilde\theta}}.
\end{array}
$$
Taking the derivative once again and simplifying leads to
\begin{align*}
    & \left( \theta+1 \right)g(y)^{\theta-2}\left\{ 1-\left( 1-p \right)^{-\theta} \right\}\left[ 1-g(y)^{\theta}+\left\{\frac{g(y)}{1-p}\right\}^{\theta} \right]^{-\frac{\left(2\theta+1\right)}{\theta}} \\
    & \quad\quad\quad\quad\quad\quad\quad\quad \times \left\{ \left( \theta-1 \right)+\left( 2\theta+1 \right)\frac{\left(1-p\right)^{\theta}-1}{\frac{\left(1-p\right)^{\theta}}{g(y)^{\theta}}-\left(1-p\right)^{\theta}+1} \right\} \\
    & \quad\quad = \left( \tilde\theta+1 \right)g(y)^{\tilde\theta-2}\left\{ 1-\left( 1-\tilde p \right)^{-\tilde\theta} \right\}\left[ 1-g(y)^{\tilde\theta}+\left\{\frac{g(y)}{1-\tilde p}\right\}^{\tilde\theta} \right]^{-\frac{\left(2\tilde\theta+1\right)}{\tilde\theta}} \\
    & \quad\quad\quad\quad\quad\quad\quad\quad\quad\quad \times \left\{ \left( \tilde\theta-1 \right)+\left( 2\tilde\theta+1 \right)\frac{\left(1-\tilde p\right)^{\tilde\theta}-1}{\frac{\left(1-\tilde p\right)^{\tilde\theta}}{g(y)^{\tilde\theta}}-\left(1-\tilde p\right)^{\tilde\theta}+1} \right\}.
\end{align*}
Dividing this equality by the previous one and solving the resulting equality with basic algebra gives $\theta=\tilde\theta$. Inserting this into the initial equality $h^{180}_{T|C, \theta}\{p|F_{C}(y)\}=h^{180}_{T|C, \tilde\theta}\{\tilde p|F_{C}(y)\}$ gives $p=\tilde p$.

Finally, we consider the Clayton(270) copula:
\begin{equation*}
    \C_{270, \theta}(u,v)=u-\left\{u^{-\theta}+\left(1-v\right)^{-\theta}-1\right\}^{-\frac{1}{\theta}}\label{eq:Clayton(270)}
\end{equation*}
for $\theta\in(0,\infty)$. Hence,
\begin{equation*}
    h^{270}_{T|C, \theta}(u|v)=\frac{\partial}{\partial v}\C_{270, \theta}(u,v)=\left\{ 1-\left(1-v\right)^{\theta}+\left(\frac{1-v}{u}\right)^{\theta} \right\}^{-\frac{\theta+1}{\theta}}, \label{eq:Clayton(270)hTC}
\end{equation*}
and similarly for $h^{270}_{C|T, \theta}(v|u)$. For the verification of (6) and (7), once again, since $\lim_{y\rightarrow 0}F_{C, \theta_C}(y)=\lim_{y\rightarrow 0}F_{U, \theta_U}(y)=0$, we directly obtain
\begin{equation*}
    \lim_{y\rightarrow 0} h^{270}_{T|C, \theta}\{pF_{U, \theta_U}(y)|F_{C, \theta_C}(y)\} = 0 = \lim_{y\rightarrow 0} h^{270}_{C|T, \theta}\{F_{C, \theta_C}(y)|pF_{U, \theta_U}(y)\},
\end{equation*}
for all $\theta, \theta_C, \theta_U, p$. To shows (8) we need to proceed in a similar way as for Clayton(180). With the same notation as above, i.e. $g(y) = 1-F_{C}(y)$, we compute and simplify $h^{270'}_{T|C, \theta}\{p|F_{C}(y))\}=h^{270'}_{T|C, \tilde\theta}\{\tilde p|F_{C}(y))\}$:
\begin{align}
    & (\theta+1)g(y)^{\theta-1}\left(1-p^{-\theta}\right)\left[ 1-g(y)^{\theta}+\left\{\frac{g(y)}{p}\right\}^{\theta} \right]^{-\frac{\left(2\theta+1\right)}{\theta}} \nonumber\\
   & \quad\quad\quad\quad\quad\quad = (\tilde\theta+1)g(y)^{\theta-1}\left(1-\tilde p^{-\tilde\theta}\right)\left[ 1-g(y)^{\tilde\theta}+\left\{\frac{g(y)}{\tilde p}\right\}^{\tilde\theta} \right]^{-\frac{\left(2\tilde\theta+1\right)}{\tilde\theta}}. \label{270I}
\end{align}
Taking, once more, the derivative at both sides of this equality and simplifying the result leads to
\begin{align}
    & \left(\theta+1\right)\left(1-p^{-\theta}\right)g(y)^{\theta-2}\left[ 1-g(y)^{\theta}+\left\{\frac{g(y)}{p}\right\}^{\theta} \right]^{-\frac{\left(2\theta+1\right)}{\theta}} \nonumber \\
  & \quad\quad\quad\quad \times\left( \left(\theta-1\right)+\left\{1-F_{C}(y)\right\}^{\theta}\left(2\theta+1\right)\left[ 1-g(y)^{\theta}+\left\{\frac{g(y)}{p}\right\}^{\theta} \right]^{-1}\left(1-p^{-\theta}\right) \right) \nonumber\\
   & \quad\quad =(\tilde\theta+1)(1-\tilde p^{-\tilde\theta})g(y)^{\tilde\theta-2}\left[ 1-g(y)^{\tilde\theta}+\left\{\frac{g(y)}{\tilde p}\right\}^{\tilde\theta} \right]^{-\frac{\left(2\tilde\theta+1\right)}{\tilde\theta}} \nonumber \\
  & \quad\quad\quad\quad \times\left( \left(\tilde\theta-1\right)+\left\{1-F_{C}(y)\right\}^{\tilde\theta}\left(2\tilde\theta+1\right)\left[ 1-g(y)^{\tilde\theta}+\left\{\frac{g(y)}{\tilde p}\right\}^{\tilde\theta} \right]^{-1}\left(1-\tilde p^{-\tilde\theta}\right) \right). \label{270II}
\end{align}

Considering the case $\tau<y<\infty$, we can divide (\ref{270I}) by $g(y)\times(\ref{270II})$, and then take the limit as $y$ goes to infinity. Solving the so-obtained equality should then lead to $p=\tilde p$ and $\theta=\tilde\theta$. This is indeed the case, since solving this equality directly leads to $\theta=\tilde\theta$ and some simple algebra based on $h^{270}_{T|C, \theta}\{p|F_{C}(y)\}=h^{270}_{T|C, \theta}\{\tilde p|F_{C}(y)\}$ shows that $p=\tilde p$.

\bibliographystyle{dcu} 
\bibliography{References}

\newpage

\addcontentsline{toc}{section}{\large Supplement B: Additional results of Veridex data analysis}

\noindent
{\Large {\bf Supplement B: Additional results of Veridex data analysis}}
\medskip

\centering
\section*{Non-truncated distribution for the variable $U$}

\begin{table}[!h]
\centering
\fontsize{7}{7}\selectfont
\begin{tabular}{l|rrrrrrrrr}
 Copula & \multirow{3}{*}{$\tau_K$} & \multirow{3}{*}{$p$} & \multirow{3}{*}{$\theta_{U1}$} & \multirow{3}{*}{$\theta_{U2}$} & \multirow{3}{*}{$\theta_{C1}$} & \multirow{3}{*}{$\theta_{C2}$} & \multirow{3}{*}{Med$_U$} & \multirow{3}{*}{-logLik} & \multirow{3}{*}{AIC} \\
  & & & & & & & & & \\
  Margins & & & & & & & & & \\
  \hline
 Independ. & & & & & & & & \\
 & & & & & & & & \\
 Weibull & & 0.38 & 36.54 & 1.64 & 115.08 & 4.62 & 29.24 & 1480.3 & 2970.6 \\ 
 & & (0.03) & (2.38) & (0.13) & (1.95) & (0.25) & & & \\ 
 Gamma & & 0.38 & 14.77 & 2.25 & 6.03 & 17.43 & 28.42 & 1476.1 & 2962.1 \\ 
 & & (0.03) & (2.47) & (0.31) & (0.63) & (1.79) & & & \\ 
 Log-normal & & 0.40 & 0.81 & 3.34 & 0.25 & 4.63 & 28.13 & 1480.3 & 2970.6 \\ 
 & & (0.03) & (0.07) & (0.10) & (0.01) & (0.02) & & & \\
 \hline
 Frank & & & & & & & & \\
 & & & & & & & & \\
 Weibull & 0.67 & 0.41 & 39.78 & 1.53 & 99.91 & 3.23 & 31.32 & 1476.5 & 2965.0 \\
 & (0.06) & (0.03) & (3.78) & (0.15) & (2.38) & (0.23) & & & \\ 
 Gamma & 0.48 & 0.40 & 16.87 & 2.10 & 7.14 & 13.30 & 30.05 & 1473.5 & 2958.9 \\ 
 & (0.12) & (0.03) & (3.44) & (0.31) & (1.08) & (2.27) & & & \\ 
 Log-normal & 0.49 & 0.44 & 0.88 & 3.47 & 0.27 & 4.52 & 32.02 & 1474.9 & 2961.8 \\ 
 & (0.09) & (0.04) & (0.10) & (0.14) & (0.02) & (0.03) & & & \\
 \hline
 Gumbel & & & & & & & & \\
 & & & & & & & & \\
 Weibull & 0.74 & 0.41 & 40.55 & 1.53 & 99.19 & 3.14 & 31.92 & 1474.8 & 2961.5 \\
 & (0.05) & (0.03) & (4.08) & (0.15) & (2.43) & (0.23) & & & \\ 
 Gamma & 0.55 & 0.40 & 17.14 & 2.09 & 7.50 & 12.52 & 30.25 & 1474.2 & 2960.4 \\ 
 & (0.12) & (0.03) & (3.59) & (0.31) & (1.32) & (2.49) & & & \\ 
 Log-normal & 0.54 & 0.45 & 0.89 & 3.47 & 0.28 & 4.51 & 32.20 & 1476.3 & 2964.6 \\ 
 & (0.10) & (0.05) & (0.10) & (0.15) & (0.02) & (0.03) & & & \\
 \hline
 \it{Joe} & & & & & & & & \\
 & & & & & & & & \\
 Weibull & 0.69 & 0.38 & 40.40 & 1.57 & 103.56 & 3.66 & 32.01 & 1479.3 & 2970.6 \\
 & (0.09) & (0.03) & (3.96) & (0.16) & (2.61) & (0.31) & & & \\ 
 \it{Gamma} & \it{0.66} & \it{0.41} & \it{18.45} & \it{2.02} & \it{7.52} & \it{12.36} & \it{31.23} & \it{1472.3} & \it{2956.7} \\ 
 & (0.07) & (0.03) & (4.16) & (0.31) & (1.11) & (1.98) & & & \\ 
 Log-normal & 0.65 & 0.47 & 0.93 & 3.55 & 0.28 & 4.50 & 34.83 & 1472.3 & 2956.7 \\ 
 & (0.06) & (0.05) & (0.11) & (0.18) & (0.02) & (0.02) & & & \\
 \hline
 Clayton(90) & & & & & & & & \\
 & & & & & & & & \\
 Weibull & -0.52 & 0.38 & 35.80 & 1.68 & 125.69 & 4.92 & 28.77 & 1476.1 & 2964.2 \\
 & (0.08) & (0.03) & (2.20) & (0.13) & (2.90) & (0.31) & & & \\ 
 Gamma & -0.00 & 0.38 & 14.80 & 2.24 & 6.05 & 17.38 & 28.43 & 1476.1 & 2964.1 \\ 
 & (0.04) & (0.03) & (0.09) & (0.08) & (0.08) & (0.13) & & & \\ 
 Log-normal & -0.00 & 0.40 & 0.81 & 3.34 & 0.25 & 4.62 & 28.17 & 1480.3 & 2972.6 \\ 
 & (0.02) & (0.03) & (0.01) & (0.03) & (0.01) & (0.02) & & & \\
 \hline
 Clayton(180) & & & & & & & & \\
 & & & & & & & & \\
 Weibull & 0.76 & 0.40 & 39.05 & 1.53 & 99.79 & 3.22 & 30.73 & 1476.8 & 2965.6 \\
 & (0.05) & (0.03) & (3.39) & (0.14) & (2.27) & (0.21) & & & \\ 
 Gamma & 0.66 & 0.41 & 18.42 & 2.02 & 7.53 & 12.35 & 31.22 & 1472.5 & 2957.1 \\ 
 & (0.08) & (0.03) & (4.16) & (0.31) & (1.12) & (2.01) & & & \\ 
 Log-normal & 0.65 & 0.47 & 0.93 & 3.55 & 0.28 & 4.50 & 34.75 & 1472.7 & 2957.3 \\ 
 & (0.07) & (0.05) & (0.11) & (0.18) & (0.02) & (0.02) & & & \\
 \hline
 Clayton(270) & & & & & & & & \\
 & & & & & & & & \\
 Weibull & -0.01 & 0.38 & 36.53 & 1.65 & 115.50 & 4.61 & 29.23 & 1480.3 & 2972.6 \\
 & (0.14) & (0.03) & (2.38) & (0.13) & (4.54) & (0.25) & & & \\ 
 Gamma & -0.26 & 0.38 & 14.38 & 2.28 & 7.79 & 14.51 & 28.15 & 1475.7 & 2963.4 \\ 
 & (0.21) & (0.03) & (2.34) & (0.31) & (2.06) & (3.00) & & & \\ 
 Log-normal & -0.41 & 0.39 & 0.78 & 3.30 & 0.30 & 4.74 & 27.08 & 1478.8 & 2969.6 \\ 
 & (0.15) & (0.03) & (0.07) & (0.09) & (0.03) & (0.05) & & & \\
 \hline
 Gaussian & & & & & & & & \\
 & & & & & & & & \\
 Weibull & 0.70 & 0.41 & 40.06 & 1.55 & 99.30 & 3.13 & 31.64 & 1475.0 & 2962.0 \\
 & (0.06) & (0.03) & (3.79) & (0.14) & (2.58) & (0.25) & & & \\ 
 Gamma & 0.32 & 0.39 & 15.81 & 2.17 & 6.62 & 14.78 & 29.24 & 1475.7 & 2963.4 \\ 
 & (0.28) & (0.03) & (3.12) & (0.32) & (1.49) & (4.16) & & & \\ 
 Log-normal & -0.64 & 0.39 & 0.76 & 3.27 & 0.32 & 4.79 & 26.39 & 1479.2 & 2970.5 \\ 
 & (0.09) & (0.03) & (0.06) & (0.08) & (0.03) & (0.03) & & & \\
\end{tabular}
\smallskip
\caption{Results for the Veridex data in the case of non-truncated distributions for $U$. For each combination of copulas and margins, the first row gives the estimated value, while the second row gives the standard error based on Fisher's information.}
\label{tblVeridex}
\end{table}

\newpage

\centering
\section*{Truncated distribution for the variable $U$}

\begin{table}[!h]
\centering
\fontsize{7}{7}\selectfont
\begin{tabular}{l|rrrrrrrrr}
  Copula & \multirow{3}{*}{$\tau_K$} & \multirow{3}{*}{$p$} & \multirow{3}{*}{$\theta_{U1}$} & \multirow{3}{*}{$\theta_{U2}$} & \multirow{3}{*}{$\theta_{C1}$} & \multirow{3}{*}{$\theta_{C2}$} & \multirow{3}{*}{Med$_U$} & \multirow{3}{*}{-logLik} & \multirow{3}{*}{AIC} \\
   & & & & & & & & & \\
  Margins & & & & & & & & & \\
  \hline
 Independ. & & & & & & & & & \\
 & & & & & & & & & \\
 Weibull & & 0.38 & 39.72 & 1.52 & 115.08 & 4.61 & 29.64 & 1477.6 & 2965.1 \\
  & & (0.03) & (3.87) & (0.15) & (1.95) & (0.25) & & & \\ 
 Gamma & & 0.38 & 18.86 & 1.98 & 6.03 & 17.42 & 29.07 & 1472.6 & 2955.1 \\ 
  & & (0.03) & (4.44) & (0.31) & (0.63) & (1.78) & & & \\ 
 Log-normal & & 0.38 & 0.95 & 3.57 & 0.25 & 4.63 & 28.11 & 1474.1 & 2958.2 \\ 
  & & (0.03) & (0.12) & (0.19) & (0.01) & (0.02) & & & \\
 \hline
 Frank & & & & & & & & & \\ 
  & & & & & & & & & \\
 Weibull & -0.36 & 0.38 & 39.16 & 1.53 & 123.55 & 4.81 & 29.38 & 1474.5 & 2961.0 \\
 & (0.10) & (0.03) & (3.65) & (0.15) & (3.39) & (0.29) & & & \\ 
 Gamma & 0.43 & 0.39 & 20.34 & 1.92 & 6.92 & 13.84 & 29.82 & 1470.8 & 2953.6 \\ 
 & (0.13) & (0.03) & (5.21) & (0.31) & (1.04) & (2.39) & & & \\ 
 Log-normal & 0.43 & 0.39 & 0.98 & 3.65 & 0.27 & 4.53 & 29.00 & 1470.7 & 2953.3 \\ 
 & (0.10) & (0.03) & (0.13) & (0.23) & (0.02) & (0.03) & & & \\ 
 \hline
 Gumbel & & & & & & & & & \\
 & & & & & & & & & \\
 Weibull & 0.73 & 0.40 & 43.73 & 1.46 & 99.39 & 3.16 & 31.14 & 1473.0 & 2958.0 \\
 & (0.05) & (0.03) & (5.97) & (0.15) & (2.45) & (0.23) & & & \\ 
 Gamma & 0.51 & 0.39 & 20.43 & 1.92 & 7.16 & 13.26 & 29.86 & 1471.6 & 2955.3 \\ 
 & (0.15) & (0.03) & (5.28) & (0.31) & (1.35) & (2.87) & & & \\ 
 Log-normal & 0.48 & 0.39 & 0.98 & 3.65 & 0.27 & 4.52 & 28.94 & 1472.2 & 2956.5 \\ 
 & (0.12) & (0.03) & (0.13) & (0.22) & (0.02) & (0.03) & & & \\ 
 \hline
 {\it Joe} & & & & & & & & & \\ 
 & & & & & & & & & \\
 Weibull & 0.77 & 0.39 & 43.61 & 1.46 & 99.55 & 3.21 & 30.98 & 1475.2 & 2962.4 \\
 & (0.05) & (0.03) & (5.89) & (0.15) & (2.28) & (0.22) & & & \\ 
 Gamma & 0.63 & 0.39 & 21.20 & 1.89 & 7.33 & 12.75 & 30.21 & 1470.4 & 2952.8 \\ 
 & (0.08) & (0.03) & (5.63) & (0.31) & (1.09) & (2.08) & & & \\ 
 \it{Log-normal} & \it{0.62} & \it{0.39} & \it{1.00} & \it{3.69} & \it{0.27} & \it{4.51} & \it{29.36} & \it{1469.5} & \it{2950.9} \\ 
 & (0.07) & (0.03) & (0.14) & (0.24) & (0.02) & (0.03) & & & \\ 
 \hline
 Clayton(90) & & & & & & & & & \\
 & & & & & & & & & \\
 Weibull & -0.53 & 0.38 & 38.93 & 1.54 & 125.79 & 4.91 & 29.29 & 1473.0 & 2958.0 \\
 & (0.08) & (0.03) & (3.56) & (0.15) & (2.90) & (0.31) & & & \\ 
 Gamma & -0.63 & 0.37 & 18.14 & 2.01 & 9.10 & 13.25 & 28.70 & 1472.4 & 2956.8 \\ 
 & (0.09) & (0.03) & (4.11) & (0.31) & (1.84) & (2.30) & & & \\ 
 Log-normal & -0.72 & 0.38 & 0.93 & 3.53 & 0.32 & 4.79 & 27.70 & 1472.7 & 2957.5 \\ 
 & (0.06) & (0.03) & (0.12) & (0.17) & (0.03) & (0.03) & & & \\ 
 \hline
 Clayton(180) & & & & & & & & & \\
 & & & & & & & & & \\
 Weibull & 0.77 & 0.39 & 43.65 & 1.46 & 99.52 & 3.21 & 30.99 & 1475.1 & 2962.2 \\
 & (0.05) & (0.03) & (5.91) & (0.15) & (2.29) & (0.22) & & & \\ 
 Gamma & 0.63 & 0.39 & 21.21 & 1.89 & 7.30 & 12.80 & 30.21 & 1470.5 & 2953.0 \\ 
 & (0.09) & (0.03) & (5.63) & (0.31) & (1.12) & (2.15) & & & \\ 
 Log-normal & 0.61 & 0.39 & 1.00 & 3.69 & 0.27 & 4.51 & 29.38 & 1469.7 & 2951.4 \\ 
 & (0.08) & (0.03) & (0.14) & (0.24) & (0.02) & (0.03) & & & \\ 
 \hline
 Clayton(270) & & & & & & & & & \\
 & & & & & & & & & \\
 Weibull & -0.02 & 0.38 & 39.72 & 1.52 & 115.74 & 4.61 & 29.63 & 1477.6 & 2967.1 \\
 & (0.14) & (0.03) & (3.88) & (0.15) & (4.67) & (0.25) & & & \\ 
 Gamma & -0.30 & 0.38 & 18.56 & 1.99 & 8.19 & 13.98 & 28.92 & 1471.9 & 2955.8 \\ 
 & (0.19) & (0.03) & (4.30) & (0.31) & (2.05) & (2.75) & & & \\ 
 Log-normal & -0.49 & 0.38 & 0.93 & 3.54 & 0.31 & 4.77 & 27.81 & 1471.4 & 2954.7 \\ 
 & (0.13) & (0.03) & (0.12) & (0.18) & (0.03) & (0.05) & & & \\ 
 \hline
 Gaussian & & & & & & & & & \\ 
 & & & & & & & & & \\
 Weibull & 0.69 & 0.40 & 43.68 & 1.47 & 99.52 & 3.16 & 31.19 & 1472.9 & 2957.9 \\ 
 & (0.06) & (0.03) & (5.92) & (0.15) & (2.60) & (0.25) & & & \\ 
 Gamma & -0.57 & 0.37 & 18.13 & 2.01 & 9.78 & 12.41 & 28.71 & 1471.6 & 2955.2 \\ 
 & (0.12) & (0.03) & (4.10) & (0.31) & (2.09) & (2.23) & & & \\ 
 Log-normal & -0.67 & 0.38 & 0.93 & 3.53 & 0.33 & 4.79 & 27.72 & 1470.9 & 2953.8 \\ 
 & (0.08) & (0.03) & (0.12) & (0.17) & (0.03) & (0.03) & & & \\
\end{tabular}
\smallskip
\caption{Results for the Veridex data in the case of truncated distributions for $U$. For each combination of copulas and margins, the first row gives the estimated value, while the second row gives the standard error based on Fisher's information.}
\label{tblVeridexTr}
\end{table}

\newpage

\addcontentsline{toc}{section}{\large Supplement C: Additional simulations}

\noindent
{\Large {\bf Supplement C: Additional simulations}}
\medskip

\centering
\section{Non-truncated distribution for the variable $U$}

\subsection{Incidence of 0.8}

\subsubsection{Percentage of censoring}

\centering
\begin{tabular}{l|c|c|c|c|c|c|c}
Copula / $\tau_{K}$ & 0.00 & 0.20 & 0.50 & 0.80 & -0.20 & -0.50 & -0.80 \\
\hline
Frank & 47.12 & 47.30 & 46.86 & 38.00 & & 45.48 & \\
\hline 
Gumbel & 46.78 & 46.78 & 46.64 & 40.28 &  &  & \\
\hline 
Joe & 46.78 & 48.30 & 48.76 & 42.00 &  &  &  \\
\hline 
Clayton(90) & 47.02 &  &  &  & 43.60 & 42.78 & 43.06 \\
\hline 
Clayton(180) & 47.58 & 47.30 & 48.94 & 40.82 &  &  &  \\
\hline 
Clayton(270) & 46.40 &  &  &  & 49.00 & 47.40 & 45.06 \\
\hline 
Gaussian & 66.70 & 71.58 & 76.58 & 83.46 &  & 60.98 & \\
\hline
\end{tabular}

\subsubsection{Sample size 200}

\centering
\begin{tabular}{l|c|l|rrrrrr}
Copula & $\tau_{K}$ & & $\tau_{K}$ & $p$ & $\theta_{U1}$ & $\theta_{U2}$ & $\theta_{C1}$ & $\theta_{C2}$ \\ 
  \hline
Frank & 0.00 & Bias & 0.03 & -0.00 & -0.01 & 0.01 & 0.05 & -0.00 \\ 
& & SD & 0.27 & 0.08 & 0.08 & 0.10 & 0.35 & 0.10 \\
& & $\widehat{\mbox{SD}}$ & 0.23 & 0.07 & 0.08 & 0.09 & 0.31 & 0.09 \\
& & RMSE & 0.27 & 0.08 & 0.08 & 0.09 & 0.36 & 0.09 \\
  \cline{2-9}
 & 0.20 & Bias & -0.07 & -0.02 & -0.01 & 0.00 & 0.20 & -0.02 \\ 
& & SD & 0.31 & 0.09 & 0.08 & 0.09 & 0.64 & 0.10 \\ 
& & $\widehat{\mbox{SD}}$ &  0.22 & 0.07 & 0.08 & 0.09 & 0.39 & 0.08 \\ 
& & RMSE & 0.32 & 0.09 & 0.08 & 0.09 & 0.67 & 0.10 \\ 
  \cline{2-9}
& 0.50 & Bias & -0.26 & -0.04 & -0.00 & -0.00 & 0.95 & -0.05 \\ 
& & SD &  0.52 & 0.11 & 0.07 & 0.09 & 2.08 & 0.15 \\ 
& & $\widehat{\mbox{SD}}$ &  0.17 & 0.05 & 0.06 & 0.08 & 0.79 & 0.08 \\ 
& & RMSE &  0.58 & 0.11 & 0.07 & 0.09 & 2.29 & 0.16 \\ 
  \cline{2-9}
& 0.80 & Bias & -0.05 & 0.00 & 0.01 & 0.01 & 0.22 & 0.02 \\ 
& & SD & 0.23 & 0.05 & 0.08 & 0.09 & 1.50 & 0.10 \\ 
& & $\widehat{\mbox{SD}}$ & 0.09 & 0.03 & 0.05 & 0.07 & 0.56 & 0.08 \\ 
& & RMSE &  0.24 & 0.05 & 0.08 & 0.09 & 1.51 & 0.11 \\ 
  \cline{2-9}
& -0.50 & Bias & 0.33 & 0.06 & -0.03 & 0.08 & -0.18 & 0.11 \\ 
& & SD & 0.49 & 0.13 & 0.12 & 0.13 & 0.32 & 0.17 \\ 
& & $\widehat{\mbox{SD}}$ & 0.20 & 0.12 & 0.13 & 0.10 & 0.20 & 0.12 \\ 
& & RMSE & 0.59 & 0.14 & 0.13 & 0.15 & 0.37 & 0.20 \\ 
\hline 
\end{tabular}

\begin{figure}[p]
\centering
\begin{tabular}{l|c|l|rrrrrr}
Copula & $\tau_{K}$ & & $\tau_{K}$ & $p$ & $\theta_{U1}$ & $\theta_{U2}$ & $\theta_{C1}$ & $\theta_{C2}$ \\ 
  \hline
Gumbel & 0.00 & Bias & 0.24 & 0.07 & 0.01 & 0.01 & -0.18 & 0.05 \\
& & SD & 0.24 & 0.08 & 0.08 & 0.09 & 0.18 & 0.09 \\
& & $\widehat{\mbox{SD}}$ & 0.18 & 0.06 & 0.08 & 0.08 & 0.15 & 0.09 \\ 
& & RMSE & 0.34 & 0.10 & 0.08 & 0.09 & 0.26 & 0.10 \\
  \cline{2-9}
& 0.20 & Bias & 0.05 & 0.01 & 0.00 & -0.00 & -0.01 & -0.00 \\ 
& & SD & 0.19 & 0.07 & 0.09 & 0.10 & 0.19 & 0.07 \\ 
& & $\widehat{\mbox{SD}}$ & 0.13 & 0.06 & 0.07 & 0.05 & 0.12 & 0.07 \\
\hline 
\end{tabular}
\end{figure}

\begin{figure}[p]
\centering
\begin{tabular}{l|c|l|rrrrrr}
Copula & $\tau_{K}$ & & $\tau_{K}$ & $p$ & $\theta_{U1}$ & $\theta_{U2}$ & $\theta_{C1}$ & $\theta_{C2}$ \\ 
  \hline
Clayton(180) & 0.00 & Bias & 0.18 & 0.06 & 0.01 & 0.03 & -0.16 & 0.07 \\ 
& & SD & 0.20 & 0.08 & 0.08 & 0.10 & 0.16 & 0.09 \\
& & $\widehat{\mbox{SD}}$ & 0.12 & 0.06 & 0.07 & 0.07 & 0.12 & 0.01 \\
& & RMSE & 0.27 & 0.10 & 0.08 & 0.10 & 0.23 & 0.11 \\
  \cline{2-9}
& 0.20 & Bias & 0.01 & 0.00 & 0.01 & 0.00 & 0.02 & -0.00 \\ 
& & SD & 0.14 & 0.06 & 0.08 & 0.10 & 0.17 & 0.07 \\ 
& & $\widehat{\mbox{SD}}$ & 0.13 & 0.07 & 0.08 & 0.09 & 0.15 & 0.07 \\
\hline 
\end{tabular}
\end{figure}

\begin{figure}[p]
\centering
\begin{tabular}{l|l|c|l|rrrrrr}
Copula & Copula & \multicolumn{8}{c}{} \\
data gen. & estim. & $\tau_{K}$ & & $\tau_{K}$ & $p$ & $\theta_{U1}$ & $\theta_{U2}$ & $\theta_{C1}$ & $\theta_{C2}$ \\
  \hline
Gumbel & Frank & 0.20 & Bias & -0.10 & -0.05 & -0.02 & 0.00 & 0.17 & -0.04 \\
& & & SD & 0.20 & 0.06 & 0.07 & 0.10 & 0.25 & 0.07 \\
& & & $\widehat{\mbox{SD}}$ & 0.21 & 0.07 & 0.07 & 0.09 & 0.29 & 0.07 \\
& & & RMSE & 0.22 & 0.08 & 0.08 & 0.10 & 0.30 & 0.08 \\
  \cline{3-10}
& & 0.50 & Bias & -0.13 & -0.05 & -0.02 & 0.02 & 0.16 & -0.01 \\
& & & SD & 0.17 & 0.05 & 0.06 & 0.09 & 0.19 & 0.07 \\
\hline
\end{tabular}
\end{figure}

\subsubsection{Sample size 500}

\centering
\begin{tabular}{l|c|l|rrrrrr}
Copula & $\tau_{K}$ & & $\tau_{K}$ & $p$ & $\theta_{U1}$ & $\theta_{U2}$ & $\theta_{C1}$ & $\theta_{C2}$ \\ 
  \hline
Frank & 0.00 & Bias & 0.00 & -0.00 & -0.00 & 0.00 & 0.03 & -0.01 \\ 
& & SD & 0.18 & 0.06 & 0.05 & 0.06 & 0.22 & 0.06 \\ 
& & $\widehat{\mbox{SD}}$ & 0.17 & 0.05 & 0.05 & 0.06 & 0.21 & 0.06 \\ 
& & RMSE & 0.18 & 0.05 & 0.04 & 0.06 & 0.22 & 0.06 \\
  \cline{2-9}
& 0.20 & Bias &  -0.03 & -0.01 & 0.00 & -0.00 & 0.08 & -0.01 \\ 
& & SD &  0.21 & 0.05 & 0.05 & 0.06 & 0.39 & 0.07 \\ 
& & $\widehat{\mbox{SD}}$ &  0.14 & 0.04 & 0.04 & 0.05 & 0.20 & 0.05 \\
\hline 
\end{tabular}

\begin{figure}[p]
\centering
\begin{tabular}{l|c|l|rrrrrr}
Copula & $\tau_{K}$ & & $\tau_{K}$ & $p$ & $\theta_{U1}$ & $\theta_{U2}$ & $\theta_{C1}$ & $\theta_{C2}$ \\ 
  \hline
Joe & 0.00 & Bias & 0.09 & 0.05 & 0.03 & 0.01 & -0.10 & 0.05 \\ 
& & SD & 0.11 & 0.06 & 0.06 & 0.06 & 0.12 & 0.07 \\ 
& & $\widehat{\mbox{SD}}$ & 0.07 & 0.05 & 0.05 & 0.06 & 0.09 & 0.06 \\
& & RMSE & 0.14 & 0.07 & 0.06 & 0.05 & 0.15 & 0.08 \\ 
  \cline{2-9}
& 0.20 & Bias & -0.01 & -0.00 & 0.00 & -0.00 & 0.03 & -0.01 \\ 
& & SD & 0.10 & 0.04 & 0.05 & 0.06 & 0.14 & 0.05 \\ 
& & $\widehat{\mbox{SD}}$ & 0.08 & 0.04 & 0.04 & 0.05 & 0.10 & 0.04 \\
\hline 
\end{tabular}
\end{figure}

\begin{figure}[p]
\centering
\begin{tabular}{l|c|l|rrrrrr}
Copula & $\tau_{K}$ & & $\tau_{K}$ & $p$ & $\theta_{U1}$ & $\theta_{U2}$ & $\theta_{C1}$ & $\theta_{C2}$ \\ 
  \hline
Clayton(270) & 0.00 & Bias & 0.12 & -0.02 & 0.01 & -0.01 & 0.21 & -0.04 \\
& & SD & 0.17 & 0.05 & 0.05 & 0.06 & 0.32 & 0.08 \\
& & $\widehat{\mbox{SD}}$ & 0.17 & 0.06 & 0.06 & 0.07 & 0.33 & 0.09 \\
& & RMSE & 0.21 & 0.04 & 0.05 & 0.06 & 0.38 & 0.09 \\
  \cline{2-9}
& -0.20 & Bias & 0.01 & -0.01 & -0.00 & 0.00 & 0.03 & 0.00 \\ 
& & SD & 0.19 & 0.06 & 0.06 & 0.06 & 0.26 & 0.09 \\ 
& & $\widehat{\mbox{SD}}$ & 0.19 & 0.06 & 0.06 & 0.07 & 0.27 & 0.10 \\
\hline 
\end{tabular}
\end{figure}

\begin{figure}[p]
\centering
\begin{tabular}{l|l|c|l|rrrrrr}
Copula & Copula & \multicolumn{8}{c}{} \\
data gen. & estim. & $\tau_{K}$ & & $\tau_{K}$ & $p$ & $\theta_{U1}$ & $\theta_{U2}$ & $\theta_{C1}$ & $\theta_{C2}$ \\
  \hline
Gumbel & Frank & 0.20 & Bias & -0.12 & -0.05 & -0.02 & -0.00 & 0.17 & -0.04 \\
& & & SD & 0.12 & 0.04 & 0.05 & 0.07 & 0.15 & 0.04 \\
& & & $\widehat{\mbox{SD}}$ & 0.14 & 0.04 & 0.04 & 0.05 & 0.17 & 0.04 \\
& & & RMSE & 0.17 & 0.05 & 0.04 & 0.06 & 0.23 & 0.06 \\
  \cline{3-10}
& & 0.50 & Bias & -0.13 & -0.05 & -0.02 & 0.02 & 0.16 & -0.01 \\
& & & SD & 0.10 & 0.03 & 0.04 & 0.06 & 0.11 & 0.04 \\
\hline
\end{tabular}
\end{figure}

\subsubsection{Sample size 1000}

\centering
\begin{tabular}{l|c|l|rrrrrr}
Copula & $\tau_{K}$ & & $\tau_{K}$ & $p$ & $\theta_{U1}$ & $\theta_{U2}$ & $\theta_{C1}$ & $\theta_{C2}$ \\ 
  \hline
Frank & 0.00 & Bias & -0.00 & -0.00 & 0.00 & -0.00 & 0.02 & -0.00 \\
& & SD & 0.12 & 0.04 & 0.03 & 0.05 & 0.14 & 0.04 \\
& & $\widehat{\mbox{SD}}$ & 0.12 & 0.04 & 0.03 & 0.04 & 0.14 & 0.04 \\
& & RMSE & 0.13 & 0.03 & 0.03 & 0.04 & 0.14 & 0.04 \\
  \cline{2-9}
& 0.20 & Bias &  -0.01 & -0.00 & 0.00 & -0.01 & 0.01 & -0.00 \\ 
& & SD & 0.11 & 0.03 & 0.03 & 0.04 & 0.11 & 0.03 \\ 
& & $\widehat{\mbox{SD}}$ & 0.11 & 0.03 & 0.03 & 0.04 & 0.11 & 0.03 \\
\hline 
\end{tabular}

\begin{figure}[p]
\centering
\begin{tabular}{l|c|l|rrrrrr}
Copula & $\tau_{K}$ & & $\tau_{K}$ & $p$ & $\theta_{U1}$ & $\theta_{U2}$ & $\theta_{C1}$ & $\theta_{C2}$ \\ 
  \hline
Joe & 0.00 & Bias & 0.05 & 0.03 & 0.02 & -0.00 & -0.06 & 0.03 \\
& & SD & 0.07 & 0.04 & 0.04 & 0.04 & 0.09 & 0.05 \\ 
& & $\widehat{\mbox{SD}}$ & 0.05 & 0.04 & 0.04 & 0.04 & 0.07 & 0.04 \\ 
& & RMSE & 0.08 & 0.05 & 0.04 & 0.04 & 0.10 & 0.06 \\ 
  \cline{2-9}
& 0.20 & Bias & -0.03 & -0.01 & 0.00 & -0.01 & 0.04 & -0.01 \\ 
& & SD & 0.08 & 0.03 & 0.03 & 0.05 & 0.11 & 0.04 \\ 
& & $\widehat{\mbox{SD}}$ & 0.06 & 0.03 & 0.03 & 0.04 & 0.07 & 0.03 \\
\hline 
\end{tabular}
\end{figure}

\begin{figure}[p]
\centering
\begin{tabular}{l|c|l|rrrrrr}
Copula & $\tau_{K}$ & & $\tau_{K}$ & $p$ & $\theta_{U1}$ & $\theta_{U2}$ & $\theta_{C1}$ & $\theta_{C2}$ \\ 
  \hline
Clayton(270) & 0.00 & Bias & 0.10 & -0.01 & 0.01 & -0.01 & 0.15 & -0.04 \\ 
& & SD & 0.13 & 0.03 & 0.03 & 0.04 & 0.23 & 0.06 \\
& & $\widehat{\mbox{SD}}$ & 0.28 & 0.04 & 0.04 & 0.05 & 0.39 & 0.10 \\
& & RMSE  & 0.16 & 0.03 & 0.03 & 0.04 & 0.27 & 0.07 \\
  \cline{2-9}
& -0.20 & Bias & 0.01 & -0.00 & -0.00 & 0.00 & 0.01 & 0.00 \\ 
& & SD & 0.16 & 0.04 & 0.04 & 0.04 & 0.20 & 0.08 \\ 
& & $\widehat{\mbox{SD}}$ & 0.17 & 0.04 & 0.04 & 0.04 & 0.22 & 0.08 \\
\hline 
\end{tabular}
\end{figure}

\begin{figure}[p]
\centering
\begin{tabular}{l|l|c|l|rrrrrr}
Copula & Copula & \multicolumn{8}{c}{} \\
data gen. & estim. & $\tau_{K}$ & & $\tau_{K}$ & $p$ & $\theta_{U1}$ & $\theta_{U2}$ & $\theta_{C1}$ & $\theta_{C2}$ \\
  \hline
Gumbel & Frank & 0.20 & Bias & -0.12 & -0.04 & -0.01 & -0.01 & 0.18 & -0.05 \\
& & & SD & 0.09 & 0.03 & 0.03 & 0.05 & 0.11 & 0.03 \\
& & & $\widehat{\mbox{SD}}$ & 0.10 & 0.03 & 0.03 & 0.04 & 0.12 & 0.03 \\
& & & RMSE & 0.15 & 0.05 & 0.03 & 0.04 & 0.21 & 0.05 \\
  \cline{3-10}
& & 0.50 & Bias & -0.13 & -0.04 & -0.02 & 0.01 & 0.16 & -0.01 \\
& & & SD & 0.07 & 0.02 & 0.03 & 0.04 & 0.08 & 0.03 \\
\hline
\end{tabular}
\end{figure}

\subsubsection{Sample size 2500}

\begin{tabular}{l|c|l|rrrrrr}
Copula & $\tau_{K}$ & & $\tau_{K}$ & $p$ & $\theta_{U1}$ & $\theta_{U2}$ & $\theta_{C1}$ & $\theta_{C2}$ \\ 
  \hline
Frank & 0.00 & Bias & -0.01 & -0.00 & 0.00 & -0.00 & 0.01 & -0.00 \\ 
& & SD & 0.08 & 0.02 & 0.02 & 0.03 & 0.08 & 0.02 \\
& & $\widehat{\mbox{SD}}$ & 0.08 & 0.02 & 0.02 & 0.03 & 0.08 & 0.03 \\
& & RMSE & 0.08 & 0.03 & 0.00 & 0.03 & 0.08 & 0.03 \\
  \cline{2-9}
& 0.20 & Bias & -0.01 & 0.00 & 0.00 & -0.00 & 0.01 & 0.00 \\ 
& & SD & 0.06 & 0.02 & 0.02 & 0.03 & 0.06 & 0.02 \\ 
& & $\widehat{\mbox{SD}}$ & 0.07 & 0.02 & 0.02 & 0.02 & 0.06 & 0.02 \\
\hline 
\end{tabular}

\begin{figure}[p]
\centering
\begin{tabular}{l|c|l|rrrrrr}
Copula & $\tau_{K}$ & & $\tau_{K}$ & $p$ & $\theta_{U1}$ & $\theta_{U2}$ & $\theta_{C1}$ & $\theta_{C2}$ \\ 
  \hline
Joe & 0.00 & Bias & 0.02 & 0.01 & 0.01 & -0.00 & -0.03 & 0.02 \\
& & SD & 0.04 & 0.03 & 0.03 & 0.03 & 0.06 & 0.03 \\
& & $\widehat{\mbox{SD}}$ & 0.03 & 0.02 & 0.02 & 0.02 & 0.05 & 0.03 \\
& & RMSE & 0.04 & 0.03 & 0.03 & 0.03 & 0.06 & 0.03 \\
  \cline{2-9}
& 0.20 & Bias & -0.02 & -0.00 & 0.01 & -0.02 & 0.03 & -0.00 \\ 
& & SD & 0.05 & 0.02 & 0.02 & 0.03 & 0.07 & 0.02 \\ 
& & $\widehat{\mbox{SD}}$ & 0.04 & 0.02 & 0.02 & 0.02 & 0.04 & 0.02 \\
\hline 
\end{tabular}
\end{figure}

\begin{figure}[p]
\centering
\begin{tabular}{l|c|l|rrrrrr}
Copula & $\tau_{K}$ & & $\tau_{K}$ & $p$ & $\theta_{U1}$ & $\theta_{U2}$ & $\theta_{C1}$ & $\theta_{C2}$ \\ 
  \hline
Clayton(270) & 0.00 & Bias & 0.08 & -0.01 & 0.01 & -0.01 & 0.11 & -0.03 \\
& & SD & 0.10 & 0.02 & 0.02 & 0.03 & 0.15 & 0.04 \\
& & $\widehat{\mbox{SD}}$ & 0.13 & 0.02 & 0.02 & 0.03 & 0.19 & 0.05 \\
& & RMSE & 0.13 & 0.00 & 0.03 & 0.03 & 0.19 & 0.05 \\
  \cline{2-9}
& -0.20 & Bias & 0.01 & -0.00 & -0.00 & 0.00 & -0.01 & 0.01 \\ 
& & SD & 0.12 & 0.03 & 0.03 & 0.03 & 0.14 & 0.06 \\ 
& & $\widehat{\mbox{SD}}$ & 0.16 & 0.03 & 0.03 & 0.03 & 0.17 & 0.07 \\
\hline 
\end{tabular}
\end{figure}

\begin{figure}[p]
\centering
\begin{tabular}{l|l|c|l|rrrrrr}
Copula & Copula & \multicolumn{8}{c}{} \\
data gen. & estim. & $\tau_{K}$ & & $\tau_{K}$ & $p$ & $\theta_{U1}$ & $\theta_{U2}$ & $\theta_{C1}$ & $\theta_{C2}$ \\
  \hline
Gumbel & Frank & 0.20 & Bias & -0.12 & -0.05 & -0.02 & -0.01 & 0.17 & -0.04 \\
& & & SD & 0.05 & 0.02 & 0.02 & 0.03 & 0.07 & 0.02 \\
& & & $\widehat{\mbox{SD}}$ & 0.06 & 0.02 & 0.02 & 0.02 & 0.07 & 0.02 \\
& & & RMSE & 0.13 & 0.04 & 0.03 & 0.03 & 0.18 & 0.04 \\
  \cline{3-10}
& & 0.50 & Bias & -0.13 & -0.05 & -0.02 & 0.01 & 0.15 & -0.01 \\
& & & SD & 0.04 & 0.01 & 0.02 & 0.03 & 0.05 & 0.02 \\
\hline
\end{tabular}
\end{figure}

\subsection{Incidence of 0.6}

\subsubsection{Percentage of censoring}

\begin{tabular}{l|c|c|c|c|c|c|c}
Copula / $\tau_{K}$ & 0.00 & 0.20 & 0.50 & 0.80 & -0.20 & -0.50 & -0.8 \\
\hline 
Frank & 60.24 & 62.84 & 67.58 & 74.00 & & 55.62 &\\
\hline 
Gumbel & 60.04 & 61.38 & 65.70 & 71.06 & & &  \\
\hline 
Joe & 59.86 & 62.74 & 67.10 & 75.22 & & &  \\
\hline 
Clayton(90) & 60.40 &  &  &  & 56.98 & 52.86 & 52.22 \\
\hline 
Clayton(180) & 60.98 & 61.84 & 68.06 & 71.74 & & &  \\
\hline 
Clayton(270) & 59.86 &  &  &  & 59.60 & 56.54 & 53.60 \\
\hline 
Gaussian & 75.26 & 81.30 & 88.16 & 97.26 & & 65.72 & \\
\hline 
\end{tabular}

\subsubsection{Sample size 200}

\begin{tabular}{l|c|l|rrrrrr}
Copula & $\tau_{K}$ & & $\tau_{K}$ & $p$ & $\theta_{U1}$ & $\theta_{U2}$ & $\theta_{C1}$ & $\theta_{C2}$ \\ 
  \hline
Frank & 0.00 & Bias & 0.04 & 0.02 & -0.01 & 0.02 & 0.02 & -0.01 \\
& & SD & 0.30 & 0.11 & 0.10 & 0.11 & 0.27 & 0.08 \\
& & $\widehat{\mbox{SD}}$ & 0.25 & 0.10 & 0.10 & 0.10 & 0.22 & 0.08 \\
& & RMSE & 0.31 & 0.11 & 0.09 & 0.11 & 0.27 & 0.08 \\
  \cline{2-9}
& 0.20 & Bias & -0.10 & -0.02 & -0.02 & 0.00 & 0.16 & -0.03 \\ 
& & SD & 0.38 & 0.12 & 0.11 & 0.12 & 0.47 & 0.10 \\ 
& & $\widehat{\mbox{SD}}$ & 0.24 & 0.09 & 0.10 & 0.10 & 0.24 & 0.07 \\ 
& & RMSE & 0.39 & 0.12 & 0.11 & 0.12 & 0.50 & 0.10 \\ 
  \cline{2-9} 
& 0.50 & Bias & -0.34 & -0.06 & -0.01 & -0.02 & 0.40 & -0.04 \\ 
& & SD & 0.58 & 0.16 & 0.17 & 0.12 & 0.69 & 0.12 \\ 
& & $\widehat{\mbox{SD}}$ & 0.20 & 0.09 & 0.11 & 0.10 & 0.23 & 0.07 \\ 
& & RMSE & 0.67 & 0.17 & 0.17 & 0.12 & 0.80 & 0.13 \\ 
  \cline{2-9}
& 0.80 & Bias & -0.24 & -0.04 & 0.03 & 0.03 & 0.16 & 0.03 \\ 
& & SD & 0.50 & 0.18 & 0.29 & 0.14 & 0.37 & 0.09 \\ 
& & $\widehat{\mbox{SD}}$ & 0.16 & 0.11 & 0.10 & 0.12 & 0.13 & 0.07 \\ 
& & RMSE & 0.55 & 0.18 & 0.29 & 0.14 & 0.40 & 0.09 \\ 
  \cline{2-9}
& -0.50 & Bias & 0.32 & 0.12 & 0.02 & 0.07 & -0.16 & 0.08 \\ 
& & SD & 0.47 & 0.16 & 0.15 & 0.13 & 0.26 & 0.14 \\ 
& & $\widehat{\mbox{SD}}$ & 0.21 & 0.12 & 0.19 & 0.11 & 0.16 & 0.11 \\ 
& & RMSE & 0.57 & 0.20 & 0.15 & 0.15 & 0.31 & 0.16 \\ 
\hline 
\end{tabular}

\begin{figure}[p]
\centering
\begin{tabular}{l|c|l|rrrrrr}
Copula & $\tau_{K}$ & & $\tau_{K}$ & $p$ & $\theta_{U1}$ & $\theta_{U2}$ & $\theta_{C1}$ & $\theta_{C2}$ \\ 
  \hline
Gumbel & 0.00 & Bias & 0.23 & 0.11 & 0.06 & 0.01 & -0.13 & 0.02 \\ 
& & SD & 0.24 & 0.12 & 0.13 & 0.11 & 0.14 & 0.07 \\
& & $\widehat{\mbox{SD}}$ & 0.21 & 0.11 & 0.12 & 0.10 & 0.13 & 0.07 \\
& & RMSE & 0.33 & 0.17 & 0.14 & 0.11 & 0.19 & 0.08 \\
  \cline{2-9}
& 0.20 & Bias & 0.06 & 0.04 & 0.03 & -0.00 & -0.01 & -0.01 \\ 
& & SD & 0.24 & 0.11 & 0.13 & 0.12 & 0.15 & 0.07 \\ 
& & $\widehat{\mbox{SD}}$ & 0.21 & 0.11 & 0.13 & 0.10 & 0.13 & 0.07 \\
\hline 
\end{tabular}
\end{figure}

\begin{figure}[p]
\centering
\begin{tabular}{l|c|l|rrrrrr}
Copula & $\tau_{K}$ & & $\tau_{K}$ & $p$ & $\theta_{U1}$ & $\theta_{U2}$ & $\theta_{C1}$ & $\theta_{C2}$ \\ 
  \hline
Clayton(180) & 0.00 & Bias & 0.52 & 0.26 & 0.02 & 0.15 & -0.34 & 0.05 \\ 
& & SD & 0.00 & 0.00 & 0.00 & 0.00 & 0.00 & 0.00 \\
& & $\widehat{\mbox{SD}}$ & 0.13 & 0.06 & 0.06 & 0.10 & 0.06 & 0.07 \\
& & RMSE & 0.52 & 0.26 & 0.00 & 0.15 & 0.34 & 0.04 \\
  \cline{2-9}
& 0.20 & Bias & 0.03 & 0.03 & 0.03 & 0.01 & -0.00 & -0.00 \\ 
& & SD & 0.19 & 0.11 & 0.16 & 0.12 & 0.14 & 0.06 \\ 
& & $\widehat{\mbox{SD}}$ & 0.17 & 0.10 & 0.13 & 0.08 & 0.13 & 0.06 \\
\hline 
\end{tabular}
\end{figure}

\begin{figure}[p]
\centering
\begin{tabular}{l|l|c|l|rrrrrr}
Copula & Copula & \multicolumn{8}{c}{} \\
data gen. & estim. & $\tau_{K}$ & & $\tau_{K}$ & $p$ & $\theta_{U1}$ & $\theta_{U2}$ & $\theta_{C1}$ & $\theta_{C2}$ \\
  \hline
Gumbel & Frank & 0.20 & Bias & -0.07 & -0.03 & -0.02 & 0.01 & 0.08 & -0.02 \\
& & & SD & 0.25 & 0.09 & 0.11 & 0.12 & 0.18 & 0.06 \\
& & & $\widehat{\mbox{SD}}$ & 0.25 & 0.09 & 0.11 & 0.10 & 0.21 & 0.07 \\
& & & RMSE & 0.26 & 0.09 & 0.11 & 0.12 & 0.19 & 0.07 \\
  \cline{3-10}
& & 0.50 & Bias & 0.30 & 0.31 & 0.67 & -0.18 & 0.06 & -0.01 \\
& & & SD & 0.00 & 0.00 & 0.00 & 0.00 & 0.00 & 0.00 \\
\hline
\end{tabular}
\end{figure}

\subsubsection{Sample size 500}

\begin{tabular}{l|c|l|rrrrrr}
Copula & $\tau_{K}$ & & $\tau_{K}$ & $p$ & $\theta_{U1}$ & $\theta_{U2}$ & $\theta_{C1}$ & $\theta_{C2}$ \\ 
  \hline
Frank & 0.00 & Bias & -0.00 & 0.00 & 0.00 & -0.00 & 0.03 & -0.01 \\
& & SD & 0.20 & 0.07 & 0.07 & 0.08 & 0.17 & 0.05 \\
& & $\widehat{\mbox{SD}}$ & 0.18 & 0.07 & 0.06 & 0.07 & 0.15 & 0.05 \\
& & RMSE & 0.20 & 0.07 & 0.06 & 0.08 & 0.18 & 0.04 \\
  \cline{2-9}
& 0.20 & Bias & -0.06 & -0.01 & -0.00 & -0.01 & 0.07 & -0.01 \\ 
& & SD & 0.25 & 0.08 & 0.07 & 0.08 & 0.28 & 0.06 \\ 
& & $\widehat{\mbox{SD}}$ & 0.17 & 0.07 & 0.06 & 0.06 & 0.14 & 0.04 \\
\hline 
\end{tabular}

\begin{figure}[p]
\centering
\begin{tabular}{l|c|l|rrrrrr}
Copula & $\tau_{K}$ & & $\tau_{K}$ & $p$ & $\theta_{U1}$ & $\theta_{U2}$ & $\theta_{C1}$ & $\theta_{C2}$ \\ 
  \hline
Joe & 0.00 & Bias & 0.10 & 0.07 & 0.06 & -0.00 & -0.07 & 0.03 \\
& & SD & 0.12 & 0.09 & 0.10 & 0.07 & 0.10 & 0.05 \\
& & $\widehat{\mbox{SD}}$ & 0.11 & 0.08 & 0.09 & 0.06 & 0.09 & 0.05 \\
& & RMSE & 0.16 & 0.11 & 0.11 & 0.07 & 0.12 & 0.05 \\
  \cline{2-9}
& 0.20 & Bias & -0.01 & 0.01 & 0.02 & -0.01 & 0.02 & -0.01 \\ 
& & SD & 0.12 & 0.07 & 0.08 & 0.08 & 0.10 & 0.04 \\ 
& & $\widehat{\mbox{SD}}$ & 0.11 & 0.07 & 0.08 & 0.06 & 0.09 & 0.04 \\
\hline 
\end{tabular}
\end{figure}

\begin{figure}[p]
\centering
\begin{tabular}{l|c|l|rrrrrr}
Copula & $\tau_{K}$ & & $\tau_{K}$ & $p$ & $\theta_{U1}$ & $\theta_{U2}$ & $\theta_{C1}$ & $\theta_{C2}$ \\ 
  \hline
Clayton(270) & 0.00 & Bias & 0.13 & -0.03 & -0.00 & -0.01 & 0.17 & -0.04 \\
& & SD & 0.16 & 0.05 & 0.06 & 0.07 & 0.24 & 0.07 \\
& & $\widehat{\mbox{SD}}$ & 0.21 & 0.06 & 0.07 & 0.07 & 0.29 & 0.09 \\
& & RMSE & 0.20 & 0.06 & 0.06 & 0.07 & 0.29 & 0.08 \\
  \cline{2-9}
& -0.20 & Bias & 0.01 & -0.00 & -0.01 & 0.01 & 0.01 & 0.00 \\ 
& & SD & 0.18 & 0.07 & 0.06 & 0.07 & 0.21 & 0.09 \\ 
& & $\widehat{\mbox{SD}}$ & 0.23 & 0.07 & 0.07 & 0.12 & 0.24 & 0.10 \\
\hline 
\end{tabular}
\end{figure}

\begin{figure}[p]
\centering
\begin{tabular}{l|l|c|l|rrrrrr}
Copula & Copula & \multicolumn{8}{c}{} \\
data gen. & estim. & $\tau_{K}$ & & $\tau_{K}$ & $p$ & $\theta_{U1}$ & $\theta_{U2}$ & $\theta_{C1}$ & $\theta_{C2}$ \\
  \hline
Gumbel & Frank & 0.20 & Bias & -0.10 & -0.04 & -0.02 & -0.00 & 0.09 & -0.02 \\
& & & SD & 0.15 & 0.06 & 0.06 & 0.08 & 0.11 & 0.04 \\
& & & $\widehat{\mbox{SD}}$ & 0.17 & 0.06 & 0.06 & 0.06 & 0.13 & 0.04 \\
& & & RMSE & 0.18 & 0.06 & 0.06 & 0.08 & 0.14 & 0.04 \\
  \cline{3-10}
& & 0.50 & Bias & -0.13 & -0.05 & -0.03 & 0.02 & 0.08 & 0.01 \\
& & & SD & 0.17 & 0.07 & 0.07 & 0.09 & 0.11 & 0.05 \\
\hline
\end{tabular}
\end{figure}

\subsubsection{Sample size 1000}

\begin{tabular}{l|c|l|rrrrrr}
Copula & $\tau_{K}$ & & $\tau_{K}$ & $p$ & $\theta_{U1}$ & $\theta_{U2}$ & $\theta_{C1}$ & $\theta_{C2}$ \\ 
  \hline
Frank & 0.00 & Bias & -0.00 & 0.00 & 0.00 & -0.01 & 0.01 & -0.00 \\
& & SD & 0.13 & 0.05 & 0.04 & 0.06 & 0.10 & 0.03 \\
& & $\widehat{\mbox{SD}}$ & 0.13 & 0.05 & 0.04 & 0.05 & 0.10 & 0.03 \\
& & RMSE & 0.13 & 0.04 & 0.04 & 0.05 & 0.10 & 0.03 \\
  \cline{2-9}
& 0.20 & Bias & -0.01 & 0.00 & 0.01 & -0.01 & 0.01 & 0.00 \\ 
& & SD & 0.12 & 0.04 & 0.04 & 0.05 & 0.08 & 0.03 \\ 
& & $\widehat{\mbox{SD}}$ & 0.13 & 0.05 & 0.04 & 0.05 & 0.09 & 0.03 \\
\hline 
\end{tabular}

\begin{figure}[p]
\centering
\begin{tabular}{l|c|l|rrrrrr}
Copula & $\tau_{K}$ & & $\tau_{K}$ & $p$ & $\theta_{U1}$ & $\theta_{U2}$ & $\theta_{C1}$ & $\theta_{C2}$ \\ 
  \hline
Joe & 0.00 & Bias & 0.06 & 0.04 & 0.04 & -0.01 & -0.05 & 0.02 \\
& & SD & 0.09 & 0.06 & 0.06 & 0.05 & 0.07 & 0.04 \\
& & $\widehat{\mbox{SD}}$ & 0.07 & 0.05 & 0.06 & 0.04 & 0.06 & 0.04 \\
& & RMSE & 0.10 & 0.07 & 0.08 & 0.05 & 0.09 & 0.04 \\
  \cline{2-9}
& 0.20 & Bias & -0.03 & -0.00 & 0.01 & -0.02 & 0.03 & -0.01 \\ 
& & SD & 0.10 & 0.05 & 0.06 & 0.06 & 0.09 & 0.03 \\ 
& & $\widehat{\mbox{SD}}$ & 0.08 & 0.05 & 0.05 & 0.04 & 0.06 & 0.03 \\
\hline 
\end{tabular}
\end{figure}

\begin{figure}[p]
\centering
\begin{tabular}{l|c|l|rrrrrr}
Copula & $\tau_{K}$ & & $\tau_{K}$ & $p$ & $\theta_{U1}$ & $\theta_{U2}$ & $\theta_{C1}$ & $\theta_{C2}$ \\ 
  \hline
Clayton(270) & 0.00 & Bias & 0.06 & -0.03 & 0.00 & 0.06 & 0.12 & -0.02 \\ 
& & SD & 0.00 & 0.00 & 0.00 & 0.00 & 0.00 & 0.00 \\
& & $\widehat{\mbox{SD}}$ & 0.21 & 0.05 & 0.04 & 0.05 & 0.25 & 0.08 \\
& & RMSE & 0.05 & 0.03 & 0.00 & 0.05 & 0.11 & 0.00 \\
  \cline{2-9}
& -0.20 & Bias & 0.02 & 0.00 & 0.00 & 0.00 & -0.00 & 0.00 \\ 
& & SD & 0.15 & 0.05 & 0.04 & 0.05 & 0.16 & 0.07 \\ 
& & $\widehat{\mbox{SD}}$ & 0.19 & 0.06 & 0.05 & 0.05 & 0.19 & 0.08 \\
\hline 
\end{tabular}
\end{figure}

\begin{figure}[p]
\centering
\begin{tabular}{l|l|c|l|rrrrrr}
Copula & Copula & \multicolumn{8}{c}{} \\
data gen. & estim. & $\tau_{K}$ & & $\tau_{K}$ & $p$ & $\theta_{U1}$ & $\theta_{U2}$ & $\theta_{C1}$ & $\theta_{C2}$ \\
  \hline
Gumbel & Frank & 0.20 & Bias & -0.11 & -0.04 & -0.02 & -0.01 & 0.09 & -0.02 \\
& & & SD & 0.10 & 0.04 & 0.04 & 0.06 & 0.08 & 0.03 \\
& & & $\widehat{\mbox{SD}}$ & 0.12 & 0.04 & 0.04 & 0.05 & 0.09 & 0.03 \\
& & & RMSE & 0.14 & 0.05 & 0.04 & 0.06 & 0.12 & 0.03 \\
  \cline{3-10}
& & 0.50 & Bias & -0.14 & -0.05 & -0.03 & 0.01 & 0.08 & 0.01 \\
& & & SD & 0.11 & 0.04 & 0.05 & 0.07 & 0.07 & 0.03 \\
\hline
\end{tabular}
\end{figure}

\subsubsection{Sample size 2500}

\begin{tabular}{l|c|l|rrrrrr}
Copula & $\tau_{K}$ & & $\tau_{K}$ & $p$ & $\theta_{U1}$ & $\theta_{U2}$ & $\theta_{C1}$ & $\theta_{C2}$ \\ 
  \hline
Frank & 0.00 & Bias & -0.00 & 0.00 & 0.01 & -0.01 & 0.00 & -0.00 \\
& & SD & 0.08 & 0.03 & 0.03 & 0.04 & 0.06 & 0.02 \\
& & $\widehat{\mbox{SD}}$ & 0.08 & 0.03 & 0.03 & 0.03 & 0.06 & 0.02 \\
& & RMSE & 0.08 & 0.03 & 0.03 & 0.03 & 0.06 & 0.00 \\
  \cline{2-9}
& 0.20 & Bias & -0.01 & -0.00 & 0.01 & -0.01 & 0.01 & 0.00 \\ 
& & SD & 0.08 & 0.03 & 0.03 & 0.04 & 0.05 & 0.02 \\ 
& & $\widehat{\mbox{SD}}$ & 0.08 & 0.03 & 0.03 & 0.03 & 0.05 & 0.02 \\
\hline 
\end{tabular}

\begin{figure}[p]
\centering
\begin{tabular}{l|c|l|rrrrrr}
Copula & $\tau_{K}$ & & $\tau_{K}$ & $p$ & $\theta_{U1}$ & $\theta_{U2}$ & $\theta_{C1}$ & $\theta_{C2}$ \\ 
  \hline
Joe & 0.00 & Bias & 0.04 & 0.03 & 0.03 & -0.01 & -0.03 & 0.01 \\
& & SD & 0.06 & 0.04 & 0.04 & 0.03 & 0.05 & 0.03 \\
& & $\widehat{\mbox{SD}}$ & 0.05 & 0.03 & 0.04 & 0.03 & 0.05 & 0.03 \\
& & RMSE & 0.07 & 0.04 & 0.04 & 0.03 & 0.05 & 0.03 \\
  \cline{2-9}
& 0.20 & Bias & -0.03 & -0.00 & 0.01 & -0.03 & 0.03 & -0.00 \\ 
& & SD & 0.07 & 0.03 & 0.04 & 0.04 & 0.06 & 0.02 \\ 
& & $\widehat{\mbox{SD}}$ & 0.05 & 0.03 & 0.04 & 0.03 & 0.04 & 0.02 \\
\hline 
\end{tabular}
\end{figure}

\begin{figure}[p]
\centering
\begin{tabular}{l|c|l|rrrrrr}
Copula & $\tau_{K}$ & & $\tau_{K}$ & $p$ & $\theta_{U1}$ & $\theta_{U2}$ & $\theta_{C1}$ & $\theta_{C2}$ \\ 
  \hline
Clayton(270) & 0.00 & Bias & 0.08 & -0.02 & 0.00 & -0.01 & 0.10 & -0.03 \\
& & SD & 0.09 & 0.03 & 0.03 & 0.03 & 0.12 & 0.04 \\
& & $\widehat{\mbox{SD}}$ & 0.14 & 0.03 & 0.03 & 0.03 & 0.17 & 0.05 \\
& & RMSE & 0.13 & 0.03 & 0.03 & 0.03 & 0.15 & 0.04 \\
  \cline{2-9}
& -0.20 & Bias & 0.02 & 0.00 & -0.00 & 0.00 & -0.01 & 0.01 \\ 
& & SD & 0.11 & 0.04 & 0.03 & 0.03 & 0.12 & 0.05 \\ 
& & $\widehat{\mbox{SD}}$ & 0.13 & 0.04 & 0.03 & 0.03 & 0.13 & 0.06 \\
\hline 
\end{tabular}
\end{figure}

\begin{figure}[p]
\centering
\begin{tabular}{l|l|c|l|rrrrrr}
Copula & Copula & \multicolumn{8}{c}{} \\
data gen. & estim. & $\tau_{K}$ & & $\tau_{K}$ & $p$ & $\theta_{U1}$ & $\theta_{U2}$ & $\theta_{C1}$ & $\theta_{C2}$ \\
  \hline
Gumbel & Frank & 0.20 & Bias & -0.11 & -0.04 & -0.02 & -0.02 & 0.09 & -0.02 \\
& & & SD & 0.06 & 0.03 & 0.03 & 0.04 & 0.05 & 0.02 \\
& & & $\widehat{\mbox{SD}}$ & 0.07 & 0.03 & 0.02 & 0.03 & 0.06 & 0.02 \\
& & & RMSE & 0.13 & 0.04 & 0.03 & 0.04 & 0.10 & 0.03 \\
  \cline{3-10}
& & 0.50 & Bias & -0.14 & -0.05 & -0.03 & 0.00 & 0.08 & 0.01 \\
& & & SD & 0.07 & 0.03 & 0.03 & 0.04 & 0.04 & 0.02 \\
\hline
\end{tabular}
\end{figure}

\subsection{Incidence of 0.4}

\subsubsection{Percentage of censoring}

\begin{tabular}{l|c|c|c|c|c|c|c}
Copula / $\tau_{K}$ & 0.00 & 0.20 & 0.50 & 0.80 & -0.20 & -0.50 & -0.80 \\
\hline 
Frank & 73.78 & 77.28 & 83.04 & 91.70 & & 67.36 & \\
\hline 
Gumbel & 73.54 & 76.30 & 80.90 & 90.18 & & &  \\
\hline 
Joe & 73.90 & 75.52 & 80.60 & 88.66 & & &  \\
\hline 
Clayton(90) & 73.04 &  &  &  & 70.62 & 66.88 & 65.36 \\
\hline 
Clayton(180) & 73.38 & 76.42 & 81.06 & 87.70 & & &  \\
\hline 
Clayton(270) & 73.70 &  &  &  & 71.66 & 67.80 & 65.26 \\
\hline 
Gaussian & 83.66 & 89.30 & 94.90 & 99.86 & & 73.08 & \\
\hline 
\end{tabular}

\subsubsection{Sample size 200}

\begin{tabular}{l|c|l|rrrrrr}
Copula & $\tau_{K}$ & & $\tau_{K}$ & $p$ & $\theta_{U1}$ & $\theta_{U2}$ & $\theta_{C1}$ & $\theta_{C2}$ \\ 
  \hline
Frank & 0.00 & Bias & 0.02 & 0.04 & -0.01 & 0.04 & 0.04 & -0.02 \\ 
& & SD & 0.40 & 0.15 & 0.13 & 0.14 & 0.24 & 0.08 \\
& & $\widehat{\mbox{SD}}$ & 0.28 & 0.12 & 0.14 & 0.13 & 0.17 & 0.07 \\
& & RMSE & 0.40 & 0.16 & 0.13 & 0.14 & 0.25 & 0.08 \\
  \cline{2-9}
& 0.20 & Bias & -0.11 & 0.03 & 0.02 & 0.02 & 0.10 & -0.03 \\ 
& & SD & 0.47 & 0.18 & 0.36 & 0.14 & 0.28 & 0.09 \\ 
& & $\widehat{\mbox{SD}}$ & 0.29 & 0.14 & 0.45 & 0.13 & 0.17 & 0.07 \\ 
& & RMSE & 0.48 & 0.19 & 0.37 & 0.14 & 0.29 & 0.09 \\ 
  \cline{2-9} 
& 0.50 & Bias & -0.41 & -0.02 & 0.02 & 0.00 & 0.20 & -0.03 \\ 
& & SD & 0.63 & 0.24 & 0.52 & 0.17 & 0.30 & 0.09 \\ 
& & $\widehat{\mbox{SD}}$ & 0.26 & 0.14 & 0.43 & 0.14 & 0.14 & 0.07 \\ 
& & RMSE & 0.75 & 0.24 & 0.52 & 0.17 & 0.35 & 0.09 \\ 
  \cline{2-9}
& 0.80 & Bias & -0.51 & -0.10 & -0.04 & 0.07 & 0.09 & 0.02 \\ 
& & SD & 0.65 & 0.25 & 0.63 & 0.24 & 0.15 & 0.07 \\ 
& & $\widehat{\mbox{SD}}$ & 0.26 & 0.33 & 1.42 & 0.21 & 0.10 & 0.06 \\ 
& & RMSE & 0.83 & 0.27 & 0.64 & 0.25 & 0.18 & 0.08 \\ 
  \cline{2-9}
& -0.50 & Bias & 0.33 & 0.15 & 0.06 & 0.08 & -0.12 & 0.05 \\ 
& & SD & 0.49 & 0.20 & 0.17 & 0.15 & 0.22 & 0.11 \\ 
& & $\widehat{\mbox{SD}}$ & 0.21 & 0.12 & 0.24 & 0.12 & 0.14 & 0.09 \\ 
& & RMSE & 0.59 & 0.25 & 0.18 & 0.17 & 0.25 & 0.12 \\ 
\hline 
\end{tabular}

\begin{figure}[p]
\centering
\begin{tabular}{l|c|l|rrrrrr}
Copula & $\tau_{K}$ & & $\tau_{K}$ & $p$ & $\theta_{U1}$ & $\theta_{U2}$ & $\theta_{C1}$ & $\theta_{C2}$ \\ 
  \hline
Gumbel & 0.00 & Bias & 0.28 & 0.16 & 0.12 & 0.02 & -0.10 & 0.00 \\
& & SD & 0.29 & 0.19 & 0.25 & 0.14 & 0.12 & 0.07 \\
& & $\widehat{\mbox{SD}}$ & 0.31 & 0.20 & 0.39 & 0.13 & 0.14 & 0.07 \\
& & RMSE & 0.40 & 0.25 & 0.28 & 0.14 & 0.15 & 0.07 \\
  \cline{2-9}
& 0.20 & Bias & 0.12 & 0.11 & 0.11 & -0.00 & -0.03 & -0.02 \\ 
& & SD & 0.30 & 0.19 & 0.30 & 0.15 & 0.13 & 0.06 \\ 
& & $\widehat{\mbox{SD}}$ & 0.32 & 0.21 & 0.78 & 0.13 & 0.15 & 0.07 \\
\hline 
\end{tabular}
\end{figure}

\begin{figure}[p]
\centering
\begin{tabular}{l|c|l|rrrrrr}
Copula & $\tau_{K}$ & & $\tau_{K}$ & $p$ & $\theta_{U1}$ & $\theta_{U2}$ & $\theta_{C1}$ & $\theta_{C2}$ \\ 
  \hline
Clayton(180) & 0.00 & Bias & 0.55 & 0.31 & 0.09 & 0.21 & -0.22 & -0.02 \\
& & SD & 0.00 & 0.00 & 0.00 & 0.00 & 0.00 & 0.00 \\
& & $\widehat{\mbox{SD}}$ & 0.15 & 0.12 & 0.10 & 0.14 & 0.07 & 0.06 \\
& & RMSE & 0.55 & 0.31 & 0.09 & 0.21 & 0.22 & 0.00 \\
  \cline{2-9}
& 0.20 & Bias & 0.34 & 0.19 & 0.02 & 0.21 & -0.16 & -0.02 \\ 
& & SD & 0.00 & 0.00 & 0.00 & 0.00 & 0.00 & 0.00 \\ 
& & $\widehat{\mbox{SD}}$ & 0.15 & 0.12 & 0.11 & 0.15 & 0.07 & 0.06 \\
\hline 
\end{tabular}
\end{figure}

\begin{figure}[p]
\centering
\begin{tabular}{l|l|c|l|rrrrrr}
Copula & Copula & \multicolumn{8}{c}{} \\
data gen. & estim. & $\tau_{K}$ & & $\tau_{K}$ & $p$ & $\theta_{U1}$ & $\theta_{U2}$ & $\theta_{C1}$ & $\theta_{C2}$ \\
  \hline
Gumbel & Frank & 0.20 & Bias & 0.01 & 0.06 & 0.06 & 0.02 & 0.02 & -0.02 \\
& & & SD & 0.34 & 0.18 & 0.42 & 0.15 & 0.15 & 0.06 \\
& & & $\widehat{\mbox{SD}}$ & 0.30 & 0.15 & 0.50 & 0.13 & 0.16 & 0.06 \\
& & & RMSE & 0.34 & 0.18 & 0.42 & 0.15 & 0.15 & 0.07 \\
  \cline{3-10}
& & 0.50 & Bias & -0.12 & 0.03 & 0.08 & 0.04 & 0.05 & 0.01 \\
& & & SD & 0.35 & 0.22 & 0.54 & 0.17 & 0.13 & 0.07 \\
\hline
\end{tabular}
\end{figure}

\subsubsection{Sample size 500}

\begin{tabular}{l|c|l|rrrrrr}
Copula & $\tau_{K}$ & & $\tau_{K}$ & $p$ & $\theta_{U1}$ & $\theta_{U2}$ & $\theta_{C1}$ & $\theta_{C2}$ \\ 
  \hline
Frank & 0.00 & Bias & 0.02 & 0.03 & 0.01 & 0.01 & 0.01 & -0.01 \\
& & SD & 0.27 & 0.10 & 0.11 & 0.10 & 0.14 & 0.04 \\
& & $\widehat{\mbox{SD}}$ & 0.23 & 0.09 & 0.10 & 0.08 & 0.12 & 0.04 \\
& & RMSE & 0.27 & 0.11 & 0.10 & 0.09 & 0.14 & 0.04 \\
  \cline{2-9}
& 0.20 & Bias & -0.04 & 0.02 & 0.01 & -0.01 & 0.04 & -0.01 \\ 
& & SD & 0.32 & 0.12 & 0.15 & 0.10 & 0.18 & 0.05 \\ 
& & $\widehat{\mbox{SD}}$ & 0.22 & 0.10 & 0.12 & 0.09 & 0.11 & 0.04 \\
\hline 
\end{tabular}

\begin{figure}[p]
\centering
\begin{tabular}{l|c|l|rrrrrr}
Copula & $\tau_{K}$ & & $\tau_{K}$ & $p$ & $\theta_{U1}$ & $\theta_{U2}$ & $\theta_{C1}$ & $\theta_{C2}$ \\ 
  \hline
Joe & 0.00 & Bias & 0.16 & 0.11 & 0.14 & -0.00 & -0.06 & 0.02 \\
& & SD & 0.18 & 0.15 & 0.20 & 0.09 & 0.09 & 0.04 \\
& & $\widehat{\mbox{SD}}$ & 0.11 & 0.13 & 0.24 & 0.08 & 0.06 & 0.04 \\
& & RMSE & 0.24 & 0.19 & 0.24 & 0.09 & 0.10 & 0.04 \\
  \cline{2-9}
& 0.20 & Bias & 0.03 & 0.06 & 0.08 & -0.00 & -0.00 & -0.01 \\ 
& & SD & 0.19 & 0.15 & 0.21 & 0.10 & 0.09 & 0.04 \\ 
& & $\widehat{\mbox{SD}}$ & 0.17 & 0.11 & 0.22 & 0.09 & 0.08 & 0.04 \\
\hline 
\end{tabular}
\end{figure}

\begin{figure}[p]
\centering
\begin{tabular}{l|c|l|rrrrrr}
Copula & $\tau_{K}$ & & $\tau_{K}$ & $p$ & $\theta_{U1}$ & $\theta_{U2}$ & $\theta_{C1}$ & $\theta_{C2}$ \\ 
  \hline
Clayton(270) & 0.00 & Bias & 0.12 & -0.03 & -0.01 & -0.01 & 0.12 & -0.04 \\
& & SD & 0.14 & 0.05 & 0.10 & 0.09 & 0.16 & 0.06 \\
& & $\widehat{\mbox{SD}}$ & 0.20 & 0.06 & 0.09 & 0.08 & 0.21 & 0.07 \\
& & RMSE & 0.19 & 0.05 & 0.10 & 0.09 & 0.20 & 0.07 \\
  \cline{2-9}
& -0.20 & Bias & 0.01 & 0.00 & -0.00 & 0.01 & 0.00 & 0.00 \\ 
& & SD & 0.17 & 0.06 & 0.09 & 0.08 & 0.15 & 0.07 \\ 
& & $\widehat{\mbox{SD}}$ & 0.20 & 0.07 & 0.09 & 0.08 & 0.17 & 0.09 \\
\hline 
\end{tabular}
\end{figure}

\begin{figure}[p]
\centering
\begin{tabular}{l|l|c|l|rrrrrr}
Copula & Copula & \multicolumn{8}{c}{} \\
data gen. & estim. & $\tau_{K}$ & & $\tau_{K}$ & $p$ & $\theta_{U1}$ & $\theta_{U2}$ & $\theta_{C1}$ & $\theta_{C2}$ \\
  \hline
Gumbel & Frank & 0.20 & Bias & -0.06 & 0.00 & 0.01 & -0.00 & 0.04 & -0.01 \\
& & & SD & 0.23 & 0.10 & 0.14 & 0.11 & 0.10 & 0.04 \\
& & & $\widehat{\mbox{SD}}$ & 0.23 & 0.10 & 0.12 & 0.09 & 0.12 & 0.04 \\
& & & RMSE & 0.24 & 0.10 & 0.14 & 0.11 & 0.11 & 0.04 \\
  \cline{3-10}
& & 0.50 & Bias & -0.11 & 0.02 & 0.05 & 0.01 & 0.04 & 0.00 \\
& & & SD & 0.27 & 0.17 & 0.32 & 0.12 & 0.09 & 0.05 \\
\hline
\end{tabular}
\end{figure}

\subsubsection{Sample size 1000}

\begin{tabular}{l|c|l|rrrrrr}
Copula & $\tau_{K}$ & & $\tau_{K}$ & $p$ & $\theta_{U1}$ & $\theta_{U2}$ & $\theta_{C1}$ & $\theta_{C2}$ \\ 
  \hline
Frank & 0.00 & Bias & 0.11 & 0.05 & 0.04 & 0.14 & -0.07 & -0.05 \\
& & SD & 0.00 & 0.00 & 0.00 & 0.00 & 0.00 & 0.00 \\
& & $\widehat{\mbox{SD}}$ & 0.16 & 0.07 & 0.05 & 0.07 & 0.07 & 0.03 \\
& & RMSE & 0.11 & 0.04 & 0.04 & 0.14 & 0.08 & 0.05 \\
  \cline{2-9}
& 0.20 & Bias & -0.00 & 0.01 & 0.03 & -0.02 & 0.01 & -0.00 \\ 
& & SD & 0.18 & 0.08 & 0.09 & 0.08 & 0.09 & 0.03 \\ 
& & $\widehat{\mbox{SD}}$ & 0.17 & 0.08 & 0.08 & 0.06 & 0.07 & 0.03 \\
\hline 
\end{tabular}

\begin{figure}[p]
\centering
\begin{tabular}{l|c|l|rrrrrr}
Copula & $\tau_{K}$ & & $\tau_{K}$ & $p$ & $\theta_{U1}$ & $\theta_{U2}$ & $\theta_{C1}$ & $\theta_{C2}$ \\ 
  \hline
Joe & 0.00 & Bias & 0.10 & 0.06 & 0.09 & -0.01 & -0.04 & 0.01 \\
& & SD & 0.12 & 0.09 & 0.13 & 0.07 & 0.06 & 0.03 \\
& & $\widehat{\mbox{SD}}$ & 0.09 & 0.07 & 0.11 & 0.06 & 0.05 & 0.03 \\
& & RMSE & 0.15 & 0.10 & 0.15 & 0.07 & 0.07 & 0.03 \\
  \cline{2-9}
& 0.20 & Bias & -0.03 & 0.01 & 0.04 & -0.02 & 0.02 & -0.01 \\ 
& & SD & 0.14 & 0.09 & 0.15 & 0.08 & 0.07 & 0.03 \\ 
& & $\widehat{\mbox{SD}}$ & 0.12 & 0.07 & 0.12 & 0.06 & 0.06 & 0.03 \\
\hline 
\end{tabular}
\end{figure}

\begin{figure}[p]
\centering
\begin{tabular}{l|c|l|rrrrrr}
Copula & $\tau_{K}$ & & $\tau_{K}$ & $p$ & $\theta_{U1}$ & $\theta_{U2}$ & $\theta_{C1}$ & $\theta_{C2}$ \\ 
  \hline
Clayton(270) & 0.00 & Bias & 0.10 & -0.02 & -0.00 & -0.02 & 0.10 & -0.03 \\
& & SD & 0.12 & 0.04 & 0.07 & 0.06 & 0.12 & 0.05 \\
& & $\widehat{\mbox{SD}}$ & 0.19 & 0.05 & 0.06 & 0.06 & 0.18 & 0.06 \\
& & RMSE & 0.15 & 0.04 & 0.06 & 0.07 & 0.15 & 0.05 \\
  \cline{2-9}
& -0.20 & Bias & 0.01 & 0.00 & 0.00 & 0.00 & -0.00 & 0.00 \\ 
& & SD & 0.14 & 0.05 & 0.06 & 0.06 & 0.12 & 0.06 \\ 
& & $\widehat{\mbox{SD}}$ & 0.16 & 0.05 & 0.06 & 0.06 & 0.14 & 0.07 \\
\hline 
\end{tabular}
\end{figure}

\begin{figure}[p]
\centering
\begin{tabular}{l|l|c|l|rrrrrr}
Copula & Copula & \multicolumn{8}{c}{} \\
data gen. & estim. & $\tau_{K}$ & & $\tau_{K}$ & $p$ & $\theta_{U1}$ & $\theta_{U2}$ & $\theta_{C1}$ & $\theta_{C2}$ \\
  \hline
Gumbel & Frank & 0.20 & Bias & -0.08 & -0.01 & -0.00 & -0.02 & 0.04 & -0.01 \\
& & & SD & 0.14 & 0.06 & 0.08 & 0.08 & 0.07 & 0.03 \\
& & & $\widehat{\mbox{SD}}$ & 0.17 & 0.07 & 0.07 & 0.06 & 0.08 & 0.03 \\
& & & RMSE & 0.16 & 0.05 & 0.08 & 0.08 & 0.08 & 0.03 \\
  \cline{3-10}
& & 0.50 & Bias & -0.11 & -0.00 & 0.03 & -0.01 & 0.03 & 0.01 \\
& & & SD & 0.20 & 0.13 & 0.20 & 0.10 & 0.06 & 0.04 \\
\hline
\end{tabular}
\end{figure}

\subsubsection{Sample size 2500}

\begin{tabular}{l|c|l|rrrrrr}
Copula & $\tau_{K}$ & & $\tau_{K}$ & $p$ & $\theta_{U1}$ & $\theta_{U2}$ & $\theta_{C1}$ & $\theta_{C2}$ \\ 
  \hline
Frank & 0.00 & Bias & 0.01 & 0.01 & 0.02 & -0.02 & -0.00 & -0.00 \\
& & SD & 0.10 & 0.04 & 0.04 & 0.05 & 0.05 & 0.02 \\
& & $\widehat{\mbox{SD}}$ & 0.11 & 0.04 & 0.04 & 0.04 & 0.05 & 0.02 \\
& & RMSE & 0.09 & 0.04 & 0.04 & 0.05 & 0.04 & 0.00 \\
  \cline{2-9}
& 0.20 & Bias & 0.00 & 0.01 & 0.03 & -0.03 & 0.00 & 0.00 \\ 
& & SD & 0.10 & 0.04 & 0.06 & 0.05 & 0.04 & 0.02 \\ 
& & $\widehat{\mbox{SD}}$ & 0.11 & 0.05 & 0.05 & 0.04 & 0.05 & 0.02 \\
\hline 
\end{tabular}

\begin{figure}[p]
\centering
\begin{tabular}{l|c|l|rrrrrr}
Copula & $\tau_{K}$ & & $\tau_{K}$ & $p$ & $\theta_{U1}$ & $\theta_{U2}$ & $\theta_{C1}$ & $\theta_{C2}$ \\ 
  \hline
Joe & 0.00 & Bias & 0.06 & 0.04 & 0.06 & -0.02 & -0.03 & 0.01 \\
& & SD & 0.08 & 0.04 & 0.06 & 0.04 & 0.04 & 0.02 \\
& & $\widehat{\mbox{SD}}$ & 0.04 & 0.04 & 0.06 & 0.03 & 0.03 & 0.02 \\
& & RMSE & 0.10 & 0.05 & 0.08 & 0.04 & 0.04 & 0.03 \\
  \cline{2-9}
& 0.20 & Bias & -0.04 & -0.00 & 0.02 & -0.04 & 0.02 & -0.00 \\ 
& & SD & 0.10 & 0.04 & 0.06 & 0.05 & 0.05 & 0.02 \\ 
& & $\widehat{\mbox{SD}}$ & 0.10 & 0.05 & 0.07 & 0.04 & 0.04 & 0.02 \\
\hline 
\end{tabular}
\end{figure}

\begin{figure}[p]
\centering
\begin{tabular}{l|c|l|rrrrrr}
Copula & $\tau_{K}$ & & $\tau_{K}$ & $p$ & $\theta_{U1}$ & $\theta_{U2}$ & $\theta_{C1}$ & $\theta_{C2}$ \\ 
  \hline
Clayton(270) & 0.00 & Bias & 0.07 & -0.01 & 0.00 & -0.03 & 0.07 & -0.02 \\
& & SD & 0.08 & 0.03 & 0.04 & 0.04 & 0.08 & 0.03 \\
& & $\widehat{\mbox{SD}}$ & 0.11 & 0.03 & 0.04 & 0.04 & 0.11 & 0.04 \\
& & RMSE & 0.11 & 0.03 & 0.04 & 0.05 & 0.10 & 0.03 \\
  \cline{2-9}
& -0.20 & Bias & 0.02 & 0.01 & 0.00 & 0.00 & -0.01 & 0.01 \\ 
& & SD & 0.10 & 0.04 & 0.04 & 0.04 & 0.09 & 0.04 \\ 
& & $\widehat{\mbox{SD}}$ & 0.11 & 0.04 & 0.04 & 0.04 & 0.09 & 0.05 \\
\hline 
\end{tabular}
\end{figure}

\begin{figure}[p]
\centering
\begin{tabular}{l|l|c|l|rrrrrr}
Copula & Copula & \multicolumn{8}{c}{} \\
data gen. & estim. & $\tau_{K}$ & & $\tau_{K}$ & $p$ & $\theta_{U1}$ & $\theta_{U2}$ & $\theta_{C1}$ & $\theta_{C2}$ \\
  \hline
Gumbel & Frank & 0.20 & Bias & -0.09 & -0.02 & -0.00 & -0.03 & 0.04 & -0.01 \\
& & & SD & 0.09 & 0.04 & 0.05 & 0.05 & 0.04 & 0.02 \\
& & & $\widehat{\mbox{SD}}$ & 0.10 & 0.04 & 0.04 & 0.04 & 0.05 & 0.02 \\
& & & RMSE & 0.12 & 0.04 & 0.04 & 0.06 & 0.06 & 0.00 \\
  \cline{3-10}
& & 0.50 & Bias & -0.13 & -0.03 & 0.01 & -0.02 & 0.04 & 0.01 \\
& & & SD & 0.13 & 0.07 & 0.09 & 0.07 & 0.04 & 0.02 \\
\hline
\end{tabular}
\end{figure}

\section{Truncated distribution for the variable $U$}

\subsection{Incidence of 0.6}

\subsubsection{Percentage of censoring}

\begin{tabular}{l|c|c|c|c|c|c|c}
Copula / $\tau_{K}$ & 0.00 & 0.20 & 0.50 & 0.80 & -0.20 & -0.50 & -0.80 \\
\hline 
Frank & 58.62 & 60.68 & 65.10 & 69.00 & & 54.12 & \\
\hline 
Gumbel & 58.48 & 59.66 & 63.58 & 67.44 & & & \\
\hline 
Joe & 58.04 & 61.04 & 64.62 & 71.60 & & & \\
\hline 
Clayton(90) & 58.66 & & & & 55.14 & 51.24 & 50.54 \\
\hline 
Clayton(180) & 59.58 & 60.10 & 65.70 & 68.32 & & & \\
\hline 
Clayton(270) & 58.06 & & & & 58.16 & 55.00 & 52.36 \\
\hline 
Gaussian & 75.02 & 81.08 & 87.66 & 96.84 & & 65.30 & \\
\hline 
\end{tabular}

\subsubsection{Sample size 200}

\begin{tabular}{l|c|l|rrrrrr}
Copula & $\tau_{K}$ & & $\tau_{K}$ & $p$ & $\theta_{U1}$ & $\theta_{U2}$ & $\theta_{C1}$ & $\theta_{C2}$ \\ 
  \hline
Frank & 0.00 & Bias & 0.03 & 0.01 & 0.15 & -0.01 & 0.03 & -0.01 \\
& & SD & 0.29 & 0.10 & 1.76 & 0.11 & 0.29 & 0.08 \\
& & $\widehat{\mbox{SD}}$ & 0.24 & 0.09 & 10.94 & 0.12 & 0.23 & 0.08 \\
& & RMSE & 0.29 & 0.09 & 1.77 & 0.11 & 0.29 & 0.08 \\
  \cline{2-9}
& 0.20 & Bias & -0.10 & -0.02 & 0.10 & -0.03 & 0.18 & -0.03 \\ 
& & SD & 0.37 & 0.10 & 0.73 & 0.12 & 0.53 & 0.10 \\ 
& & $\widehat{\mbox{SD}}$ & 0.23 & 0.08 & 4.26 & 0.12 & 0.26 & 0.07 \\ 
& & RMSE & 0.39 & 0.10 & 0.74 & 0.12 & 0.56 & 0.10 \\ 
  \cline{2-9} 
& 0.50 & Bias & -0.42 & -0.09 & 0.10 & -0.05 & 0.58 & -0.05 \\ 
& & SD & 0.60 & 0.13 & 1.16 & 0.12 & 0.88 & 0.13 \\ 
& & $\widehat{\mbox{SD}}$ & 0.20 & 0.07 & 5.88 & 0.12 & 0.29 & 0.08 \\ 
& & RMSE & 0.73 & 0.15 & 1.16 & 0.13 & 1.05 & 0.14 \\ 
  \cline{2-9}
& 0.80 & Bias & -0.30 & -0.08 & 0.22 & -0.01 & 0.26 & 0.02 \\ 
& & SD & 0.55 & 0.11 & 2.20 & 0.13 & 0.56 & 0.10 \\ 
& & $\widehat{\mbox{SD}}$ & 0.14 & 0.06 & 3.86 & 0.12 & 0.17 & 0.08 \\ 
& & RMSE & 0.63 & 0.14 & 2.21 & 0.13 & 0.61 & 0.10 \\ 
  \cline{2-9}
& -0.50 & Bias & 0.09 & -0.03 & 0.02 & 0.10 & -0.00 & -0.06 \\ 
& & SD & 0.00 & 0.00 & 0.00 & 0.00 & 0.00 & 0.00 \\ 
& & $\widehat{\mbox{SD}}$ & 0.13 & 0.07 & 0.15 & 0.13 & 0.15 & 0.08 \\ 
& & RMSE & 0.09 & 0.03 & 0.00 & 0.10 & 0.00 & 0.06 \\ 
\hline 
\end{tabular}

\begin{figure}[p]
\centering
\begin{tabular}{l|c|l|rrrrrr}
Copula & $\tau_{K}$ & & $\tau_{K}$ & $p$ & $\theta_{U1}$ & $\theta_{U2}$ & $\theta_{C1}$ & $\theta_{C2}$ \\ 
  \hline
Gumbel & 0.00 & Bias & 0.00 & 0.02 & 0.03 & 0.04 & 0.22 & -0.02 \\
& & SD & 0.00 & 0.00 & 0.00 & 0.00 & 0.00 & 0.00 \\
& & $\widehat{\mbox{SD}}$ & 0.03 & 0.04 & 0.11 & 0.11 & 0.12 & 0.07 \\
& & RMSE & 0.00 & 0.00 & 0.03 & 0.03 & 0.22 & 0.00 \\
  \cline{2-9}
& 0.20 & Bias & 0.05 & 0.01 & 0.14 & -0.03 & -0.01 & -0.01 \\ 
& & SD & 0.23 & 0.09 & 0.54 & 0.11 & 0.15 & 0.07 \\ 
& & $\widehat{\mbox{SD}}$ & 0.25 & 0.09 & 2.38 & 0.12 & 0.19 & 0.08 \\
\hline 
\end{tabular}
\end{figure}

\begin{figure}[p]
\centering
\begin{tabular}{l|c|l|rrrrrr}
Copula & $\tau_{K}$ & & $\tau_{K}$ & $p$ & $\theta_{U1}$ & $\theta_{U2}$ & $\theta_{C1}$ & $\theta_{C2}$ \\ 
  \hline
Clayton(180) & 0.00 & Bias & 0.18 & 0.07 & 0.22 & -0.00 & -0.12 & 0.04 \\
& & SD & 0.20 & 0.09 & 0.76 & 0.12 & 0.14 & 0.07 \\
& & $\widehat{\mbox{SD}}$ & 0.14 & 0.08 & 2.54 & 0.14 & 0.12 & 0.08 \\
& & RMSE & 0.27 & 0.11 & 0.80 & 0.12 & 0.18 & 0.08 \\
  \cline{2-9}
& 0.20 & Bias & 0.02 & 0.01 & 0.15 & -0.01 & 0.00 & -0.00 \\ 
& & SD & 0.18 & 0.08 & 0.55 & 0.12 & 0.14 & 0.06 \\ 
& & $\widehat{\mbox{SD}}$ & 0.16 & 0.07 & 1.92 & 0.12 & 0.13 & 0.07 \\
\hline 
\end{tabular}
\end{figure}

\begin{figure}[p]
\centering
\begin{tabular}{l|l|c|l|rrrrrr}
Copula & Copula & \multicolumn{8}{c}{} \\
data gen. & estim. & $\tau_{K}$ & & $\tau_{K}$ & $p$ & $\theta_{U1}$ & $\theta_{U2}$ & $\theta_{C1}$ & $\theta_{C2}$ \\
  \hline
Gumbel & Frank & 0.20 & Bias & -0.08 & -0.03 & 0.06 & -0.02 & 0.09 & -0.02 \\
& & & SD & 0.23 & 0.08 & 0.24 & 0.11 & 0.19 & 0.06 \\
& & & $\widehat{\mbox{SD}}$ & 0.14 & 0.05 & 0.50 & 0.12 & 0.11 & 0.06 \\
& & & RMSE & 0.24 & 0.08 & 0.25 & 0.11 & 0.21 & 0.07 \\
  \cline{3-10}
& & 0.50 & Bias & -0.14 & -0.05 & 0.13 & -0.01 & 0.10 & 0.01 \\
& & & SD & 0.23 & 0.08 & 1.22 & 0.12 & 0.17 & 0.08 \\
\hline
\end{tabular}
\end{figure}

\subsubsection{Sample size 500}

\begin{tabular}{l|c|l|rrrrrr}
Copula & $\tau_{K}$ & & $\tau_{K}$ & $p$ & $\theta_{U1}$ & $\theta_{U2}$ & $\theta_{C1}$ & $\theta_{C2}$ \\ 
  \hline
Frank & 0.00 & Bias & -0.01 & -0.00 & 0.02 & -0.01 & 0.03 & -0.01 \\
& & SD & 0.19 & 0.06 & 0.11 & 0.07 & 0.19 & 0.05 \\
& & $\widehat{\mbox{SD}}$ & 0.17 & 0.06 & 0.14 & 0.07 & 0.16 & 0.05 \\
& & RMSE & 0.19 & 0.06 & 0.11 & 0.08 & 0.19 & 0.05 \\
  \cline{2-9}
& 0.20 & Bias & -0.07 & -0.02 & 0.02 & -0.01 & 0.10 & -0.01  \\ 
& & SD & 0.26 & 0.07 & 0.09 & 0.07 & 0.35 & 0.07 \\ 
& & $\widehat{\mbox{SD}}$ & 0.16 & 0.06 & 0.09 & 0.07 & 0.15 & 0.04 \\
\hline 
\end{tabular}

\begin{figure}[p]
\centering
\begin{tabular}{l|c|l|rrrrrr}
Copula & $\tau_{K}$ & & $\tau_{K}$ & $p$ & $\theta_{U1}$ & $\theta_{U2}$ & $\theta_{C1}$ & $\theta_{C2}$ \\ 
  \hline
Joe & 0.00 & Bias & 0.09 & 0.04 & 0.08 & -0.01 & -0.07 & 0.03 \\
& & SD & 0.11 & 0.06 & 0.12 & 0.07 & 0.10 & 0.05 \\
& & $\widehat{\mbox{SD}}$ & 0.09 & 0.05 & 0.14 & 0.07 & 0.09 & 0.05 \\
& & RMSE & 0.14 & 0.07 & 0.14 & 0.07 & 0.12 & 0.05 \\
  \cline{2-9}
& 0.20 & Bias & -0.00 & -0.00 & 0.04 & -0.01 & 0.01 & -0.00 \\ 
& & SD & 0.11 & 0.05 & 0.10 & 0.07 & 0.09 & 0.04 \\ 
& & $\widehat{\mbox{SD}}$ & 0.11 & 0.05 & 0.12 & 0.07 & 0.09 & 0.04 \\
\hline 
\end{tabular}
\end{figure}

\begin{figure}[p]
\centering
\begin{tabular}{l|c|l|rrrrrr}
Copula & $\tau_{K}$ & & $\tau_{K}$ & $p$ & $\theta_{U1}$ & $\theta_{U2}$ & $\theta_{C1}$ & $\theta_{C2}$ \\ 
  \hline
Clayton(270) & 0.00 & Bias & 0.13 & -0.03 & 0.03 & -0.02 & 0.17 & -0.04 \\
& & SD & 0.16 & 0.05 & 0.09 & 0.07 & 0.25 & 0.07 \\
& & $\widehat{\mbox{SD}}$ & 0.17 & 0.05 & 0.10 & 0.08 & 0.27 & 0.08 \\
& & RMSE & 0.20 & 0.05 & 0.09 & 0.07 & 0.30 & 0.08 \\
  \cline{2-9}
& -0.20 & Bias & 0.01 & -0.00 & 0.03 & -0.01 & 0.01 & 0.00 \\ 
& & SD & 0.18 & 0.06 & 0.09 & 0.07 & 0.21 & 0.09 \\ 
& & $\widehat{\mbox{SD}}$ & 0.24 & 0.06 & 0.11 & 0.08 & 0.26 & 0.11 \\
\hline 
\end{tabular}
\end{figure}

\begin{figure}[p]
\centering
\begin{tabular}{l|l|c|l|rrrrrr}
Copula & Copula & \multicolumn{8}{c}{} \\
data gen. & estim. & $\tau_{K}$ & & $\tau_{K}$ & $p$ & $\theta_{U1}$ & $\theta_{U2}$ & $\theta_{C1}$ & $\theta_{C2}$ \\
  \hline
Gumbel & Frank & 0.20 & Bias & -0.09 & -0.03 & -0.00 & -0.00 & 0.09 & -0.02 \\
& & & SD & 0.14 & 0.05 & 0.07 & 0.07 & 0.11 & 0.04 \\
& & & $\widehat{\mbox{SD}}$ & 0.16 & 0.06 & 0.08 & 0.07 & 0.13 & 0.04 \\
& & & RMSE & 0.16 & 0.06 & 0.08 & 0.07 & 0.14 & 0.04 \\
  \cline{3-10}
& & 0.50 & Bias & -0.14 & -0.05 & -0.01 & 0.01 & 0.09 & 0.01 \\
& & & SD & 0.14 & 0.05 & 0.07 & 0.07 & 0.10 & 0.05 \\
\hline
\end{tabular}
\end{figure}

\subsubsection{Sample size 1000}

\begin{tabular}{l|c|l|rrrrrr}
Copula & $\tau_{K}$ & & $\tau_{K}$ & $p$ & $\theta_{U1}$ & $\theta_{U2}$ & $\theta_{C1}$ & $\theta_{C2}$ \\ 
  \hline
Frank & 0.00 & Bias & -0.00 & -0.00 & 0.01 & -0.00 & 0.01 & -0.00 \\
& & SD & 0.13 & 0.04 & 0.05 & 0.05 & 0.10 & 0.03 \\
& & $\widehat{\mbox{SD}}$ & 0.13 & 0.05 & 0.06 & 0.05 & 0.11 & 0.03 \\
& & RMSE & 0.13 & 0.04 & 0.05 & 0.05 & 0.10 & 0.03 \\
  \cline{2-9}
& 0.20 & Bias & -0.00 & -0.00 & 0.02 & -0.01 & 0.01 & 0.00 \\ 
& & SD & 0.11 & 0.04 & 0.05 & 0.05 & 0.09 & 0.03 \\ 
& & $\widehat{\mbox{SD}}$ & 0.12 & 0.04 & 0.06 & 0.05 & 0.09 & 0.03 \\
\hline 
\end{tabular}

\begin{figure}[p]
\centering
\begin{tabular}{l|c|l|rrrrrr}
Copula & $\tau_{K}$ & & $\tau_{K}$ & $p$ & $\theta_{U1}$ & $\theta_{U2}$ & $\theta_{C1}$ & $\theta_{C2}$ \\ 
  \hline
Joe & 0.00 & Bias & 0.06 & 0.03 & 0.04 & -0.00 & -0.05 & 0.02 \\
& & SD & 0.08 & 0.04 & 0.07 & 0.05 & 0.07 & 0.04 \\
& & $\widehat{\mbox{SD}}$ & 0.07 & 0.04 & 0.07 & 0.05 & 0.06 & 0.04 \\
& & RMSE & 0.09 & 0.04 & 0.08 & 0.05 & 0.08 & 0.04 \\
  \cline{2-9}
& 0.20 & Bias & -0.01 & -0.00 & 0.02 & -0.00 & 0.01 & -0.00 \\
& & SD & 0.09 & 0.04 & 0.06 & 0.05 & 0.07 & 0.03 \\ 
& & $\widehat{\mbox{SD}}$ & 0.08 & 0.04 & 0.06 & 0.05 & 0.07 & 0.03 \\
\hline 
\end{tabular}
\end{figure}

\begin{figure}[p]
\centering
\begin{tabular}{l|c|l|rrrrrr}
Copula & $\tau_{K}$ & & $\tau_{K}$ & $p$ & $\theta_{U1}$ & $\theta_{U2}$ & $\theta_{C1}$ & $\theta_{C2}$ \\ 
  \hline
Clayton(270) & 0.00 & Bias & 0.10 & -0.02 & 0.02 & -0.02 & 0.13 & -0.03 \\ 
& & SD & 0.12 & 0.04 & 0.06 & 0.05 & 0.18 & 0.06 \\ 
& & $\widehat{\mbox{SD}}$ & 0.15 & 0.04 & 0.07 & 0.04 & 0.22 & 0.07 \\
& & RMSE & 0.16 & 0.04 & 0.05 & 0.05 & 0.22 & 0.06 \\
  \cline{2-9}
& -0.20 & Bias & -0.00 & -0.02 & -0.04 & 0.09 & 0.06 & -0.00 \\
& & SD & 0.00 & 0.00 & 0.00 & 0.00 & 0.00 & 0.00 \\ 
& & $\widehat{\mbox{SD}}$ & 0.14 & 0.04 & 0.04 & 0.05 & 0.17 & 0.07 \\
\hline 
\end{tabular}
\end{figure}

\begin{figure}[p]
\centering
\begin{tabular}{l|l|c|l|rrrrrr}
Copula & Copula & \multicolumn{8}{c}{} \\
data gen. & estim. & $\tau_{K}$ & & $\tau_{K}$ & $p$ & $\theta_{U1}$ & $\theta_{U2}$ & $\theta_{C1}$ & $\theta_{C2}$ \\
  \hline
Gumbel & Frank & 0.20 & Bias & -0.10 & -0.03 & -0.02 & -0.00 & 0.09 & -0.02 \\
& & & SD & 0.09 & 0.03 & 0.04 & 0.05 & 0.08 & 0.03 \\
& & & $\widehat{\mbox{SD}}$ & 0.11 & 0.04 & 0.05 & 0.05 & 0.09 & 0.03 \\
& & & RMSE & 0.13 & 0.04 & 0.04 & 0.04 & 0.11 & 0.03 \\
  \cline{3-10}
& & 0.50 & Bias & -0.14 & -0.04 & -0.03 & 0.02 & 0.09 & 0.01 \\
& & & SD & 0.09 & 0.03 & 0.04 & 0.05 & 0.07 & 0.03 \\
\hline
\end{tabular}
\end{figure}

\subsubsection{Sample size 2500}

\begin{tabular}{l|c|l|rrrrrr}
Copula & $\tau_{K}$ & & $\tau_{K}$ & $p$ & $\theta_{U1}$ & $\theta_{U2}$ & $\theta_{C1}$ & $\theta_{C2}$ \\ 
  \hline
Frank & 0.00 & Bias & 0.04 & -0.00 & -0.00 & -0.00 & 0.03 & 0.03 \\
& & SD & 0.00 & 0.00 & 0.00 & 0.00 & 0.00 & 0.00 \\
& & $\widehat{\mbox{SD}}$ & 0.08 & 0.03 & 0.03 & 0.03 & 0.06 & 0.02 \\
& & RMSE & 0.04 & 0.00 & 0.00 & 0.00 & 0.03 & 0.03 \\
  \cline{2-9}
& 0.20 & Bias & -0.00 & -0.00 & 0.00 & -0.00 & 0.00 & 0.00 \\  
& & SD & 0.07 & 0.03 & 0.03 & 0.03 & 0.05 & 0.02 \\ 
& & $\widehat{\mbox{SD}}$ & 0.08 & 0.03 & 0.03 & 0.03 & 0.05 & 0.02 \\
\hline 
\end{tabular}

\begin{figure}[p]
\centering
\begin{tabular}{l|c|l|rrrrrr}
Copula & $\tau_{K}$ & & $\tau_{K}$ & $p$ & $\theta_{U1}$ & $\theta_{U2}$ & $\theta_{C1}$ & $\theta_{C2}$ \\ 
  \hline
Joe & 0.00 & Bias & 0.04 & 0.01 & 0.02 & -0.00 & -0.03 & 0.01 \\
& & SD & 0.05 & 0.03 & 0.04 & 0.03 & 0.05 & 0.03 \\
& & $\widehat{\mbox{SD}}$ & 0.05 & 0.03 & 0.04 & 0.03 & 0.05 & 0.03 \\
& & RMSE & 0.06 & 0.03 & 0.04 & 0.03 & 0.05 & 0.03 \\
  \cline{2-9}
& 0.20 & Bias & -0.00 & -0.00 & 0.00 & -0.00 & 0.00 & 0.00 \\ 
& & SD & 0.05 & 0.02 & 0.04 & 0.03 & 0.04 & 0.02 \\ 
& & $\widehat{\mbox{SD}}$ & 0.05 & 0.03 & 0.04 & 0.03 & 0.04 & 0.02 \\
\hline 
\end{tabular}
\end{figure}

\begin{figure}[p]
\centering
\begin{tabular}{l|c|l|rrrrrr}
Copula & $\tau_{K}$ & & $\tau_{K}$ & $p$ & $\theta_{U1}$ & $\theta_{U2}$ & $\theta_{C1}$ & $\theta_{C2}$ \\ 
  \hline
Clayton(270) & 0.00 & Bias & 0.00 & -0.02 & -0.04 & 0.02 & -0.07 & 0.00 \\
& & SD & 0.00 & 0.00 & 0.00 & 0.00 & 0.00 & 0.00 \\
& & $\widehat{\mbox{SD}}$ & 0.16 & 0.03 & 0.03 & 0.03 & 0.16 & 0.05 \\
& & RMSE & 0.00 & 0.00 & 0.03 & 0.00 & 0.07 & 0.00 \\
  \cline{2-9}
& -0.20 & Bias & 0.02 & 0.00 & 0.00 & 0.00 & -0.01 & 0.01 \\ 
& & SD & 0.11 & 0.03 & 0.03 & 0.03 & 0.12 & 0.05 \\ 
& & $\widehat{\mbox{SD}}$ & 0.16 & 0.04 & 0.04 & 0.04 & 0.16 & 0.06 \\
\hline 
\end{tabular}
\end{figure}

\begin{figure}[p]
\centering
\begin{tabular}{l|l|c|l|rrrrrr}
Copula & Copula & \multicolumn{8}{c}{} \\
data gen. & estim. & $\tau_{K}$ & & $\tau_{K}$ & $p$ & $\theta_{U1}$ & $\theta_{U2}$ & $\theta_{C1}$ & $\theta_{C2}$ \\
  \hline
Gumbel & Frank & 0.20 & Bias & -0.09 & -0.03 & -0.02 & 0.00 & 0.08 & -0.02 \\
& & & SD & 0.06 & 0.02 & 0.03 & 0.03 & 0.05 & 0.02 \\
& & & $\widehat{\mbox{SD}}$ & 0.07 & 0.03 & 0.03 & 0.03 & 0.06 & 0.02 \\
& & & RMSE & 0.11 & 0.03 & 0.03 & 0.03 & 0.09 & 0.03 \\
  \cline{3-10}
& & 0.50 & Bias & -0.14 & -0.04 & -0.04 & 0.03 & 0.08 & 0.01 \\
& & & SD & 0.06 & 0.02 & 0.02 & 0.03 & 0.04 & 0.02 \\
\hline
\end{tabular}
\end{figure}